\documentclass[ppcf]{iopjournal}

\usepackage{multirow}
\usepackage{mhchem}
\usepackage{graphicx}
\usepackage[numbers]{natbib}
\begin{document}

\articletype{Paper} 

\title{The fixed boundary plasma equilibrium basis for a one Gigawatt electric stellarator power plant}

\author{Samuel A. Lazerson$^{1*}$\orcid{0000-0001-8002-0121}, Ant\'onio J. Coelho$^1$\orcid{0000-0002-8889-7259}, Dana Douqa$^1$\orcid{0000-0001-8245-7618}, Adrian Asmund Fessler$^{1,2}$\orcid{0009-0002-5469-0759}, Luisa H\"ubner$^{1,3}$\orcid{0009-0001-8373-1161}, Matteo Moscheni$^1$\orcid{0000-0002-6355-7274}, Eoin Hodge$^1$, Richard Kembleton$^4$, Joona Sissonen$^{5}$\orcid{0000-0002-8662-7351}, Konsta S\"arkim\"aki$^{5}$\orcid{0000-0001-9156-2559}, and Antti Snicker$^{5}$\orcid{0000-0001-9604-9666}}

\affil{$^1$Gauss Fusion GmbH, Garching bei München, Germany}\\
\affil{$^2$Karlsruhe Institute of Technology, Karlsruhe, Germany}\\
\affil{$^3$Technical University of Denmark, Kgs. Lyngby, Denmark}\\
\affil{$^4$Research Instruments GmbH, Bergisch Gladbach, Germany}\\
\affil{$^5$VTT Technical Research Centre of Finland Ltd., Espoo, Finland}

\affil{$^*$Author to whom any correspondence should be addressed.}

\email{samuel.lazerson@gauss-fusion.com}


\begin{abstract}
A fixed boundary stellarator equilibrium capable of producing 3 GW of fusion power (1 GW-electric) is presented as the design basis for the GIGA fusion power plant being developed by Gauss Fusion GmbH.
The stellarator concept provides a steady-state, transient free, low recirculating power approach to a fusion power plant, which builds on 50 years of progress in plasma physics.
A set of requirements for a fixed boundary equilibrium were determined through application of 0.5 D modeling.
Optimization of a modified Wendelstein 7-X (W7-X) equilibrium was performed to achieve these requirements including alpha power confinement greater than 85\%, neoclassical effective ripple below 0.01, bootstrap current below 50 kA, and reduced turbulent heat fluxes.
In order to fix the plasma volume of $1500~\textrm{m}^3$ during optimization, the VMEC code was modified to renormalize the boundary coefficient to the desired plasma volume.
The STELLOPT stellarator optimization code was modified as well to include new bootstrap current targets, a new target for the radial electric field, and the capability to hold the magnetic field on axis at a fixed value.
An intermediary conceptual design plasma and final evolved fixed boundary equilibria are compared to the original modified W7-X equilibrium.
The final evolved equilibrium is shown to achieve all the necessary requirements for the GIGA fusion power plant through more detailed modeling of stability, fast ion confinement, and transport.
\end{abstract}

\section{Introduction}
In this work, the fixed boundary plasma equilibrium basis for a Gigawatt class stellarator fusion power plant (GIGA), being developed by Gauss Fusion GmbH, is presented through the language of systems engineering \cite{GFG_CDR_exec_2025}.
The stellarator concept \cite{spitzer_stellarator_1958} provides a stable, steady-state, transient-free path to power plant operation, where decades of progress in plasma physics and design have shown consistent progress toward this goal.
This choice however implies that the plasma physics (and plasma performance) is fundamentally a function of the plasma shape (the plasma design) \cite{boozer_stellarator_2015}, whereas the challenge of the tokamak is fundamentally one of control.
Throughout this text we will use the term performance to imply the ratio between plasma parameters (temperatures and densities) at a given total heating power.
In the design of this fixed boundary equilibrium an analogy can be made to engineering design, where the geometric and material properties of a component determine the performance of the system. 
In much the same way, the plasma design (shape) determines the performance of the stellarator.
Therefore in this work we invoke a systems engineering approach for the design and description of the GIGA fixed boundary equilibrium.

Magnetic confinement fusion is the most technically mature concept for fusion energy production with the stellarator being the most compatible with the requirements of a power plant.
Superconducting magnet performance has already reached the technical maturity needed for a power plant system as compared to the technical maturity of laser systems \cite{donne_beyond_2025}. 
Additionally, the majority of systems at low levels of technical maturity are common among all types of fusion power plant devices (such as tritium breeding).
At the system level, the design maturity for magnetically confined devices is fairly high, with decades of power plant design studies having been performed.
In particular, the ARIES-CS \cite{najmabadi_aries-cs_2008,ku_physics_2008}, EU-DEMO \cite{kemp_dealing_2017} and HELIAS \cite{grieger_helias_1994,warmer_system_2016} design studies provide a significant basis on which to make informed technical scoping decisions.
The steady-state, transient-free nature of stellarators make them a favorable choice for a power plant.
This is in contrast to the tokamak which is inherently pulsed and has only been able to mitigate transients through additional systems adding complexity to an already complex system \cite{maris_impact_2024}.
The major argument against stellarators is their complexity of design, requiring significant up-front effort to achieve good plasma performance.
This can be proven in literate with the large number of publications clearly demonstrating the large effort required to design a stellarator plasma \cite{sanchez_quasi-isodynamic_2023,goodman_quasi-isodynamic_2024,lion_stellaris_2025,anderson_comprehensive_2025,sanchez_ciemat-qi4x_2026,swanson_overview_2026}.
These studies provide state-of-the-art analysis confirming their design points, within assumptions. 
In choosing the stellarator we must tackle the up front cost of designing a more complex system, avoiding the later cost of a more demanding plasma control environment.

The design of stellarators is dictated by the need to find a plasma shape which achieves a set of plasma physics goals (among which is the plasma performance or confinement).
Stellarator equilibrium theory states that, in the presence of nested flux surfaces, the shape of the flux surface, pressure profile, and toroidal current profile determine the magnetic field everywhere inside a given surface.
This is known as the ideal limit, whereas in the resistive limit flux surfaces are not guaranteed and the problem is much more complex. 
For the purposes of a power plant, a low (or zero) toroidal plasma current with nested flux surfaces is desired, as this significantly improves particle confinement and the robustness of the plasma edge. 
Toroidal current has been shown to shift strike lines in Wendelstein 7-X (W7-X), which if left unmitigated would overload divertor components \cite{gao_effects_2019}.
Experimentally the minimization of bootstrap current through plasma shaping has been demonstrated on the W7-X device \cite{geiger_effects_2010,neuner_measurements_2021} (along with the nested flux surfaces \cite{pedersen_confirmation_2016}), where in the absence of external current drive only the bootstrap current remains \cite{erckmann_current_1992}.
The pressure profile itself arises from the plasma temperatures and densities, which also determine the total fusion power.
Experimental evidence, scaling laws, and fusion power requirements provide a basis by which to determine the volume averaged pressure profile.
If one assumes a pressure profile based on experimental evidence, only the plasma shape needs to be specified.
The process by which the stellarator shape is chosen to achieve a given set of plasma characteristics is known as stellarator optimization.
The optimization process makes use of predictive simulations where possible. 
Where such simulations are too computational expensive to incorporate into the optimization, proxy functions are utilized.
Over the decades many transport optimized stellarators have been proposed with the HSX \cite{anderson_overview_2006,talmadge_experimental_2008}, W7-X \cite{wolf_major_2017,klinger_overview_2019,grulke_overview_2026} and CFQS \cite{cheng_construction_2025} devices being the current experimental realizations.
Once a stellarator shape is determined more sophisticated simulation models can then be applied to confirm plasma performance.

Stellarator plasma design can be nicely described with the language of systems engineering. 
For this reason, this paper is structured using such language. 
In the system description (section \ref{sec:systemdesc}), the background physics information, plasma functionality, and functionality breakdown of the plasma are discussed. 
Such information falls between introductory material and problem methodology of a traditional scientific paper. 
In the design basis (section \ref{sec:designbasis}), the tools, requirements capture, and system interdependencies are presented. 
While aligned with a methodology section, this section goes beyond problem definition to consider how the plasma affects and is affected by other systems. 
The concept definition (section \ref{sec:conceptdef}) presents the plasma design highlighting any critical parameters arising from design choices, making an analogy between engineering component design and stellarator plasma design.
As the shape of an engineered component can define its properties (fatigue life, maximum supported forces, maximum temperatures), so does the stellarator plasma shape define its properties (confinement, stability, rotational transform, etc.).
The qualification record (section \ref{sec:qualirec}) presents the more detailed physical analysis of plasma design, quantifying the extent to which the plasma has met its requirements (as determined in the design basis). 
In this stage, models not capable of being run in the stellarator optimization loop are run to verify that the plasma meets the design requirements. 
In the qualification plan (section \ref{sec:qualiplan}), the current state of the technical maturity of the plasma design is presented along with the plan for development in terms of its technical readiness level (TRL). 
Translation of engineering TRLs into the terminology of physics is presented to better define the scope of future test facilities. 
Finally, the paper is ended with a conclusions section.

\section{System Description} \label{sec:systemdesc}

In the most basic terms, a stellarator is a toroidal magnetic confinement device in which a plasma composed of charged particles is confined by a magnetic field which arises solely from a set of currents external to the plasma.
The challenge in stellarator design arises from the fact that an arbitrary set of external currents does not necessarily give rise to a magnetic field which confines charged particles in a given volume, let alone a plasma.
Even when magnetic fields can be found which confine a fraction of the charged particles to a plasma volume, not all classes of particles are guaranteed to be confined.
Further complicating the picture are the non-linear interactions between the plasma and magnetic fields which result in heat and particle transport outside the confining volume at rates far exceeding simple diffusive transport estimates.
The problem of stellarator design is generally broken into two design stages.  
In the first stage, the shape of magnetic flux surfaces which give rise to favorable plasma properties is found.
In the second stage, a set of external currents which give rise to an adequate approximation of the desired plasma shape is found.
This work concerns itself with the first stage, finding a plasma shape with desirable plasma properties in the context of a fusion power plant.
It should be noted that so-called 'single-stage' approaches attempt both problems simultaneously but at added computational complexity \cite{wechsung_single-stage_2022,giuliani_single-stage_2022,jorge_single-stage_2023,smiet_efficient_2025}.

\subsection{Stellarator Plasma Physics}
A core step in stellarator optimization work is solving the three-dimensional equilibrium problem, which allows us to evaluate the magnetic field for a given plasma shape, pressure profile, and current (or rotational transform profile). 
Such computations underpin the calculation of nearly all plasma physics properties. 
The magnetohydrodynamic (MHD) stability of the plasma provides an evaluation of how perturbations of the plasma are either damped or grow in time \cite{kovrizhnykh_mhd_1983,shafranov_magnetohydrodynamic_1983,bauer_nonlinear_1981}. 
As particles mirror in the device, collisional transport associated with this motion (and the presence of temperature and density gradients) is described as neoclassical \cite{ho_neoclassical_1987,maassberg_neoclassical_1993}.
This is as opposed to classical transport, which deals with how gradients in temperature and density drive motion across a uniform magnetic field in the presence of collisions \cite{hinton_collisional_1983}. 
In addition to driving transport, the mirroring motion of particles can couple to the passing particle population collisionally giving rise to a toroidal current called the bootstrap current \cite{shaing_bootstrap_1989,erckmann_current_1992,hsu_bootstrap_1992,ware_bootstrap_2006,helander_bootstrap_2009}.
This current is generally considered undesirable in stellarators as it can destabilize the plasma and negatively affect plasma-wall interactions.
Unlike in tokamaks the ambipolarity condition is not implicitly fulfilled in stellarators \cite{landreman_effects_2011,landreman_neoclassical_2011}.
The ambipolarity condition must be satisfied through the presence of a radial electric field arising from neoclassical effects. Recent work has suggested that a core positive electric field may be achievable under reactor conditions, expelling impurities from the core while pulling fuel ions in \cite{lee_direct_2024,beidler_reduction_2024}. In addition to neoclassical transport, drift waves in the plasma can grow in time resulting in so-called turbulent transport \cite{horton_drift_1999,dimits_comparisons_2000, brizard_foundations_2007,dominguez_dissipative_1992}. Turbulent transport ultimately sets plasma performance in all toroidal magnetic confinement devices by limiting the growth of temperature and density gradients. In the burning plasma state of a reactor, the energetic alphas must be confined long enough to transfer their kinetic energy via collisions to the fuel ions \cite{lotz1990optimization,lazerson_simulating_2021,bader_modeling_2021}. Additionally, unconfined energetic particles can overload and damage first wall structures \cite{kurki-suonio_protecting_2016,cornelissen_identification_2022,kulla_placement_2022}. Thus confinement of the energetic particles is a key metric for design of a fusion power plant. Stellarator optimization involves computing the equilibrium, evaluating these physical properties, and using the algorithms of numerical optimization to find improved configurations. 

The theory of MHD equilibria (or more accurately magnetohydrostatic, MHS, equilibria) fundamentally involves solving for the ideal plasma force balance. This is written as:
\begin{equation}
\vec{j}\times\vec{B} - \nabla p =0
\end{equation}
where $\vec{j}$ is the current density, $\vec{B}$ the magnetic field (technically the magnetic induction), and $p$ the plasma pressure. Generally, solutions to this equation can be grouped by the assumption of ideal and non-ideal physics. In ideal MHS, terms which allow for diffusion of the magnetic field are assumed zero. This generally implies that pressure gradients are purely radial and magnetic flux surfaces are volume filling. Codes such as VMEC \cite{hirshman_steepest-descent_1983}, DESC \cite{dudt_desc_2020}, GVEC \cite{hindenlang_gvec_2026}, and NSTAB \cite{taylor_high_1994} solve for such solutions given a three dimensional plasma boundary shape. In the non-ideal case, resistive terms are considered which allow for the formation of magnetic islands and stochastic regions. Code such as SIESTA \cite{hirshman_siesta_2011}, HINT2 \cite{suzuki_development_2006}, and M3D-C1S \cite{zhou_approach_2021} can provide such solutions, employing a variety of algorithmic, physical, and numerical models. Bridging the gap of the two models is multi-region relaxed magnetohydrodynamics (MRxMHD) which assumes a finite set of interfaces exist (flux surfaces), supporting pressure gradients across them \cite{hole_stepped_2006}. Between such regions one finds Beltrami-like fields which satisfy the equation
 \begin{equation}
\nabla\times\vec{B} = \mu\vec{B}
\end{equation}
where $\mu$ is a scalar defining the current density parallel to the magnetic field. The SPEC code is the only code which currently solves for MRxMHD equilibrium states \cite{hudson_computation_2012}. The ideal equilibrium codes have been used in the vast majority of optimization tools, encoding the magnetic field as plasma shape.

The validity of the ideal nested flux surface model of the plasma is worth calling into question before moving on. 
Questions regarding the validity of ideal MHD in three dimensions have motivated some aspect of MRxMHD research \cite{loizu_magnetic_2015,loizu_existence_2015,loizu_verification_2016}.
Similarly, these questions have also motivated extensive validation exercises of the VMEC code \cite{ramasamy_modeling_2022,ramasamy_how_2023}.
Experimental evidence from the Large Helical Device (LHD) suggests that the presence of islands may in fact require neoclassical physics to be taken into account \cite{hegna_kinetic_2011,hegna_healing_2011,hegna_plasma_2012,narushima_observations_2017}. 
MRxMHD predicts that jump discontinuities in the magnetic field can result in (formerly rational) magnetic surfaces supporting a pressure gradient through discontinuous rotational transform ($\iota$). 
This essentially allows the rotational transform to be multi-valued and thus no longer rational.
Simulations attempting to explore such phenomena in VMEC suggest that while the code lacks an explicit multi-valued rotational transform feature, a 3D current density forms at the rational surface consistent with a shielding response \cite{lazerson_verification_2016}. 
Attempts to validate against experiments in tokamaks seem to suggest that a variety of physical models are experimentally indistinguishable \cite{king_experimental_2015}. 
Most stellarator designs attempt to avoid low order rational magnetic surfaces, thereby avoiding a necessary condition for the formation of magnetic islands in the plasma. 
Additionally, rotational transform shear ($d\iota/d\rho$) is maximized in order to control island sizes should they appear in the plasma.
In the context of this work, $\rho=r/a$ is the normalized minor radius of the plasma and $s=\Phi/\Phi_{edge}=\rho^2$ is the normalized toroidal flux ($\Phi$ is the toroidal flux and $\Phi_{edge}$ is the toroidal flux at equilibrium boundary).
The $r$ variable is the flux surface averaged distance from the magnetic axis and $a$ is this value at the plasma boundary.
The flux surface average of a quantity $X$ is given by:
\begin{equation}
<X>=\frac{\int_0^{2\pi}\int_0^{2\pi}\sqrt{g}Xd\theta d\zeta}{\int_0^{2\pi}\int_0^{2\pi}\sqrt{g}d\theta d\zeta}
\end{equation}
where $\theta$ is the poloidal angle, $\zeta$ is the toroidal angle, and $\sqrt{g}$ is Jacobian.

The stability of an equilibrium plasma state is generally determined from application of perturbation theory. 
Two of the most elementary metrics for plasma stability are the Mercier criterion and magnetic well \cite{mercier_equilibrium_1964,greene_brief_1998,landreman_magnetic_2020}.  
While both provide some theoretical underpinnings for plasma stability, neither can guarantee the stability of a three-dimensional plasma to ballooning \cite{tang_kinetic-ballooning-mode_1980,cooper_spectrum_1996}, kink \cite{kruskal_equilibrium_1958}, or Alv\'enic modes \cite{cheng_lown_1986,fu_excitation_1989}. 
Ballooning modes are internal pressure-driven instabilities that appear in unfavorable magnetic field curvature regions. 
The so-called infinte-n modes are generally assumed to be the plasma beta ($\beta = 2\mu_0p/B^2$, where $\mu_0$ is the permittivity of free space) limiting phenomena in current-free (or low current) stellarators \cite{ohdachi_observation_2017}.
Computation of the stability of these modes involves solving a 2nd order ordinary differential equation of the form: 
\begin{equation}
\left[L_0\left(y\right)+\lambda R\left(y\right)\right]=0,
\end{equation}
\begin{equation}
L_0=\frac{d}{dy}\left[P\left(y\right)\frac{d}{dy}\right]+Q\left(y\right).
\end{equation}
where $y$ is a coordinate along a field line, $P\left(y\right)$ is the field line bending term, $Q\left(y\right)$ the pressure curvature drive term, $R\left(y\right)$ an inertial drive term involving geometry and mass, and $\lambda$ the eigenvalue.
Equations of this form can be computationally expensive to evaluate.
The COBRA code \cite{sanchez_cobra_2000} makes use of a variational principle to speed up computation and has been verified and validated against full solvers \cite{ham_modelling_2014,willensdorfer_dynamics_2019}. 
Kink modes are those with finite toroidal mode numbers, and are generally computed from a linearized variational form of the equilibrium equation:
\begin{equation}
\delta W_p +\delta W_v -\omega^2\delta W_k = 0
\end{equation}
where the $W_p$, $W_v$, and $W_k$ are the perturbed plasma, vacuum and kinetic energies respectively. 
In axisymmetric systems, these modes couple in the poloidal direction but not toroidally (hence their finite toroidal nature). 
In stellarators, the toroidal modes ($n$) couple into families based on the field periodicity ($N_{fp}$) of the plasma ($n\pm N_{fp}$). 
For example a device with 4 field periods, such as GIGA, has $n=0$, $n=1$, and $n=2$ mode families.  
Codes such as TERPSICHORE \cite{anderson_terpsichore_1990} and CAS3D \cite{schwab_ideal_1993} provide computation of the stability of such modes for stellarator equilibria. 
The drive for Alv\'enic modes is usually associated with energetic particle populations and is thus a concern for fusion reactors (although turbulence has also been shown to destabilize such modes as well \cite{rahbarnia_alfvenic_2021,vaz_mendes_broadband_2023,riemann_excitation_2025}). 
Computation of the stability of such modes requires significant computational effort with codes such as AE3D \cite{2016doe_soft_89S} and CKA-EUTERPE \cite{slaby_perturbative_2024}. 
However, computation of the gap structure of modes is quite tractable with codes such as STELLGAP \cite{spong_shear_2003}. 
Thus the general metric for stellarators is to ensure no large core to edge gaps exist, which would allow large radial redistribution of fast ions. 
 
\begin{figure}
 \centering
        \includegraphics[width=\textwidth]{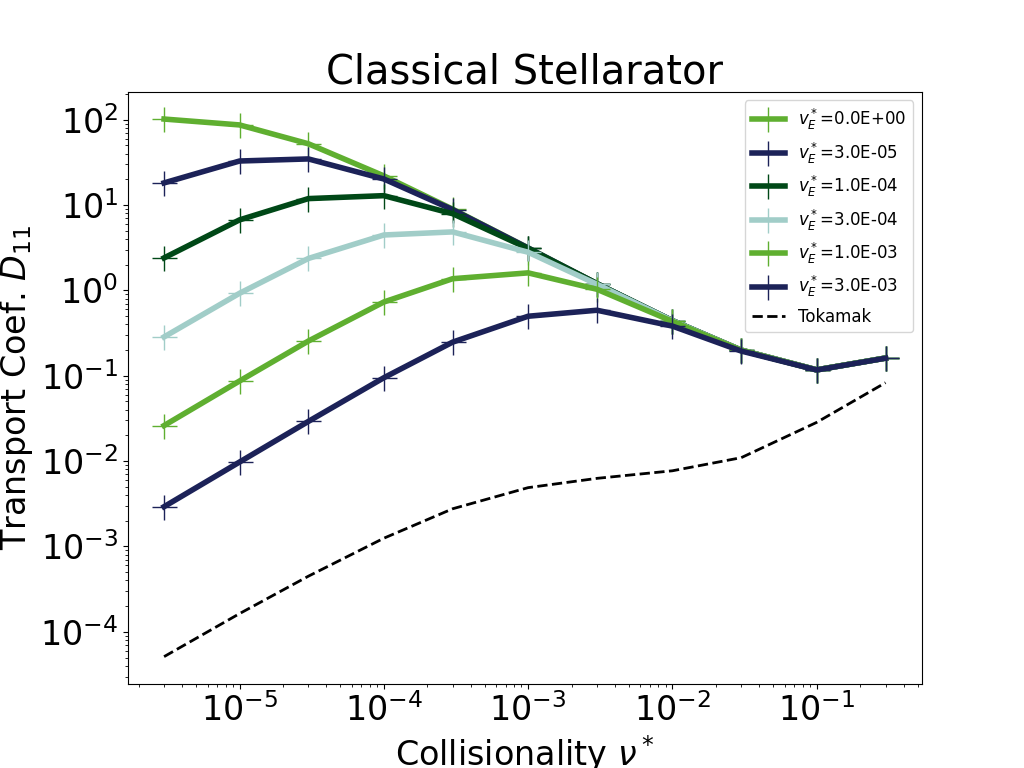}
 \caption{Comparison of mono-energetic neoclassical particle transport coefficients ($D_{11}$) for a classical stellarator to an equivalent tokamak. The x-axis is depicted as normalized collision frequency ($\nu^*=R_0\nu/\iota v$), where $R_0$ is the reference major radius, $\nu$ the collisionality, $\iota$ the rotational transform, and $v$ the particle velocity. Multiple lines are plotted for the stellarator at various choices of normalized electric field $\nu^*_E=E_r/vB_0$, where $E_r$ is the radial electric field, $v$ the particle velocity, and $B_0$ the reference magnetic field strength. The stellarator clearly shows higher transport coefficients indicating reduced confinement at low collisionality.}
\label{fig:neo_example}
\end{figure}

Neoclassical transport refers to transport associated with the mirroring motion of particles in a toroidal magnetic field \cite{galeev_transport_1968}. 
Tokamaks have low levels of neoclassical transport, as the bounce averaged radial motion of particles is zero (in the absence of toroidal ripple).
Stellarators generally do not posses such a property with bounce averaged drifts generally resulting in a large radially averaged motion of particles \cite{ho_neoclassical_1987}.
It is this cross-field transport which one wishes to minimize.
This phenomenon limited early stellarator performance where a $1/\nu$ regime at low collisionality ($\nu$) made a stellarator reactor appear impossible.
Figure \ref{fig:neo_example} depicts the neoclassical particle transport coefficients ($D_{11}$) for a classical stellarator showing this effect.
The neoclassical transport in stellarators is also sensitive to electric fields, resulting in the multiple lines (solutions).
However, it was recognized that magnetic fields could be produced in 3D configurations which had good bounce averaged drifts, or at least approximately so (quasi-omnigenity) \cite{galeev_plasma_1969}.
This has led to stellarators with quasi-axisymmetric \cite{garabedian_quasi-axially_1998}, quasi-helically symmetric \cite{boozer_quasi-helical_1995}, quasi-poloidally symmetric \cite{spong_physics_2001} designs with significantly reduced neoclassical transport.
Furthermore, direct optimization of neoclassical transport was made possible with the NEO \cite{nemov_evaluation_1999} and DKES codes \cite{van_rij_variational_1989}.
The experimental realization of neoclassical optimization has been achieved in the HSX, W7-X, and CFQS devices.

Key to the development of neoclassically optimized stellarators has been the development of the magnetic symmetries.
It was posited that the poor neoclassical confinement of stellarators could be attributed to the drift orbits of trapped particles not closing inside a given stellarator.
Thus if magnetic fields could be produced where such drift orbits close inside the device, confinement would improve.
For axisymmetric configurations such orbits close by construction.
A magnetic field where the bounce averaged radial drift for all trapped particles vanishes is known as omnigenous \cite{cary_omnigenity_1997} or piecewise omnigenous \cite{velasco_piecewise_2024}.
One way such conditions can be achieved is through magnetic fields which depend on a single angular variable such that $B=B\left(\psi,M\theta-N\zeta\right)$ in straight field lines coordinates \cite{boozer_establishment_1982}.
Here $\psi$ is the flux surface label, $\theta$ is the poloidal angle, $\zeta$ a toroidal angle, and $(M,N)$ are fixed integers.
Fields which obey this relation are known as quasi-symmetric \cite{boozer_transport_1983} with $N=0$ being quasi-axisymmetric \cite{garabedian_stellarators_1996}, $M=0$ quasi-poloidal symmetry \cite{spong_physics_2001}, and $(M,N\neq0)$ being quasi-helical symmetry \cite{nuhrenberg_quasi-helically_1988}.
A more stringent requirement than omnigeneity is that the radial drifts are identically zero everywhere on a surface, known as isodynamicity \cite{palumbo_considerations_1968}.
For omnigeneous configurations where the contours of constant magnetic field strength close poloidally the term quasi-isodynamicity (QI) has been used \cite{gori_beta_1997,subbotin_integrated_2006}.
More recently, methods for obtaining such magnetic fields through direct optimization have been demonstrated \cite{goodman_quasi-isodynamic_2024}.

Associated with neoclassical theory is the bootstrap current which arises from the collisional interaction of trapped particles with passing particles in the presence of gradients in the density and temperature \cite{bickerton_diffusion_1971}. 
The trapped particles moving co and counter to the direction of the magnetic field have different sized radial excursions. 
This results in an asymmetry in the distribution function which is opposite in sign for the ions and electrons (the banana current). 
The motion of the ions and electrons is dominated by ion collisions, with collisions driving asymmetry in the passing population as well. 
The asymmetry in the distribution function results in a net momentum along field lines, and a net parallel current (the bootstrap current).
This current is desirable in tokamaks as it minimizes the amount of external current drive required to maintain the plasma.
In stellarators the bootstrap current can destabilize equilibria and change plasma-wall interactions motivating its minimization. 
The sign of the current is found to be dependent on the magnetic spectrum with quasi-axisymmetric systems driving a bootstrap current parallel to the magnetic field (iota raising) and quasi-helically symmetric systems driving current anti-parallel to the magnetic field (iota lowering). 
Through a balance of these two effects stellarators with low bootstrap current can be achieved.
Computation of the bootstrap current in stellarators can be performed using BOOTSJ \cite{shaing_bootstrap_1989}, PENTA \cite{spong_generation_2005}, and SFINCS \cite{landreman_comparison_2014} codes. 
The time evolution of the plasma current requires a more sophisticated code like THRIFT \cite{van_ham_modeling_2025}, providing a self-consistency between the equilibrium and toroidal current density.

In stellarators the ambipolarity condition must be satisfied through the presence of a radial electric field ($E_r$).
This electric field arises from neoclassical effects where particle fluxes must be balanced.
The balance is captured by the following equation
\begin{equation}
\frac{\partial E_r}{\partial t}-\frac{1}{\partial V/\partial r}\frac{\partial}{\partial r}\left[ \frac{\partial V}{\partial r}D_E\left(\frac{\partial E_r}{\partial r}-\frac{E_r}{r}\right)\right]=\frac{|e|}{\epsilon_\perp}\left(\Gamma_e-Z_i\Gamma_i\right)
\end{equation}
where $V$ the plasma volume, $D_E$ is the electric field diffusion coefficient, $\epsilon_\perp$ is the dielectric constant, $\Gamma_e$ the radial electron particle flux, $Z_i$ the ion charge number, and $\Gamma_i$ the ion radial particle flux. 
We note that the fluxes themselves are functions of the electric field.
In practice, the left hand side is usually set to zero, and the roots to the resulting equation give the electric field.
The roots of this equation can be both stable and unstable, with stable roots associated to the ion or electron fluxes.
In the outer regions of the plasma, and usually in the core, the plasma is in the ion-root state with an inward pointing radial electric field (negative). 
However, in conditions where the electron temperature exceeds the ion temperature an electron root condition can be established in the core of the plasma with an outward pointing radial electric field (positive). 
Such a positive radial electric field is advantageous as the particle flux scales strongly with charge, flushing high charge number impurities from the core. 
Recently, stellarator configurations where this condition is possible at nearly equal ion and electron temperatures have been discovered \cite{fujita_study_2021, lee_direct_2024}. 
The SFINCS and PENTA codes provide a means to compute the radial electric field in a stellarator, given a set of temperature and density profiles.

Turbulence has been identified as the driving term for transport in both tokamaks \cite{liewer_measurements_1985,wootton_fluctuations_1990} and optimized stellarators \cite{wolf_major_2017,klinger_overview_2019,beurskens_ion_2021,carralero_role_2022}. 
Fundamentally, turbulence in toroidal devices deals with how perturbations experiencing drift motion, also known as drift waves, can become destabilized by temperature and density gradients, nonlinearly evolving into net radial particle and heat transport. 
Computation of such effects requires a gyro-kinetic treatment of the plasma which is computationally expensive and not without some level of fluctuation. 
Tools such as GENE \cite{jenko_electron_2000} and stella \cite{barnes_stella_2019} provide a means for computing the non-linear saturated states of turbulence in stellarators.
In such tools, computational complexity increases with increasing model sophistication. 
To address this, proxy functions for stellarator turbulence have been developed which have shown some promise in reducing turbulent transport through stellarator optimization \cite{mynick_optimizing_2010,xanthopoulos_controlling_2014,proll_tem_2016,lazerson_ion_2019,landreman_how_2025}.
These proxies focus on transport driven by ion temperature gradient and trapped electron mode instabilities.
Experimental realization of a turbulence optimized stellarator configuration has yet to be achieved. 
However, steepening of density gradients in W7-X by core fueling has shown suppression of turbulence \cite{bozhenkov_high-performance_2020}. 
Such suppression of turbulence by strong density gradients has been predicted for W7-X using the aforementioned tools \cite{xanthopoulos_turbulence_2020,thienpondt_influence_2024}.

The confinement of energetic particles in stellarators is critical as early designs had unacceptably large losses. 
The issue stems from the trapped particle orbits and non-conservation of toroidal angular momentum in three dimensional fields. 
Quasi-symmetric devices tend to solve this problem through the radial averaging of drift motion, essentially having contours of the second adiabatic invariant which close inside the plasma. 
While implementation of fully Monte-Carlo simulations inside an optimization loop have been proposed, such concepts tend to be computationally expensive and suffer from Monte-Carlo noise. 
Much like turbulence, a set of proxy functions which correlate well with trapped particle confinement have been proposed \cite{nemov_poloidal_2008,velasco_model_2021,leviness_energetic_2022}. 
Once an optimized configuration is established, confinement can be checked by codes like ASCOT \cite{varje_high-performance_2019}, BEAMS3D \cite{mcmillan_beams3d_2014}, and SCENIC \cite{jucker_integrated_2011}, with the first two allowing for particles to be traced to first walls of arbitrary complexity.
Recent work to include the edge plasma in these models has been performed \cite{kiviniemi_role_2025}. 
However, it should be noted that the reactor problem must treat energetic helium, where models for charge exchange (which can play a significant role \cite{ollus_simulating_2022}) have yet to be implemented in these codes.

The stellarator optimization problem is traditionally parameterized in terms of a $\chi^2$ functional
\begin{equation}
\chi^2\left(\vec{x}\right) = \sum_k \frac{\left(f_k\left(\vec{x}\right)-f_k^{target}\right)^2}{\sigma_k^2}
\end{equation}
where $\vec{x}$ are the independent variables in the optimization, $f_k$ are physical figures of merit, $f_k^{target}$ are the desired figure of merit values, $\sigma_k$ are inverse weights, and each $k$ represents a different figure of merit.
In the context of fixed boundary equilibrium optimization, the x-vector is composed of the boundary coefficients which define the plasma shape. 
The figures of merit are the various physical quantities one wishes minimized and the sigmas are relative inverse weighting factors. 
In this formulation the mathematics of non-linear curve fitting can be used to minimize $\chi^2$. 
In this work, the modified Levenberg-Marquardt \cite{lazerson_three-dimensional_2015} and the Genetic Algorithm with Differential Evolution \cite{genetic,mynick_exploration_2002} are utilized for minimization of $\chi^2$.

Missing from this discussion has been the plasma edge, or region between the last closed flux surface and first wall structures.
This has been on purpose as to address this edge region requires a self-consistent coil-plasma design, which is usually the second stage of the stellarator design process. 
As such, the design and assessment of the GIGA divertor is the subject of future work. 
However, the plasma design must at least be consistent with a given divertor concept. 
The three leading stellarator divertor concepts are the helical divertor \cite{ohyabu_large_1994}, island divertor \cite{feng_review_2022}, and non-resonant divertor \cite{bader_hsx_2017}. For the purposes of this work the island divertor concept is pursued as it is accessible with a modular coil design and provides a private flux / scrape off layer paradigm necessary for particle pumping. 
Thus the plasma equilibrium must have an edge rotational transform compatible with a low order rational outside the equilibrium last closed flux surface.

\subsection{The plasma functionality}

\begin{figure}
 \centering
        \includegraphics[width=0.49\textwidth]{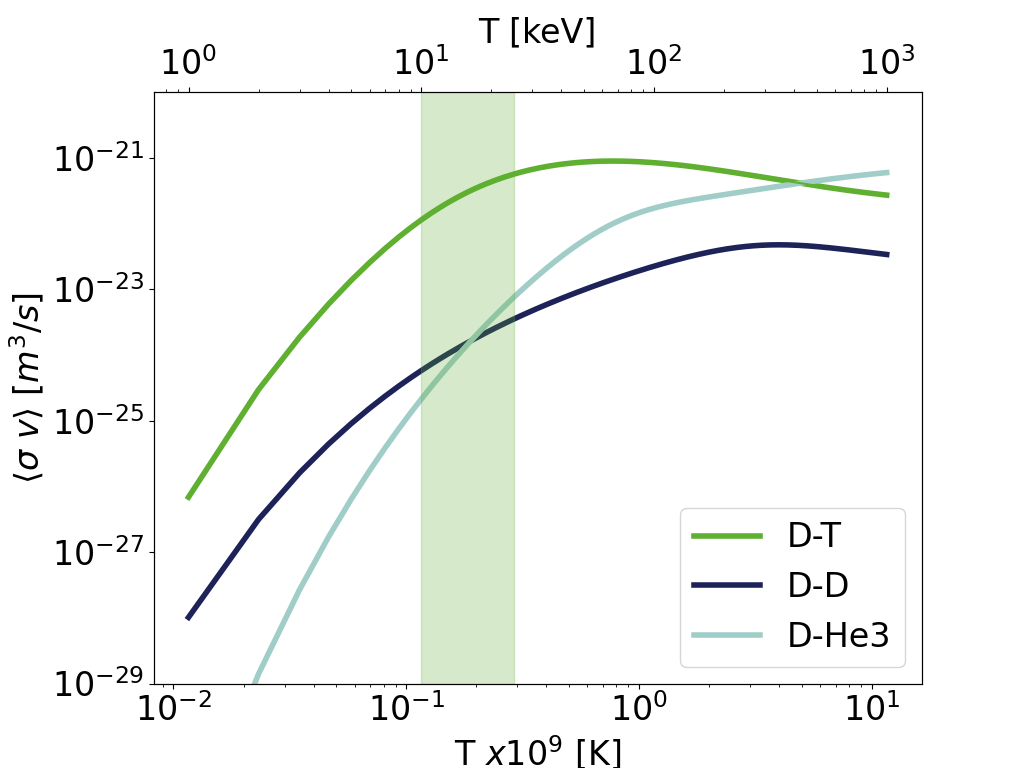}\includegraphics[width=0.49\textwidth]{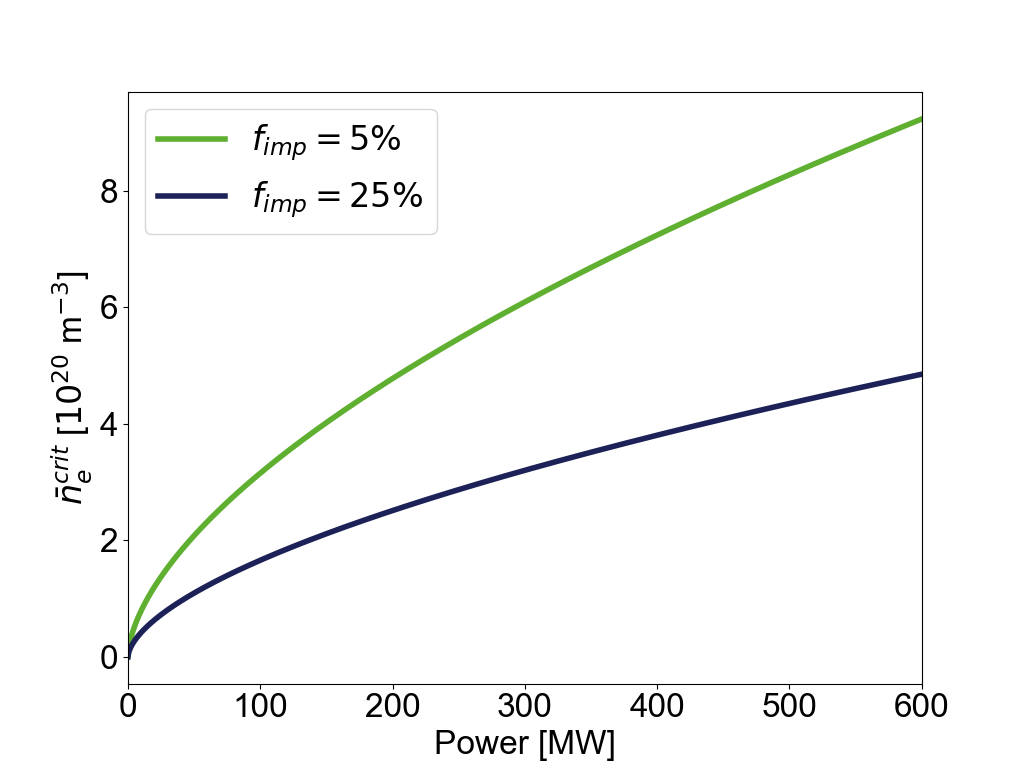}
 \caption{Fusion reaction rate cross sections for various fusion fuels (left) and radiative density limit (right) based on W7-X scalings for GIGA. The shaded region indicates the expected ion temperature regime of first generation toroidal fusion devices.}
\label{fig:fusion_temp}
\end{figure}

The primary purpose of the plasma, in the context of a power plant, is to be the reacting (confining) volume for the deuterium and tritium fuel, thereby producing neutrons. 
To achieve this, the magnetic fields must provide sufficient confinement such that the necessary densities and temperatures for thermonuclear fusion are achieved.
The fundamental reaction being
\begin{equation}
\textrm{D}+\textrm{T}\rightarrow \ce{^4_2He}~(\textrm{3.52 MeV})+\textrm{n}~(\textrm{14.06 MeV}).
\end{equation}
Secondary functionalities include the production of gammas, neutrons, and magnetic fields for the plasma control systems, along with providing a target for the auxiliary heating systems needed for plasma startup.
These are considered secondary functionalities as they are the not the reason we need the plasma, but rather consequences of the plasma presence and D-T reaction.
In this work it is assumed that the balance of plant implies a $\sim33\%$ conversion of total fusion power into electrical energy, an assumption broadly used in many fusion power plant studies \cite{tillack_use_2015}. 
Thus a 1 GW-electric (GWe) plant requires 3 GW of fusion power.  
As 80\% of the fusion energy is bound up in the 14.1 MeV neutrons, this leaves 600 MW of alpha power for plasma heating.

The primary functional requirement to produce neutrons implies certain density and temperature regimes be met. 
Figure \ref{fig:fusion_temp} shows the reaction cross sections for various fusion fuels as a function of fuel ion temperature. 
In first generation devices, ion temperatures in the range of 10 to 25 keV are to be expected. At higher temperatures the Bremsstrahlung and synchrotron radiation begin to play a dominant role, limiting plasma temperature. 
The range of plasma densities achievable depends on the plasma physics, in particular the impurity density and edge radiation behavior. 
One advantage of stellarators over tokamaks is the lack of a disruptive radiative density limit in stellarators \cite{gates_thermo-resistive_2016,helander_stellarator_2012}. 
Still, upper limits on plasma density in stellarators do exist. 
Recently work on W7-X has provided an empirical scaling regarding the average density and plasma impurity concentration \cite{fuchert_increasing_2020}.
In the empirical fit $f_{imp}=n_{imp}/n_e$ is the impurity concentration as defined in terms of the electron density ($n_e$).
While developed for a carbon walled machine, this limit provides a first upper estimate on achievable average plasma density. 
Using this empirical density limit model, we see that at full power (600 MW) densities between $2\times10^{20}~\textrm{m}^{-3}$ and $4\times10^{20}~\textrm{m}^{-3}$ are achievable even for the more pessimistic assumption of impurity concentration.

\section{Design Basis} \label{sec:designbasis}
In this section, the design basis for the GIGA power plant including tools, requirements capture and system interdependencies is discussed. 
The discussion of tools highlights not only the simulation tools used in the design of the GIGA plasma but also the optimization methodology. 
The requirements capture provides an overview of the plasma requirements by which the design is judged to be successful.
This includes quantification based on 0.5 D scaling. 
A brief discussion regarding the interfaces to other systems of the GIGA device is provided to help explain the scope of the problem.

\subsection{Tools}
The design of a stellarator plasma leverages integrated numerical simulations to produce a plasma which will meet a set of requirements. 
While many tools now exist for stellarator plasma optimization, the STELLOPT code \cite{doecode_12551} is employed in this work to determine a plasma shape which meets the plasma requirements. 
In cases where direct numerical simulation is too expensive or complex to include in an optimization loop, proxy functions are employed. 
For this reason, {\it a posteriori} checks are needed by more complex tools.

\begin{table}
\caption{Functionals used in the development of the GIGA equilibria along with their associated sources. The table is divided into those functionals always present (top) and those sometimes present (bottom).}
\centering
\begin{tabular}{l c }
\hline
Target & Source \\
\hline
Ballooning Stability & COBRAVMEC \cite{sanchez_cobra_2000} \\
Fast Ion Confinement & $\Gamma_C$ metric \cite{nemov_poloidal_2008} \\
Rotational Transform & VMEC  \cite{hirshman_steepest-descent_1983}\\
Turbulent Transport & $g^{rr}$ proxy  \cite{mynick_reducing_2011}\\
\hline
Total Bootstrap Current & BOOTSJ \cite{shaing_bootstrap_1989} \\
$1/\nu$ Transport & NEO \cite{nemov_evaluation_1999} \\
Quasi-isodynamic metric & STELLOPT \cite{doecode_12551} \\
Quasi-poloidal symmetry metric & STELLOPT \cite{doecode_12551} \\
Bootstrap Current Density & DKES proxy / PENTA \cite{van_rij_variational_1989,spong_generation_2005} \\
Radial Electric Field & PENTA \cite{spong_generation_2005} \\
\hline
\end{tabular}
\label{tab:targets}
\end{table}

The STELLOPT stellarator optimization code minimizes a user defined $\chi^2$ functional through a variety of optimization algorithms. In the context of this work, the x-vector which is being optimized is composed of the Garabedian ($\Delta_{mn}$) boundary modes \cite{garabedian_quasi-axially_1998} which define the plasma. Table \ref{tab:targets} provides an overview of functionals which were included throughout the various optimization steps. The optimizations were performed through successive applications of STELLOPT varying the weights, physics targeted, optimization algorithms, and included functionals through the design process. Both the genetic algorithm with differential evolution and the modified Levenberg-Marquardt algorithm were utilized.  Throughout this process the VMEC equilibrium code was used to compute the fixed boundary MHD equilibrium.

A few modifications were made to the STELLOPT code during the course of this work in order to better address this stellarator design problem. 
First, it was determined early on that BOOTSJ often under-predicted the bootstrap current as compared to the PENTA code. 
This motivated the implementation of a DKES based proxy function based on how parallel flows are computed.
While neoclassical parallel flows are obtained by inverting a matrix system \cite{sugama_how_2002}, such flows can be shown to be proportional to an energy convolution of $D^*_{31}/D^*_{33}$, where the star indicates corrected DKES mono energetic coefficients.
This then takes the form
\begin{equation}
j_{proxy} = \frac{D^*_{31}}{D^*_{33}} = \frac{D_{31}<B^2>}{(2/3)<B^2>(1/\nu^*) - D_{33}<B^2>}= \frac{D_{31}}{(2/3)(1/\nu^*) - D_{33}}
\end{equation}
where $D_{31}$ and $D_{33}$ are the DKES mono-energetic transport coefficients, and $\nu^*$ is the normalized collision frequency (normalized to the particle velocity) \cite{van_rij_variational_1989}.
The neoclassical parallel flows can be shown to be proportional to an energy convolution of $D_{31}/D_{33}$ \cite{lore_measurement_2010}.
In order to directly target the possibility of a positive $E_r$ in the plasma core, the PENTA code was directly interfaced to STELLOPT. 
As a side-effect, the PENTA computed bootstrap current density could be directly targeted without additional computational effort and later replaced the proxy during final optimizations.

The ability to target the vacuum magnetic field on axis was added to STELLOPT by having the code first compute a zero pressure, zero current VMEC equilibrium.
This equilibrium is then used to estimate the required enclosed toroidal flux (PHIEDGE) to achieve a target magnetic field value (in our case 6 T). 
Such a modification was found necessary to avoid the optimizer increasing the toroidal field through shape changes.
The increase in toroidal field was helping the optimizer achieve ballooning stability by artificially lowering the plasma beta (as the pressure profile was held fixed).
Finally, an internal scaling on the boundary Fourier modes is applied of the form $a^{scaled}_{mn} = a_{mn} e^{\alpha \textrm{max}(|m|,|n|)}$, where $a_{mn}$ are the Fourier modes defining the boundary and $\alpha=2$. 

Throughout the design process the VMEC equilibrium code is utilized as the equilibrium model. 
The optimized boundary modes are in the range of $m=[0,4]$ and $n=[-4,4]$, while the equilibrium domain extends from $m=[0,7]$ and $n=[-8,8]$. 
The majority of optimization work was performed with 128 radial grid points (in toroidal flux), switching to 256 for the electron root optimizations. 
The pressure profile was assumed fixed based on a 3000 MW fusion power operating point (2400 MW of neutrons and 600 MW of alphas) and a plasma volume of $1500~\textrm{m}^3$.
VMEC was modified to renormalize the boundary harmonics on input to achieve a target volume.
This removed the plasma volume from the optimization all together, making the total fusion power a fixed quantity.

Initially the current density in the equilibrium was assumed zero.
Optimization attempted to minimize the predicted bootstrap as computed by BOOTSJ.
It was found that the lack of equilibrium-bootstrap self-consistency was problematic.
Even when the BOOTSJ bootstrap was predicted to be small, once self-consistency was sought the bootstrap became large again.
It was found necessary to include the predicted bootstrap from the previous optimization step in the next, in order to properly minimize the bootstrap current.
Comparisons between BOOTSJ and PENTA found that even when BOOTSJ predicted small bootstrap, PENTA predicted finite bootstrap currents.
At that point PENTA was substituted for BOOTSJ when computing the self-consistent equilibrium bootstrap current.

In order to evaluate quantities which depend on the straight field line angle, the BOOZ\_XFORM software was utilized to perform a Boozer transformation of the VMEC equilibrium. 
This transformation initially spanned $m=[0,48]$ and $n=[-64,64]$ Boozer modes. 
Later simulations extended the number of poloidal modes up to $m=[0,64]$. 
The computation of the quasi-isodynamic metric, quasi-poloidal symmetry metric, DKES coefficients, and total bootstrap current by BOOTSJ all depend on such a transformation.
The quasi-isodynamic metric follows the squash-stretch-shuffle methodology \cite{goodman_constructing_2023}.
The quasi-poloidal symmetry metric penalizes the Fourier modes of magnetic field strength in Boozer coordinates whose amplitude is greater than zero and whose poloidal mode number ($m$) is greater than zero.

Computation of ballooning stability by the COBRAVMEC code was performed on a subset of surfaces spanning the radial equilibrium domain for optimization.
The outboard mid-plane flux tube at the toroidal angle $\phi=0$ cross section was chosen as this was stereotypically found to be the most unstable flux tube based on previous works.
A full radial profile of ballooning stability was evaluated after optimization, scanning multiple flux tube.
These simulations confirmed the choice of outboard mid-plane flux tube to in fact be the most unstable.

Evaluation of neoclassical transport effects (transport, bootstrap and radial electric field) are provided by a host of codes including BOOTSJ, NEO, DKES, and PENTA.
The BOOTSJ code provided initial evaluations of bootstrap current in the collisionless limit.
The NEO code was used to compute epsilon effective ($\epsilon_{eff}^{3/2}$), which is directly proportional to particle transport in the $1/\nu$ regime. 
It should be noted that in the $1/\nu$ regime, neoclassical particle transport is directly proportional to $\epsilon_{eff}^{3/2}$.
The DKES code computes the mono-energetic transport coefficients ($D_{11},~D_{31},~D_{33}$) for given values of normalized electric field and collisionality.
The $D_{11}$ coefficient is a directly proxy for neoclassical particle transport, while the other coefficients were used in the proxy for bootstrap current. 
The PENTA code computes integrals over the DKES coefficients to compute the local particle transport, energy transport, bootstrap current density, and radial electric field. 
Finally, it should also be mentioned that the THRIFT code has recently been extended with DKES and PENTA to provide self-consistent bootstrap and VMEC equilibrium solutions. Detailed discussion of THRIFT modeling is covered in a forthcoming work.

Turbulent transport is evaluated {\it a posteriori} for select equilibria using the stella code. 
This code provides radially local flux-tube estimates of turbulent heat and particle fluxes under the assumption of an electrostatic, collisionless plasma with kinetic electron and ion species. 
Optimization was performed using the $g^{rr}$ metric (also known as prox-1d), at mid radius. 
This proxy has been shown in previous works to significantly reduce turbulent ion heat fluxes while maintaining overall good plasma properties during optimization \cite{mynick_reducing_2011,mynick_turbulent_2014}.
In particular, the proxy was shown to correlate well with gyrokinetic heat flux estimates for a W7-X configuration.

The $\Gamma_C$ metric \cite{nemov_poloidal_2008} (evaluated at mid-radius) was used to improve fast ion confinement during optimization, with {\it a-posteriori} analysis of confinement by the BEAMS3D and ASCOT5 codes. As passing particle populations are generally well confined, collisionless simulations focusing on deeply trapped particle populations are used to verify that fast ion confinement has improved. Collisional simulations for the full slowing down time were performed for an interim configuration helping to gauge improvement in overall fast particle confinement. For the ASCOT5 simulations, BEAMS3D provides a birth population of ~4 million fusion born alpha markers, while ASCOT5 is used to evaluate the slowing-down and particle loss physics.

Although not explicitly targeted in the optimization, global MHD stability is also considered in the design analysis.
Equilibrium quantities such as the Mercier criterion and magnetic well are computed for each stage of optimization. 
Ideal kink stability is computed {\it a posteriori} using the TERPSICHORE code for the $n=0$, $n=1$, and $n=2$ mode families in GIGA. 
In these simulations the fields are extrapolated out by 30\% in the radial dimension to a conformal wall based on the equilibrium boundary \cite{turnbull_ideal_2011}. 
This is done as TERPSICHORE assumes an ideal conducting wall boundary condition. 
While this distance does not appear large, it should be sufficient to avoid the stabilizing effects of a close fitting wall. 
A preliminary assessments of Alfvén gap structure was made using the STELLGAP code.

\subsection{Requirements Capture}

\begin{figure}
 \centering
        \includegraphics[width=\textwidth]{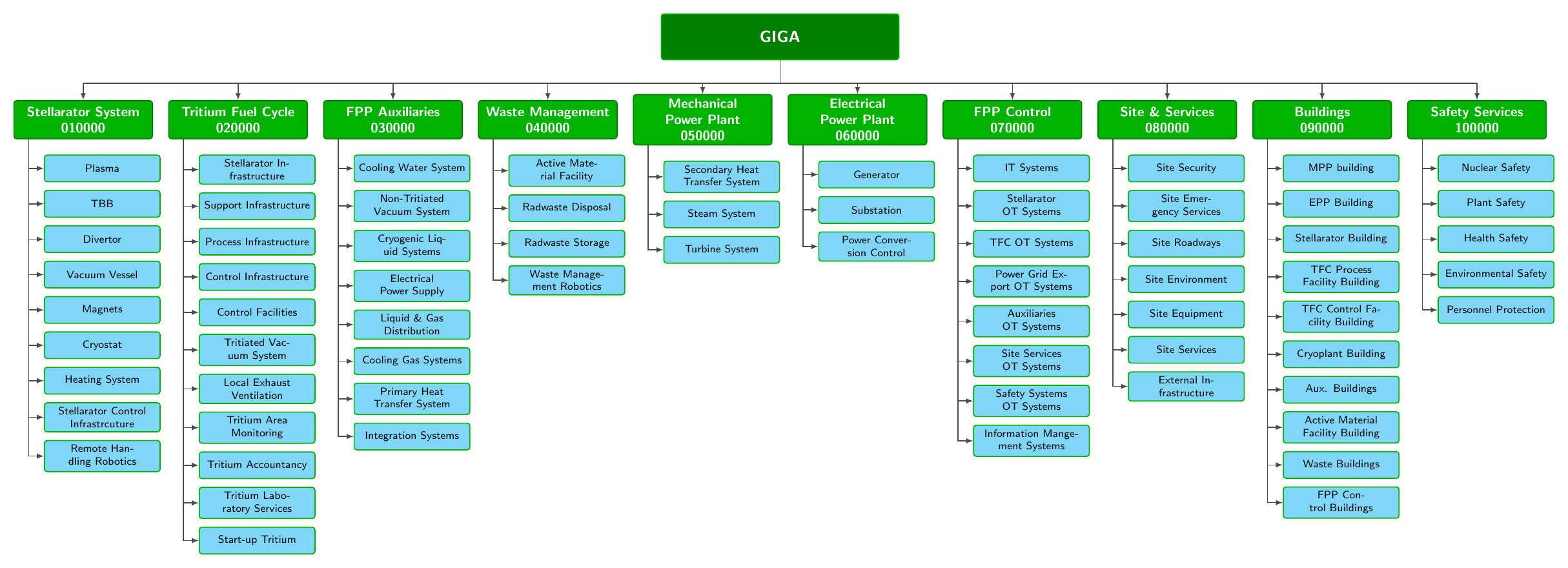}
 \caption{The systems breakdown architecture for the GIGA power plant showing how various systems are grouped.}
\label{fig:GIGA_BS}
\end{figure}

The requirements for the plasma flow down from the more general requirements of the entire GIGA system.
A 0.5-D scoping of the plasma operating scenarios provides more quantitative assessments of such requirements.
The term 0.5-D is used as radially dependent profile shapes for densities and temperatures are assumed.
The breakdown structure of the GIGA project has the stellarator system group at level 1 and the plasma as a level 2 system of the stellarator (figure \ref{fig:GIGA_BS}). 
A requirements capture for the GIGA power plant (level 0) flows down into requirements for the stellarator system (level 1). 
The requirements on the plasma system (level 2) flow down from these requirements as general statements. 
While some can easily be quantified in terms of design values, others require some modeling to place limits of acceptable performance. 
Here 0.5D modeling is performed to define acceptable parameters. 
In this way, quantified requirements can be determined with which to gauge the success of the plasma design. 
Additionally, these can highlight where additional research avenues should be focused.

While many requirements exist for the full power plant, two essentially flow down from GIGA to the stellarator system:
\begin{itemize}
\item The fusion power plant shall be optimized around a fusion reactor comprising a feasible stellarator based approach to magnetic confinement fusion.
\item The fusion power plant shall export continuous net power of 1 GWe, 7.75 TWh/year
\end{itemize}
The second requirement assumes an approximate 90\% availability. At the level of the stellarator system (level 1) these requirements are captured by the following set of requirements:
\begin{itemize}
\item The stellarator system shall provide 3 GW fusion generated energy output in steady state operation
\item The stellarator system shall develop a self-heating scenario with a minimum of initial electron cyclotron resonance heating
\item The stellarator system shall establish a stable, efficient, highly confined plasma
\item The stellarator system shall achieve plasma magnetic confinement with peak magnetic fields on axis of 6 T
\item The stellarator system shall develop and operate an optimized 4 field period plasma
\item The stellarator system shall operate with high efficiency throughout operational modes
\end{itemize}
Such requirements are the subset of full stellarator system requirements, which flow down to the plasma system.  At the level of the plasma system, these requirements are captured in the following plasma requirements:
\begin{itemize}
\item The plasma shall produce {\bf 3 GW} of fusion power
\item The plasma shall be compatible with an island divertor
\item The plasma shall confine fast ions
\item The plasma shall be ballooning stable
\item The plasma shall be kink stable ($n=0,~n=1,~n=2$ mode families)
\item The plasma shall possess no large core-edge Alfvén gaps
\item The plasma shall be neoclassically optimized
\item The plasma shall be turbulence optimized
\item The plasma shall avoid impurity accumulation
\item The plasma shall have low bootstrap current
\item The plasma shall avoid low order rational surfaces
\item The plasma shall have {\bf 4 field periods}
\item The plasma shall have {\bf 6 T} on axis at the $\phi=0$ toroidal plane
\item The plasma should achieve core electron root confinement (CERC) conditions
\end{itemize}
where explicitly quantized values have been highlighted. These 14 statements define a successfully designed plasma, but do so in a very qualitative way. Some level of modeling is required to quantify these statements. It should be noted that the choice of 4 field periods does appear arbitrary at this point along with the 6 T requirement. The 6 T requirement comes from a desire to use existing 170 GHz gyrotron technology for plasma startup. The 4 field periods come from a desire to reduce the total number of coils. A more detailed analysis of this assumption is left to future works where coil optimization will be explored.

\begin{figure}
 \centering
        \includegraphics[width=\textwidth]{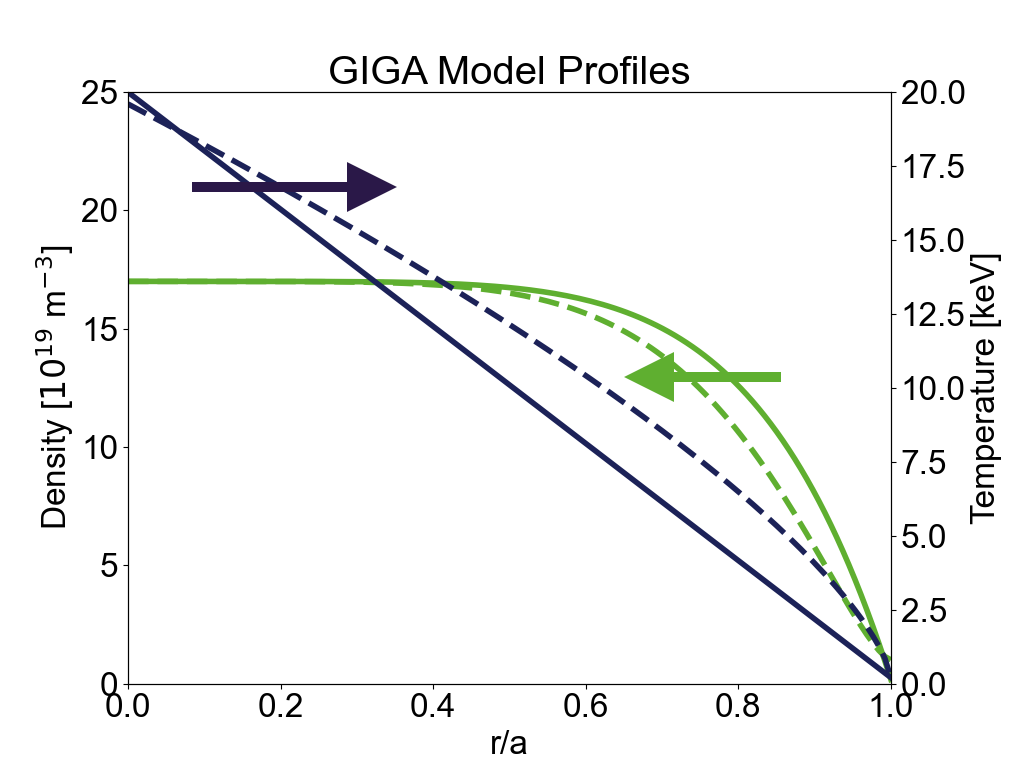}
 \caption{Temperature (purple) and density profile (green) shapes utilized in 0.5D modeling (solid) of the GIGA plasma and those used later for stellarator optimization (dashed). Arrows have been added to guide the readers eye to the left and right y-axis accordingly.}
\label{fig:profile_example}
\end{figure}

In order to quantify the plasma requirements capture, 0.5 D modeling was performed for the GIGA plasma. 
The term 0.5 D is used here as 0 D analysis assumes average values for temperature and densities, while in our case we have assumed profile shapes. 
The shape of the temperature profile of the plasma species $k$ is assumed to be of the form $T_k\left(\rho\right) = (T^{core}_k-T^{edge}_k)(1-\rho) + T^{edge}_k$ while the species $k$ density profile is of the form $n_k\left(\rho\right) = (n^{core}_k-n^{edge}_k)(1-\rho^6) + n^{edge}_k$. 
These profiles only serve as a basis for initial computation and are similar to those achieved in W7-X electron cyclotron resonance heating (ECRH) discharges (figure \ref{fig:profile_example}). 
The necessary auxiliary heating power to sustain the GIGA plasma can then be computed from the ISS04 scaling law \cite{yamada_characterization_2005}, the Bosch-Hale model for fusion birth rates \cite{bosch_improved_1992}, Bremsstrahlung \cite{Richardson19nrl}, and synchrotron radiation \cite{albajar_improved_2001,fidone_synchrotron_2001}. 
For reference, the ISS04 scaling law reads
\begin{equation}
\tau_{ISS04} = 0.134f_{ren}a^{2.28}R^{0.64}P^{-0.61}\bar{n}_e^{0.54}B^{0.84}\iota_{2/3}^{0.41}
\end{equation}
where $f_{ren}$ is a renormalization factor, $a$ the plasma minor radius, $R$ the plasma major radius, $P$ the total heating power, $\bar{n}_e$ the line-averaged density, $B$ the magnetic field strength, and $\iota_{2/3}$ is the rotational transform at $r/a=2/3$ (normalized minor radius).
Additionally, the effect of impurities can be included in the model by assuming they are fixed percentages of the electron density, reducing the deuterium and tritium density accordingly (assumed to be equal parts).  The presence of impurities changes the computation of Bremstrahlung in this model reducing the available power for sustaining the plasma. The effect of alpha losses is also included through a scaling factor on the fusion power. Unless it is otherwise stated, the ion and electron temperatures are equal.

\begin{table}
\caption{Nominal model assumptions for the 0.5D scaling law study. Impurity concentrations are in terms of the electron density.}
\centering
\begin{tabular}{l l }
\hline
Plasma Volume & 1500 $\textrm{m}^{3}$ \\
Major Radius & 20.0 m \\
Minor Radius & 1.95 m \\
Magnetic Field on axis & 6.0 T \\
Rotational Transform & 0.85 \\
Fusion Power & 3 GW \\
ISS04 Scaling Factor ($f_{ren}$) & 1.0 \\
Alpha Power Loss & 5 \% \\
He Concentration & 5 \% \\
Protium Concentration & 1 \% \\
Tungsten Concentration & 0.01 \% \\
\hline
\end{tabular}
\label{tab:0.5D}
\end{table}

\begin{figure}
 \centering
        \includegraphics[width=\textwidth]{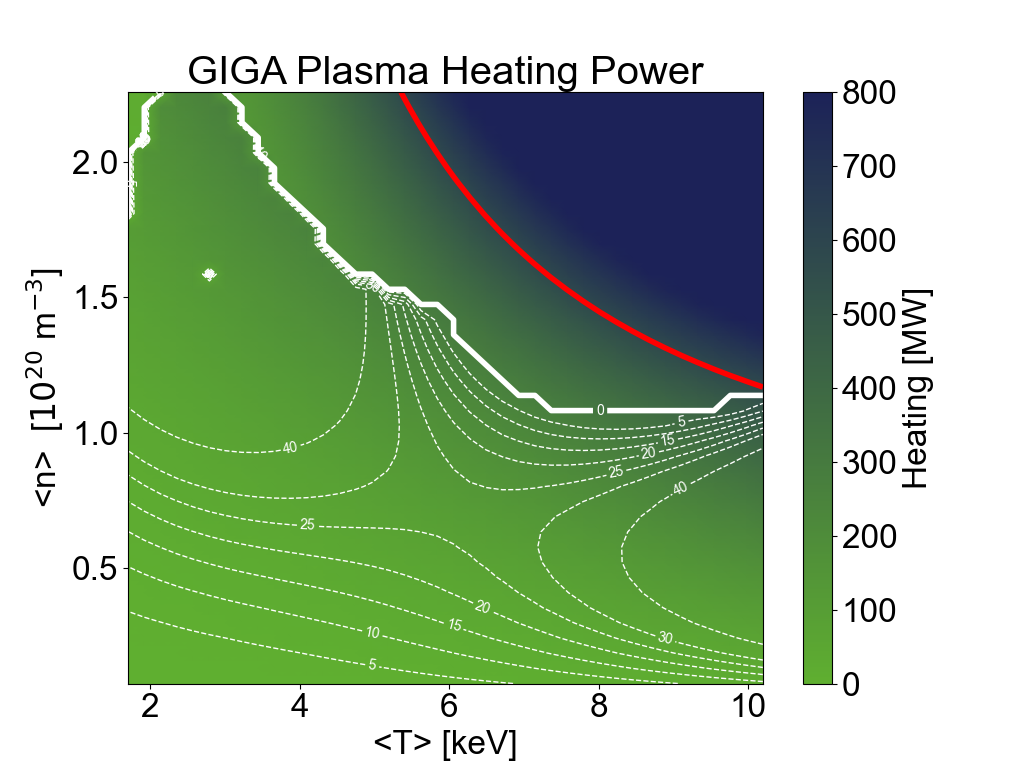}
 \caption{POPCON plot for the nominal GIGA parameters as depicted in table \ref{tab:0.5D}. The color scale shows alpha power. The white lines indicate the required auxiliary heating power. The red line indicates 3 GW total fusion power. Volume averages $<...>$ performed over profiles.}
\label{fig:POPCON}
\end{figure}

In table \ref{tab:0.5D}, we see the nominal assumptions for the 0.5D modeling and the resultant plasma operating contours (POPCON) plot in figure \ref{fig:POPCON}.  The inclusion of protium, helium, and tungsten impurities dilutes the plasma, reducing fusion power at a given electron density. It is immediately evident that even for these rather modest choices of parameters, scenarios producing full power at burn are available for a multitude of density and temperature scenarios. The saddle point in the auxiliary heating requirements (also known as the Cordey pass) suggests that between 25 and 30 MW of auxiliary heating are necessary to achieve ignition. This plot also seems to favor the high density regime over the high temperature regime.

A sensitivity study of the input parameters in table \ref{tab:0.5D} was performed to help understand the lower limits of the assumed parameters. While a few choices exist for determining a lower bound of the assumed parameters, in this study the distance between the 3 GW fusion power line and the burn condition (zero auxiliary power) was chosen as a credibility metric. When these two contours begin to overlap significantly, we assume an ignited plasma state would no longer be achievable at 3GW. This is of course a weak condition as the plasma could still function to produce energy but at degraded output related to the need for recirculating power in the auxiliary heating. Assessment of such a situation would require inclusion of a balance of plant analysis.

\begin{figure}
 \centering
        \includegraphics[width=0.49\textwidth]{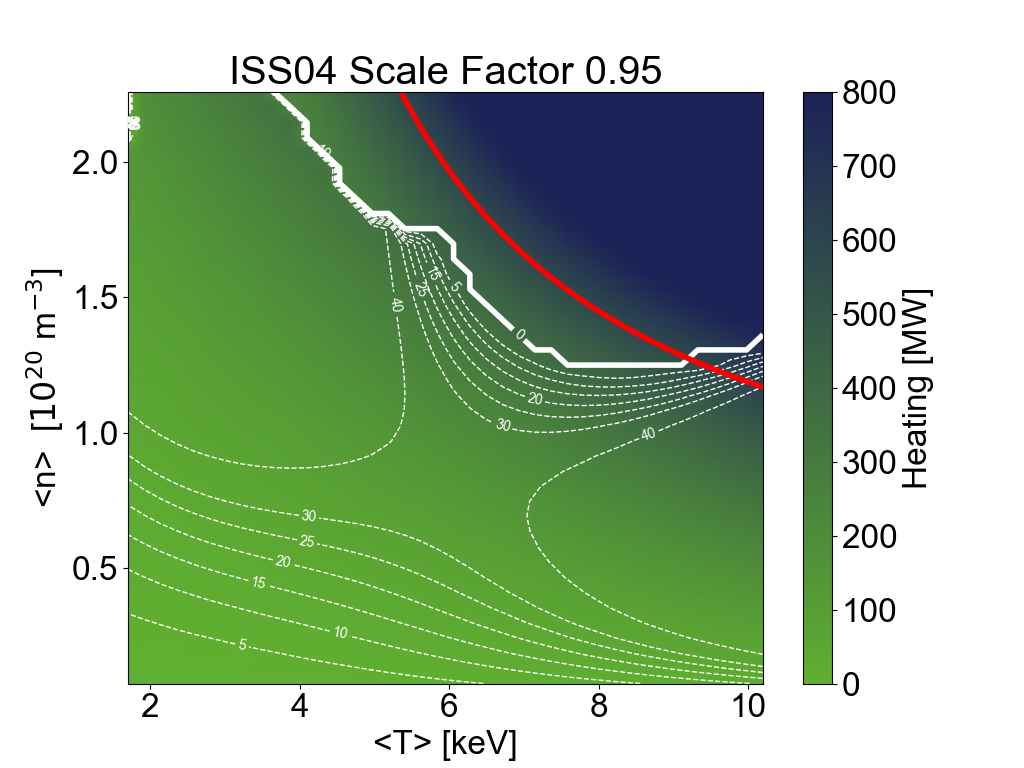}\includegraphics[width=0.49\textwidth]{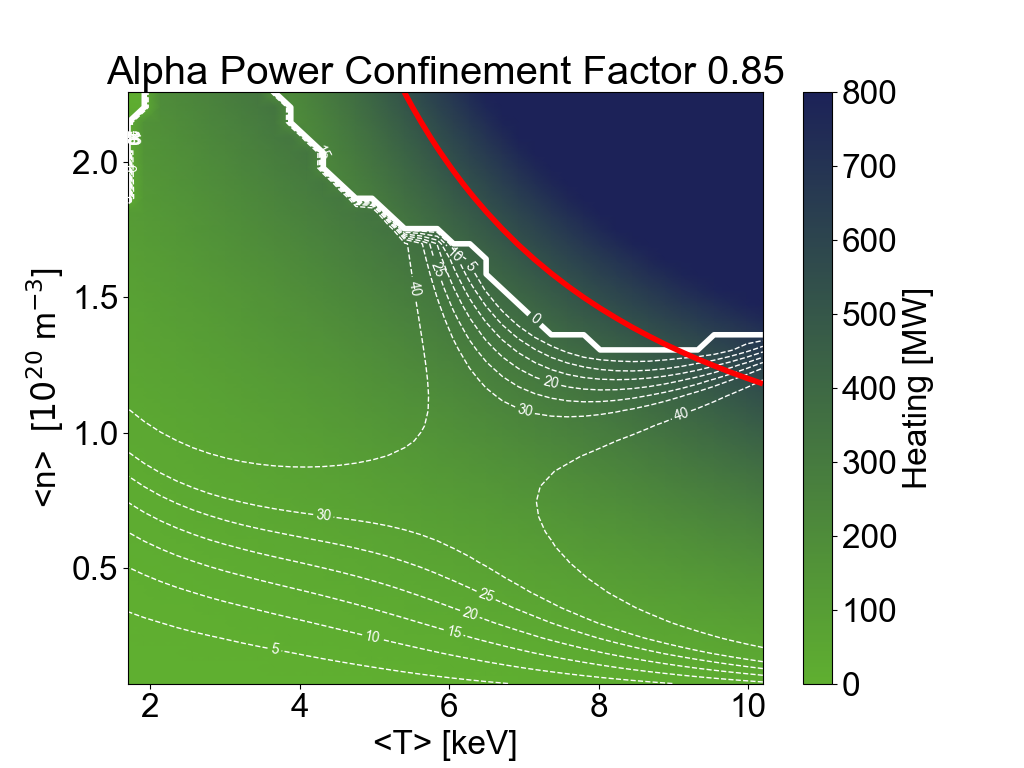}\\
        \includegraphics[width=0.49\textwidth]{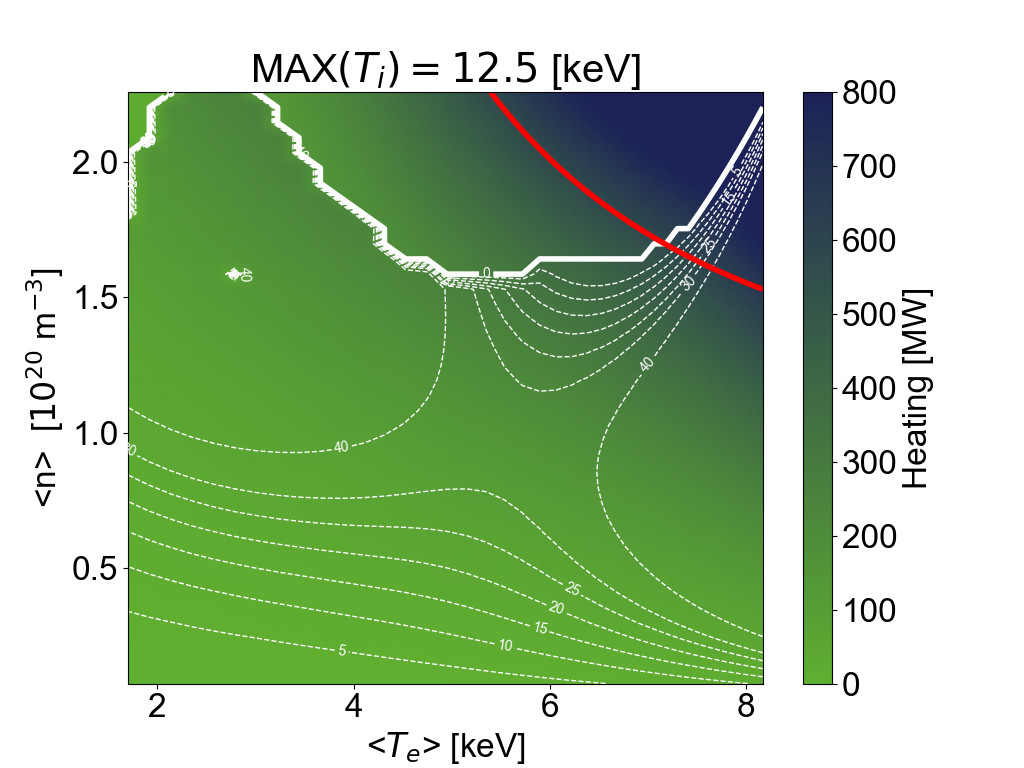}\includegraphics[width=0.49\textwidth]{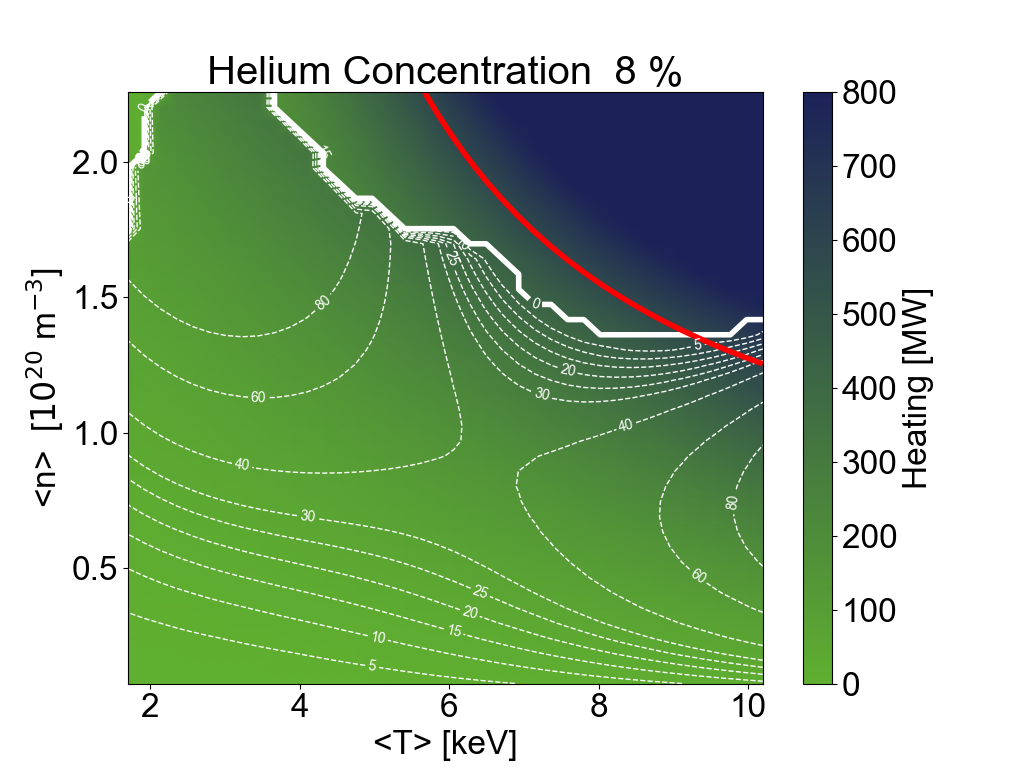}\\
 \caption{POPCON plot scanning the sensitivity to input parameter variation. ISS04 scaling factor (upper left), alpha power confinement (upper right), ion temperature clamping (lower left), and Helium concentration (lower right) are depicted.}
\label{fig:POPCON_sensativity}
\end{figure}

Figure \ref{fig:POPCON_sensativity} depicts four examples of profile variation which were performed.
A common feature of each sensitivity study is the degradation of the Cordey path. Degradation implying an increase in necessary auxiliary heating.
We can see even at these degraded performance parameters, the gap between ignition (solid white line) and full power (solid red line) still permits variation of both temperature and density, while maintaining an ignited plasma. 
Albeit in a much narrower operating range than for the nominal parameters. 
Thus improvements in the plasma physics are less about the feasibility of the plasma to reach burning conditions but rather about reductions in required auxiliary power to start up. 
This has direct financial implications for the overall cost of the power plant (reduced installed heating power).

It is important to note that in the development of GIGA, there has been no explicit assumption that $f_{ren}$ will exceed unity. While it has been demonstrated that $f_{ren}=1$ can be exceeded transiently in machines like W7-X, data for steady-state operation suggests that such conditions are only maintained at low plasma temperature (HDH-mode in W7-AS for example \cite{hirsch_major_2008}). Additionally, such results are usually only achieved with strong ion heating in the case of W7-AS HDH-mode, or strong density peaking as in the case of W7-X high performance discharges. What is clear from such experimental data is that improvements in $f_{ren}$ are strongly correlated with a reduction in turbulence. Hence, a stellarator power plant will require a turbulence optimized configuration where magnetic field shaping reduces turbulence. One could ask the question what effect achieving $f_{ren}=1.2$ or $f_{ren}=1.4$ would have on the GIGA plasma.  Such choices reduce the Cordey pass to less than 10 and 5 MW respectively, a wholly unrealistic number for the required auxiliary heating power. However, theoretical evidence does suggest that when the plasma is solely heated by fusion born alphas $f_{ren}$ may exceed unity \cite{warmer_limits_2015}. One should be skeptical of  the claim $f_{ren}>1$ when $P_{aux}>0$ until experimental evidence to the contrary is presented.

One advantage of assuming the profile shapes in such a study is the ability to assess the effect of profile variation on the parameters. Generally, small variations within the limits of believable profile shapes show small effects in the POPCON plots. However, beyond profile shaping, one can study the effect of ion temperature clamping on power plant performance (lower left panel of figure \ref{fig:POPCON_sensativity}). It is clear from this study that ion temperatures higher than 10 keV are necessary to achieve full power operation in burning plasma conditions. However, it was interesting that even for ion temperatures clamped to 12.5 keV we find a large operating window.  Of course full power operation moves to high electron temperatures at fixed density, but burning plasmas are still possible. This may be a favorable situation overall given the impurity and fueling characteristics of such clamped ion temperature plasmas in W7-X. Additionally, such plasmas may more easily access core electron root conditions which also have favorable particle transport. Still, we develop GIGA with the assumption that such clamping can be suppressed through transport optimization.

\begin{table}
\caption{Parameters for the design of the GIGA plasma. These show up as inputs to the plasma equilibrium with the first three explicit inputs and the others implicit assumptions regarding the profiles. Concentrations are relative to the electron density.}
\centering
\begin{tabular}{ l c r}
\hline
Parameter & Unit & Value \\
\hline
Plasma Volume & $\textrm{m}^3$ & 1500 \\
Field Periodicity & --- & 4 \\
Magnetic Field on axis & T & $\le 6$ \\
Core Fuel Ion Temperature & keV & $>10$ \\
Core Electron Density & $\textrm{m}^{-3}$ & $>2\times10^{20}$ \\
Fusion Neutron Power &  MW & 2400 \\
Fusion Alpha Power &  MW  & 600 \\
Auxiliary Heating Power &  MW  & $<100$ \\
Proton Concentration & \% & $<5$ \\
Helium Concentration & \% & $<8$ \\
Tungsten Concentration & \% & $<0.01$ \\
\end{tabular}
\label{tab:param}
\end{table}

\begin{table}
\caption{Target ranges for design of the GIGA plasma. These are quantities are derived from analysis of the plasma.}
\centering
\begin{tabular}{ l c r}
\hline
Target & Unit & Value \\
\hline
Edge Rotational Transform & --- & $\iota_{core}<\iota_{edge}<1.0$ \\
Core Rotational Transform & --- & $0.8<\iota_{core}<\iota_{edge}$ \\
Alpha Power Confinement & MW & 510 \\
Effective Helical Ripple ($\epsilon^{3/2}_{effective}$) & --- & $<0.01$ \\
Maximum Heat Flux & $\textrm{kW/m}^2$ & $<600$ \\
Net Toroidal Current & kA & $<50$ \\
Ballooning Stability & --- & Stable \\
Kink Stability ($n=0$) & --- & Stable \\
Kink Stability ($n=1$) & --- & Stable \\
Kink Stability ($n=2$) & --- & Stable \\
Alfv\'enic Stability & --- & No core-edge gaps \\
Core Electric Field & kV/m & $>0$
\end{tabular}
\label{tab:targets}
\end{table}

Beyond the input parameters to the 0.5 D analysis, certain parameters still need upper limits to produce a fully quantized set of requirements, in particular the transport levels.
We fundamentally assume a transport optimized stellarator with neoclassical transport sub-dominant when compared to turbulence.
We generally capture such low neoclassical transport through a requirement on the effective helical ripple.
Generally, we require that $\epsilon^{3/2}_{effective}< 0.01$ with many stellarator designs easily achieving this.
Quantifying an acceptable level of turbulent transport is not trivial, as 0D modeling determines acceptable minimum core ion temperatures ($T_{ion}>10$ keV).
Still based on assumptions about the fusion power, and transport, we find that the total heat flux should be $<600~\textrm{kW/m}^2$. 
Such estimates are very approximate with full 1D transport modeling being required to verify that a given fusion relevant scenario is achievable.
Table \ref{tab:param} presents the required parameters, or requirements which show up as inputs (both implicit and explicit) to the design of the plasma.
Table \ref{tab:targets} presents the required targets for the plasma design, quantities which are intrinsic to the design of the plasma requiring optimization to achieve the target values.

\subsection{System Interdependencies}

It is worthwhile to shortly discuss the interdependencies of the plasma on some of the other systems which comprise the stellarator system. 
The stellarator system as defined for GIGA is composed of the plasma, tritium breeding blankets, divertor, vacuum vessel, magnets, cryostat, heating system, stellarator control interface, and remote handling robotics systems. 
This will not be an exhaustive discussion but will focus on how the plasma can affect these systems through changes to the plasma shape. 
The divertor comprises the region between the first wall and plasma boundary and is responsible for power and particle exhaust. 
The desire for an island divertor in GIGA implies that changes to the edge rotational transform in GIGA can have an effect on the edge island. 
Thus control and minimization of the effects of the toroidal current are desirable in GIGA. 
The tritium breeding blankets and vacuum vessel must enclose the plasma and are often part of what is known as the radial build. 
The radial build is a set of concentric radial regions around the plasma which are used to approximate regions such as the breeding blankets, plasma vessel, and thermal shield. 
Here considerations are geometric with regard to the plasma, although the plasma is also an emitter of neutrons and photons which create thermal loads on these systems. 
While the neutron source is volumetric, it is volumetric about the magnetic axis of the device, with most emission coming from inside $r/a=0.2$.  
The magnets give rise to the plasma with coil optimization being the process by which coil-plasma self-consistency is achieved. 
Here the plasma is theorized to encode some information about the magnetic coils \cite{kappel_magnetic_2024}.
This encoding includes the rotational transform which can determine mode structure of an edge island outside the plasma, the coil complexity, and the coil-plasma distance.
Additionally, plasma field strengths must be consistent with the choice of coil technology. 
Similarly, the plasma becomes the target for the heating system, requiring that the technology for auxiliary heating be compatible with the plasma. 
In the case of GIGA, 170 GHz electron cyclotron heating is envisioned, compatible with the 6 T on axis magnetic field. 
The stellarator control infrastructure must measure the plasma and control the impurity seeding, heating and fueling actuators. 
Thus the sensor systems used to measure the plasma must be compatible with the operating parameter range. 
The remote handling robotics are not active while plasma is present, while the cryostat simply must be compatible with the emissions of the plasma.

The interaction of the plasma neutrons with the first wall should also be discussed before moving on.  
While detailed modeling of the neutron environment is required to understand the local variation in wall loads, it is generally assumed that neutron loads above $1~\textrm{MW/m}^2$ should be avoided, as values much higher than this imply material damage rates higher than acceptable. 
In this context, acceptable implies dpa levels in accordance with a four to five year blanket replacement interval. 
This implies a relation between the neutron power and the first wall area $q_{neutron} = 0.8P_{fusion}/A_{wall} \le 1~\textrm{MW/m}^2$. 
Meanwhile, the fusion power generally scales like the plasma profiles and the plasma volume.  
While one can argue that the choice of total fusion power is a free parameter, in general it is a requirement of the power plant. 
This then implies that one cannot simply make a given stellarator smaller at fixed power output without increasing wall loads beyond what are assumed to be acceptable limits.  
For GIGA the plasma has a surface area of  $\sim2100~\textrm{m}^2$ and a neutron power of $2400~\textrm{MW}$.  
This implies that the average wall load is below $1~\textrm{MW/m}^2$ with a 30 cm plasma-wall gap. Thus for a given desired electrical capacity, it is the materials which are setting limits on device size, not plasma physics.
Increases in plasma performance thus do not enable more compact devices, but rather open a possible path to power plants with lower power output.

\section{Concept Definition} \label{sec:conceptdef}
The GIGA plasma shape was determined through numerical optimization of a modified W7-X fixed boundary configuration, which served as the theoretical basis for the development of W7-X. A high-iota high-mirror W7-X boundary definition served as a starting point for application of the STELLOPT code \cite{nuhrenberg_global_1996}. The boundary harmonics, field periodicity, and pressure profiles were all modified to match the requirements of GIGA (referred to as the `Initial' equilibrium in this work, GIGA\_v500). The optimization proceeded through successive applications of both the genetic algorithm with differential evolution along with the modified Levenberg-Marquardt algorithm. This resulted in two configurations, a conceptual design configuration (GIGA\_v515) and an evolved configuration (GIGA\_v549). The evolved configuration serves as the basis for GIGA and builds upon the lessons learned in the development of the conceptual design configuration \cite{GFG_CDR_exec_2025}.

\subsection{Development of the Conceptual Design Plasma}

\begin{figure}
 \centering
        \includegraphics[width=0.49\textwidth]{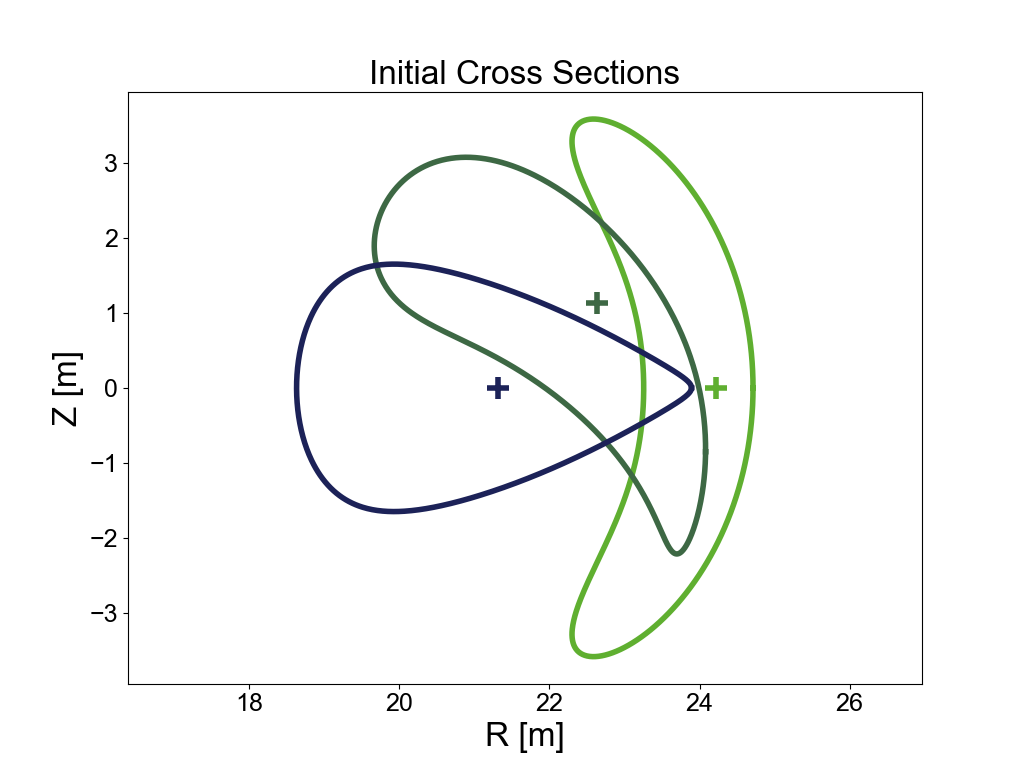}\includegraphics[width=0.49\textwidth]{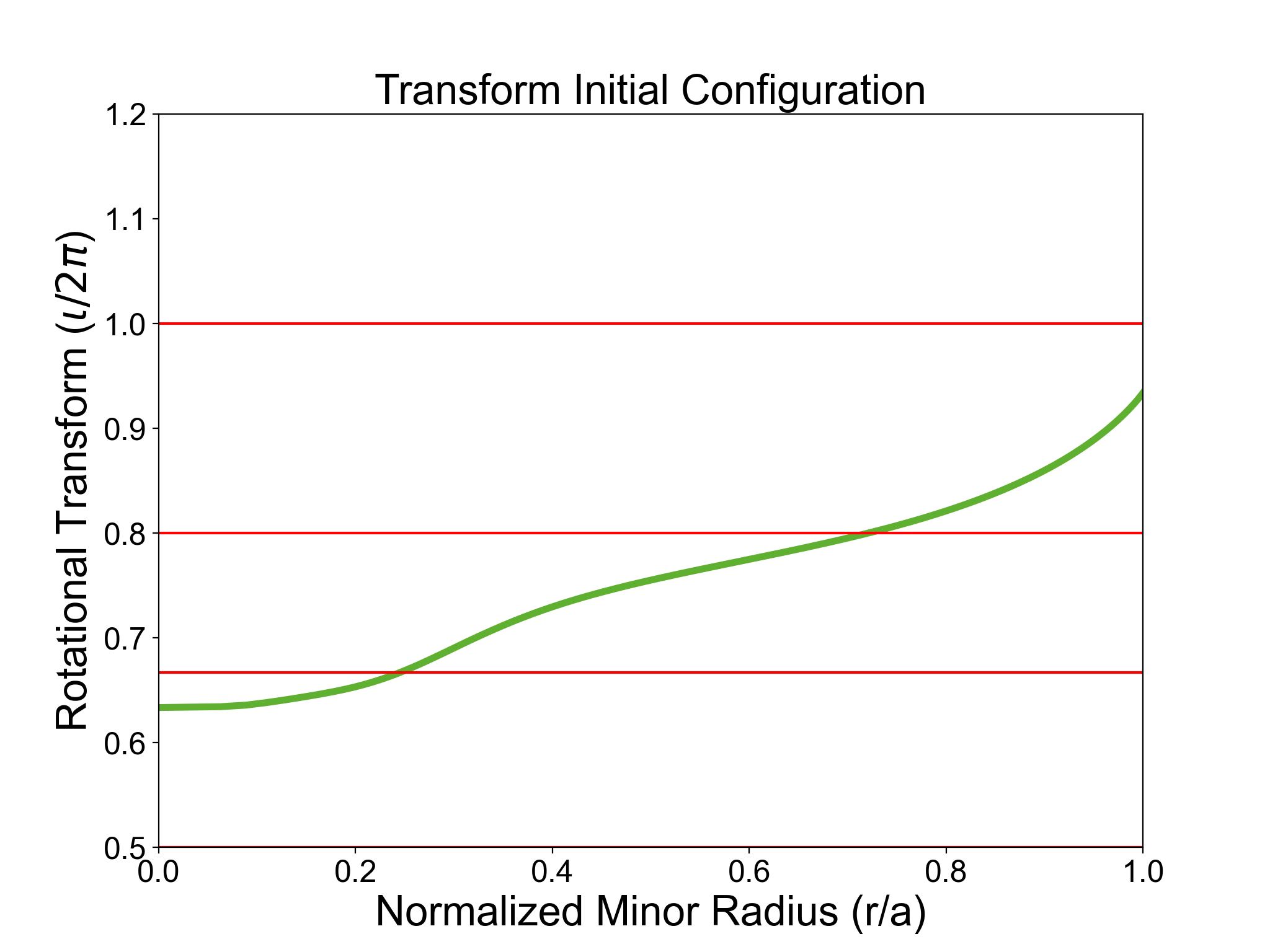}\\
        \includegraphics[width=0.49\textwidth]{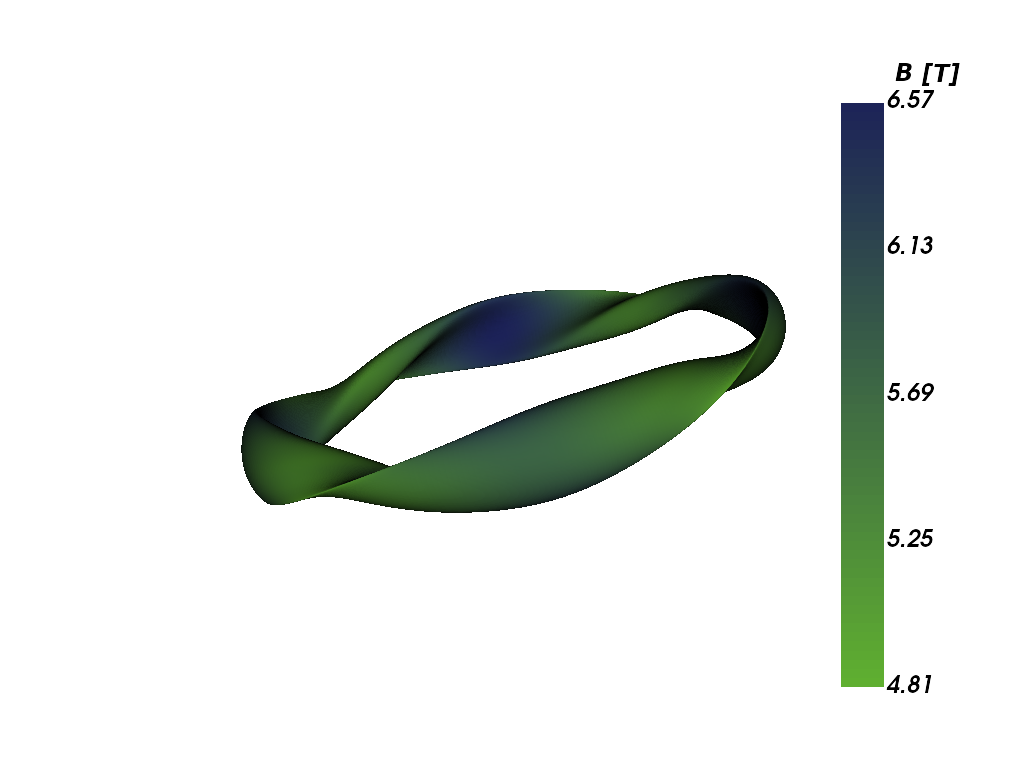}\includegraphics[width=0.49\textwidth]{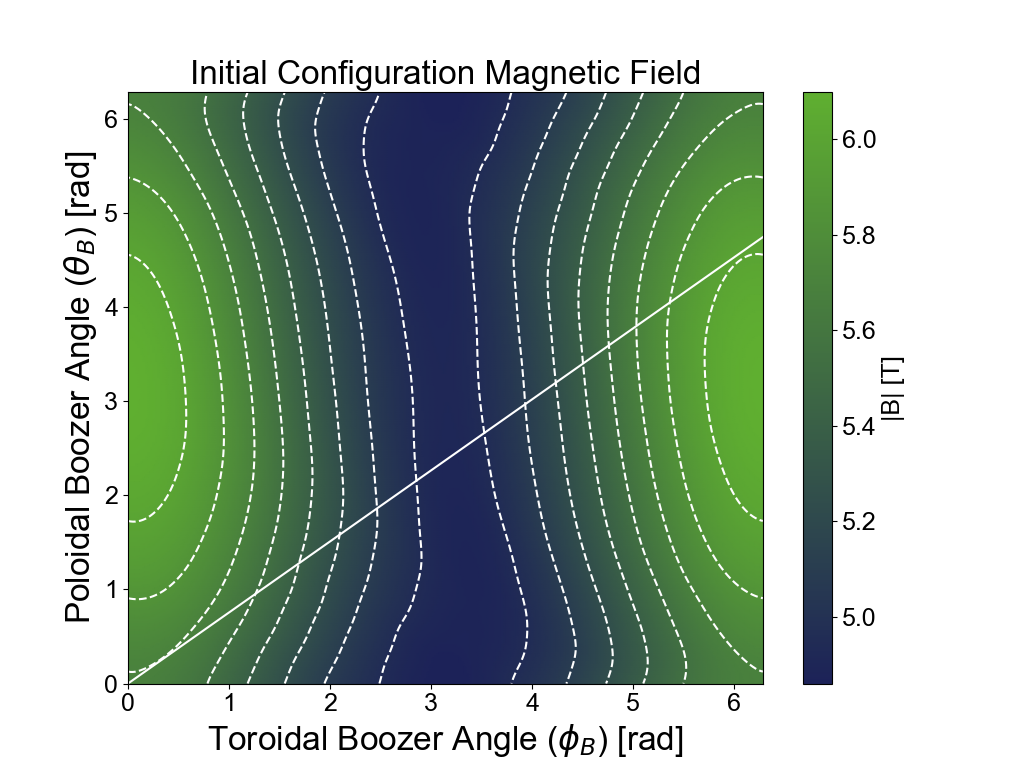}\\
 \caption{Boundary shape (upper left), rotational transform (upper right), 3D B-field pattern (lower left), and B-field pattern at mid radius (lower right) for the initial equilibrium. Cross sections are shown at zero, quarter, and half field period. Red lines in rotational transform plot depict low order rationals. Solid white line in Boozer coordinate plot depicts the field line.}
\label{fig:GIGA_v500}
\end{figure}

\begin{figure}
 \centering
        \includegraphics[width=0.32\textwidth]{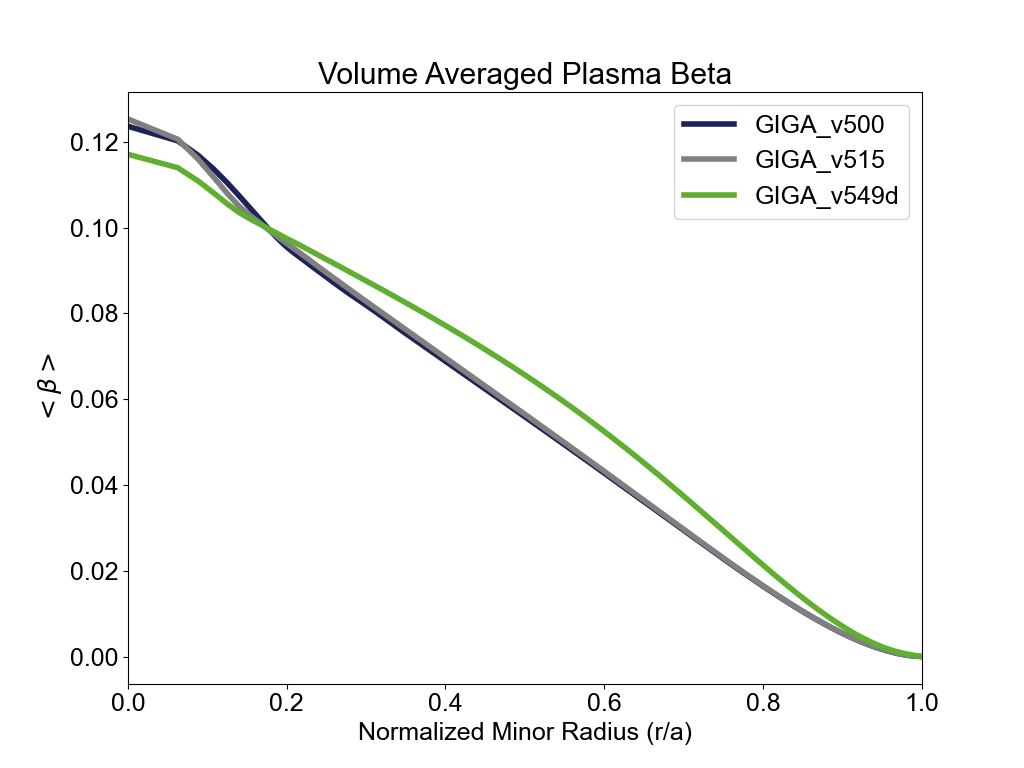}\includegraphics[width=0.32\textwidth]{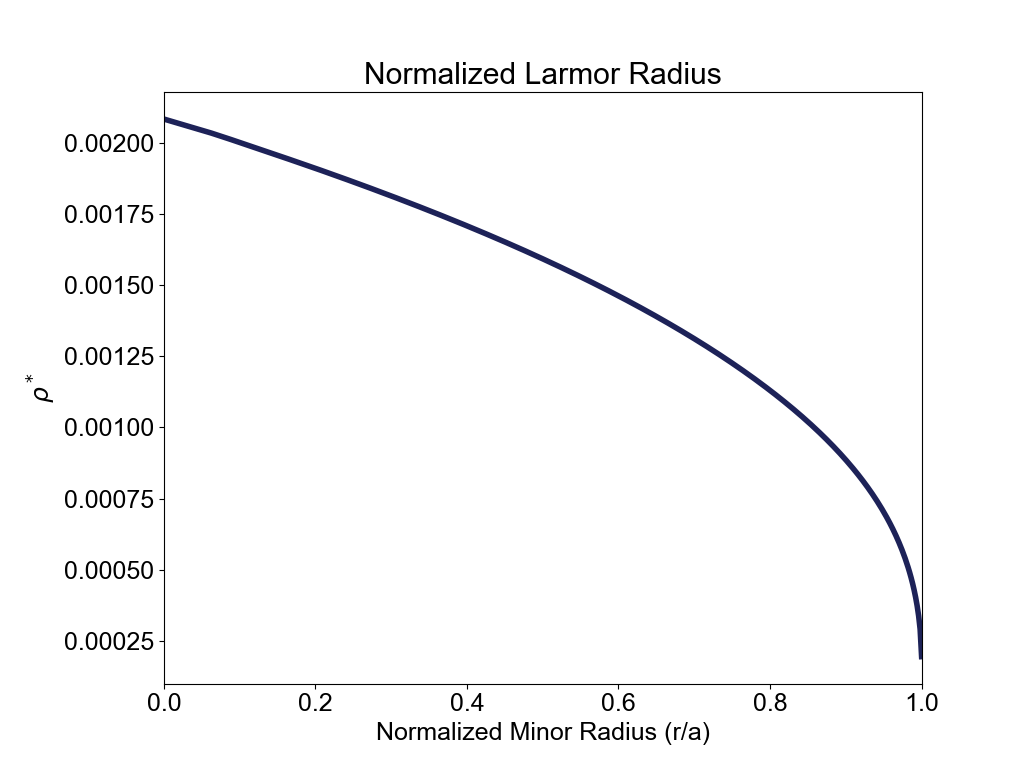}\includegraphics[width=0.32\textwidth]{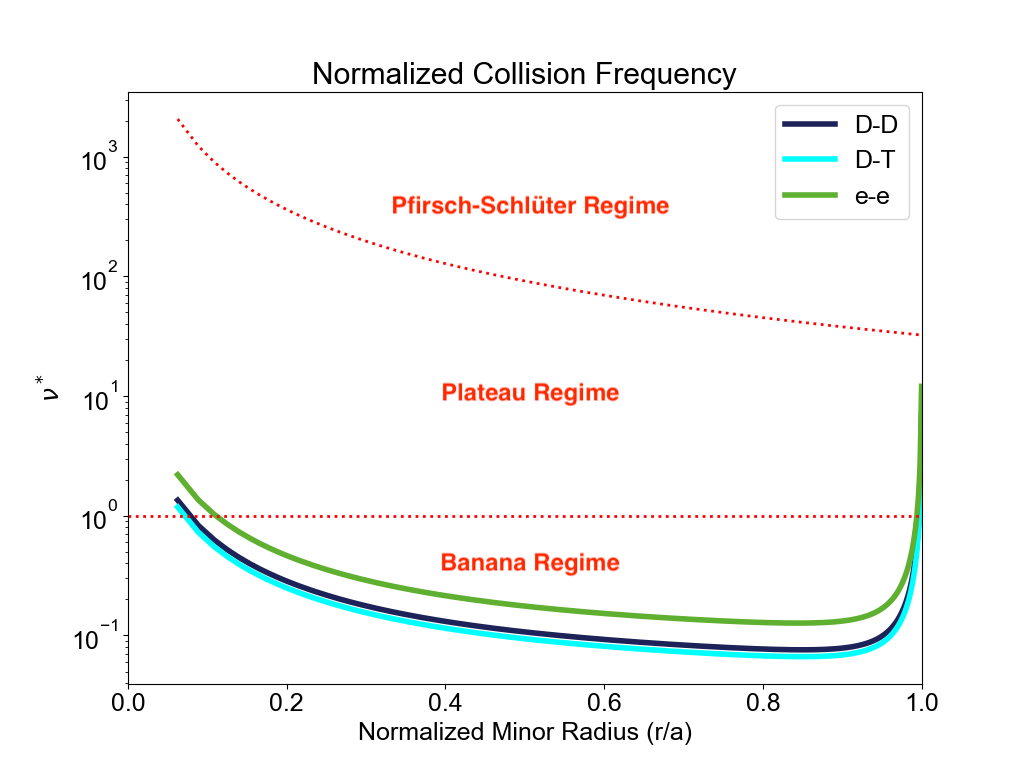}
 \caption{Radial profiles of the plasma beta, normalized Larmor radius, and normalized collisionality. The GIGA\_v549 configuration is used for the Larmor radius and collisionality plots. The collisionality plot indicates the three transport regimes common to tokamaks.  }
\label{fig:GIGA_betarhonu}
\end{figure}

The process of optimization began with an initial equilibrium based on a set of modified fixed boundary W7-X equilibrium harmonics (figure \ref{fig:GIGA_v500}).
The modification performed involved rescaling the plasma volume to $1500~\textrm{m}^3$, adjusting the field periodicity from 5 down to 4, and setting the enclosed toroidal flux such that a  6 T field was present at the magnetic axis (at the $\phi=0$ toroidal angle). 
While high-, low-, and standard-iota configurations were available, the need to reduce the field periodicity and have the edge rotational transform close to unity motivated the choice of the high iota configuration. 
The change in field periodicity essentially dropped the edge rotational transform from $n/m=5/4$ to $n/m=4/4$.
The electron density profile was taken to be $n_e(s)=n_{e0}(1-s^{2.8})^{1.5}+n_{eE}$ with $n_{e0}=1.7\times 10^{20}~\textrm{m}^{-3}$ and $n_{eE}=0.1\times 10^{20}~\textrm{m}^{-3}$, while the electron temperature was taken to be $T_e(s)=T_{e0}(1-s^{0.5})^{0.7}+T_{eE}$ with $T_{e0}=19.6~\textrm{keV}$ and $T_{eE}=0.2~\textrm{keV}$.
The choice of exponents was made based on fits to preliminary transport modeling of this configuration.
These choices produce a total fusion power which is 12\% higher than the required 3000 MW.
The profile shapes used during plasma boundary optimization were based on fits to 1D simulations of plasma transport using fusion alpha heating profiles and simplified transport equations.
The resulting plasma beta, normalized Larmor radius, and normalized collisionality can be seen in figure \ref{fig:GIGA_betarhonu}.
For the equilibrium optimization, the plasmas were assumed purely composed of deuterium and tritium, with no impurities.
The current profile was assumed zero unless otherwise stated. 
The conceptual phase (including the development of GIGA\_v515) made use of 128 radial grid points (equidistant in toroidal flux) with a poloidal Fourier space of $m=[0,7]$ and a toroidal Fourier space of $n=[-8,8]$. 
The actual boundary harmonics themselves incompletely filled the space $m=[0,4]$ and $n=[-3,4]$.

The development of the conceptual design plasma involved successive application of the Genetic Algorithm with Differential Evolution. Each optimization was performed for 20 generations with populations of 1000 members. In each generation 90\% of the genes (input vector $\vec{x}$) were mutated for each member. A mutation strategy of the form $\vec{x}_{new} = \vec{x}_{best}+0.5(\vec{x}_1-\vec{x}_2)$ was used with exponential gene cross-over.  The $\vec{x}_1$ and $\vec{x}_2$ refer to two other members of the population, while $\vec{x}_{best}$ refers to the member with the lowest $\chi^2$. The optimized vector $\vec{x}$ was composed of the boundary harmonics transformed into a Garabedian representation ($\rho$-like).  In each optimization the $\vec{x}$-vector was bounded to $\pm10\%$ of the harmonics obtained from the previous optimization run. The optimizations targeted ballooning stability, effective helical ripple, fast ion confinement ($\Gamma_c$), the rotational transform, turbulent transport (prox1d), quasi-poloidal symmetry, and the quasi-isodynamic metric. 

The sigmas for the target functionals were adjusted from one optimization to the next in order to bias minimization towards desired properties. This was often done by using the `renormalization' feature of STELLOPT, where each group of targets would have their sigmas automatically renormalized so that the total chi-squared for a given target was unity. This also allowed manual biasing of sigmas so that specific targets would dominate the reduction in chi-squared.

\begin{figure}
 \centering
        \includegraphics[width=0.49\textwidth]{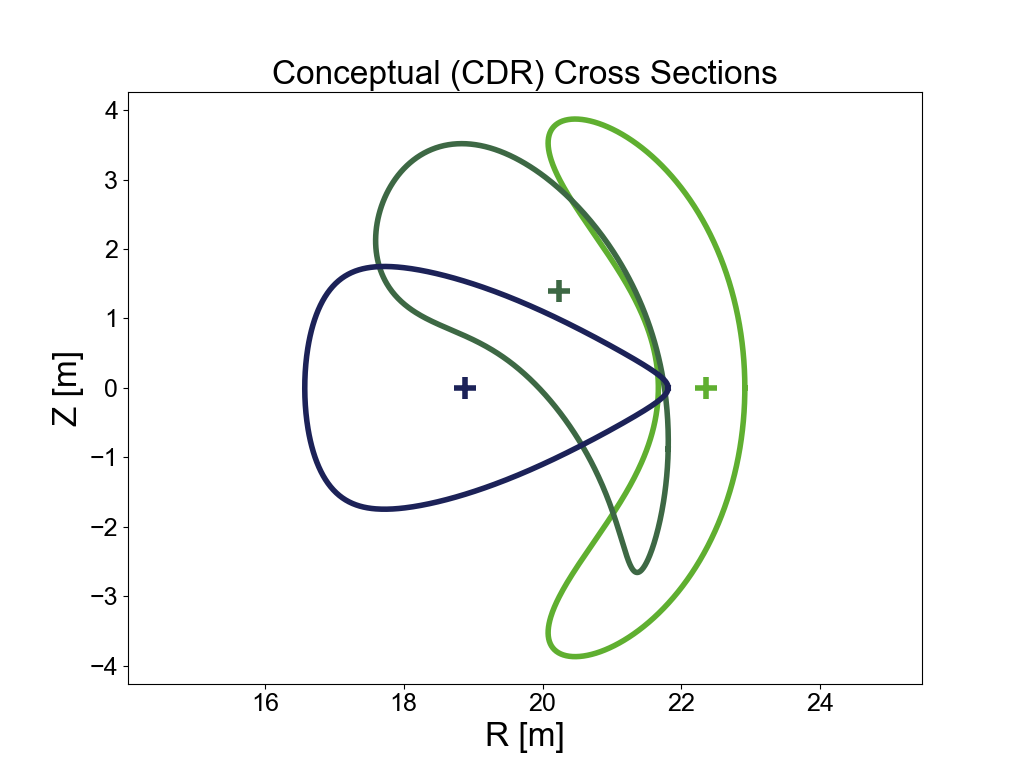}\includegraphics[width=0.49\textwidth]{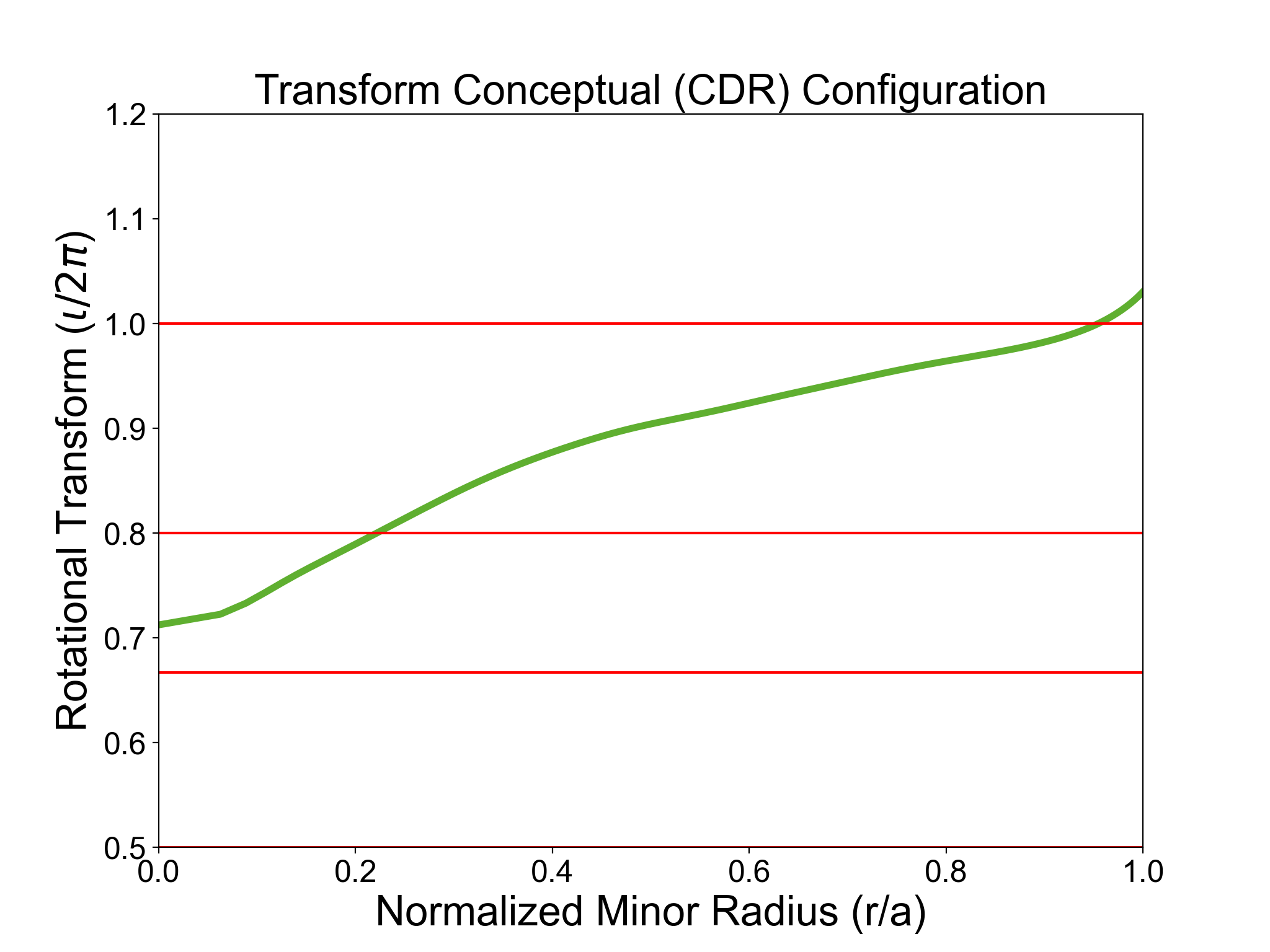}\\
        \includegraphics[width=0.49\textwidth]{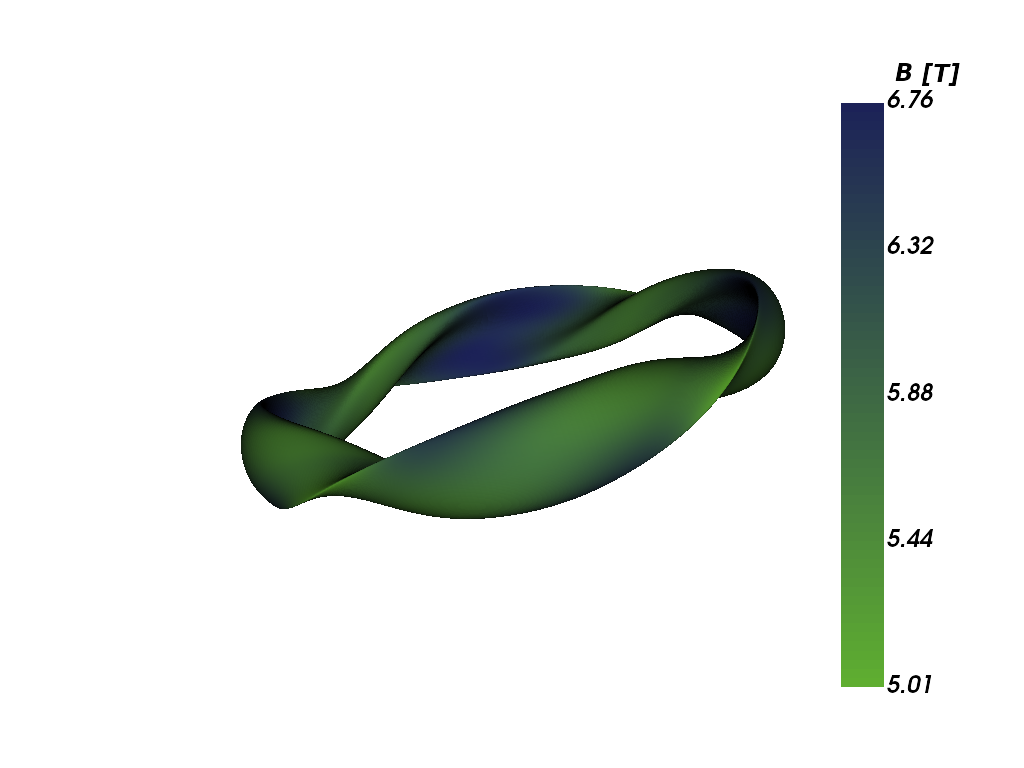}\includegraphics[width=0.49\textwidth]{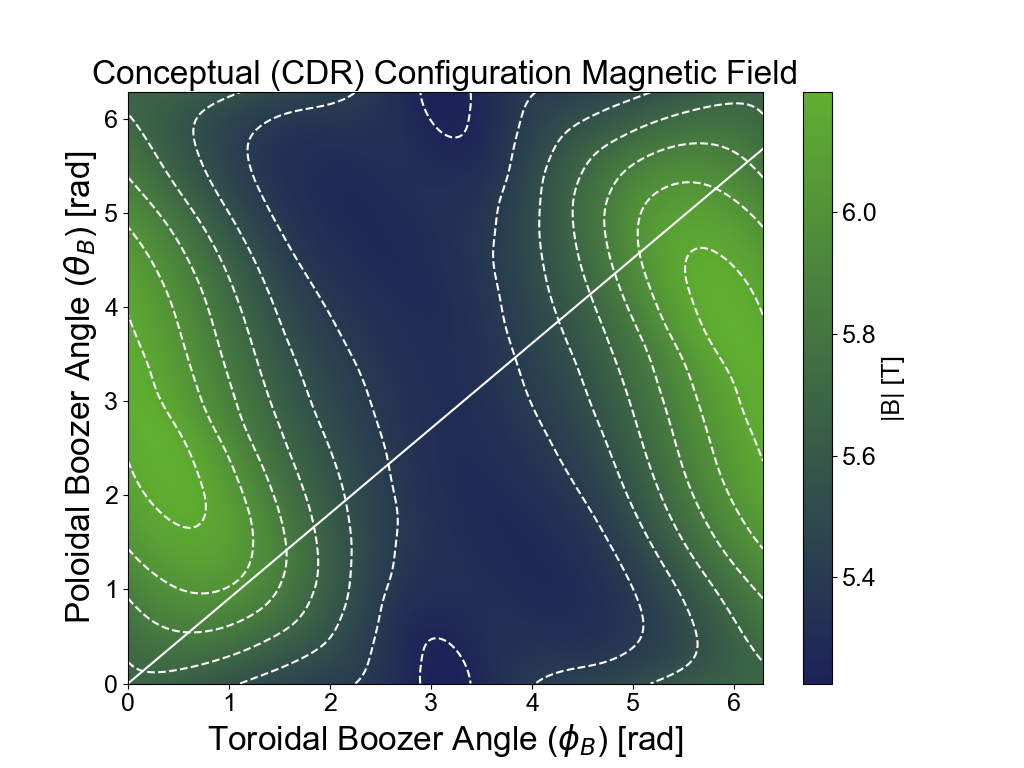}\\
 \caption{Boundary shape (upper left), rotational transform (upper right), 3D B-field pattern (lower left), and B-field pattern at mid radius (lower right) for the conceptual design equilibrium (GIGA\_v515). Cross sections shown at zero, quarter, and half field period. Red lines in rotational transform plot depict low order rationals. Solid white line in Boozer coordinate plot depicts the field line.}
\label{fig:GIGA_v515}
\end{figure}

These optimizations culminated in the conceptual design plasma (CDR plasma) which served as the basis for the engineering conceptualization of the GIGA power plant (figure \ref{fig:GIGA_v515}). This equilibrium fulfilled many of the geometric and magnetic requirements of the GIGA power plant, while requiring additional effort to fulfill all plasma physics requirements. In particular it was found that the configuration suffered from very large bootstrap current while the rotational transform was too close to the $n/m=4/5$ resonance in the core. The configuration also showed ballooning stability, magnetic well, and Mercier stability. However, it suffered from kink instabilities when the bootstrap current was accounted for in the equilibrium. The fast ion confinement was found to be only marginal for a reactor scenario, although significantly improved over the initial equilibrium. The $85\%$ power confinement requirement was only just met in detailed slowing down simulations. Assessment of the transport levels did suggest that turbulence had been significantly reduced and neoclassical transport levels were low. Additional neoclassical assessments did not find core electron root confinement conditions (CERC). The details of these statements will be shown in section \ref{sec:qualirec}.

\subsection{Development of the Evolved Plasma Configuration}

In order to address the shortcomings of the conceptual design plasma, optimization was continued with new focus on reducing bootstrap, improving fast ion confinement, and achieving CERC conditions. Initial attempts to reduce the bootstrap current made use of the interface between BOOTSJ and STELLOPT. This was found to be insufficient in reducing the bootstrap current.  A combination of the DKES based proxy function and direct optimization through the inclusion of PENTA in STELLOPT were performed. Improvement of the fast ion confinement was achieved through continued minimization of the $\Gamma_C$ metric. It was found that below a value of 0.04, all deeply trapped particles were collisionlessly well confined. Once the bootstrap had been reduced, initial attempts were made to achieve CERC by targeting a finite effective helical ripple while minimizing a set DKES $D_{11}$ coefficients. This did not prove successful, motivating the interfacing of the PENTA code to STELLOPT. With PENTA interfaced to STELLOPT both the radial electric field in the core and bootstrap current density could be targeted. Successive optimization resulted in development of the evolved GIGA equilibrium, which has met all the requirements in table \ref{tab:targets}.

\begin{figure}
 \centering
        \includegraphics[width=0.49\textwidth]{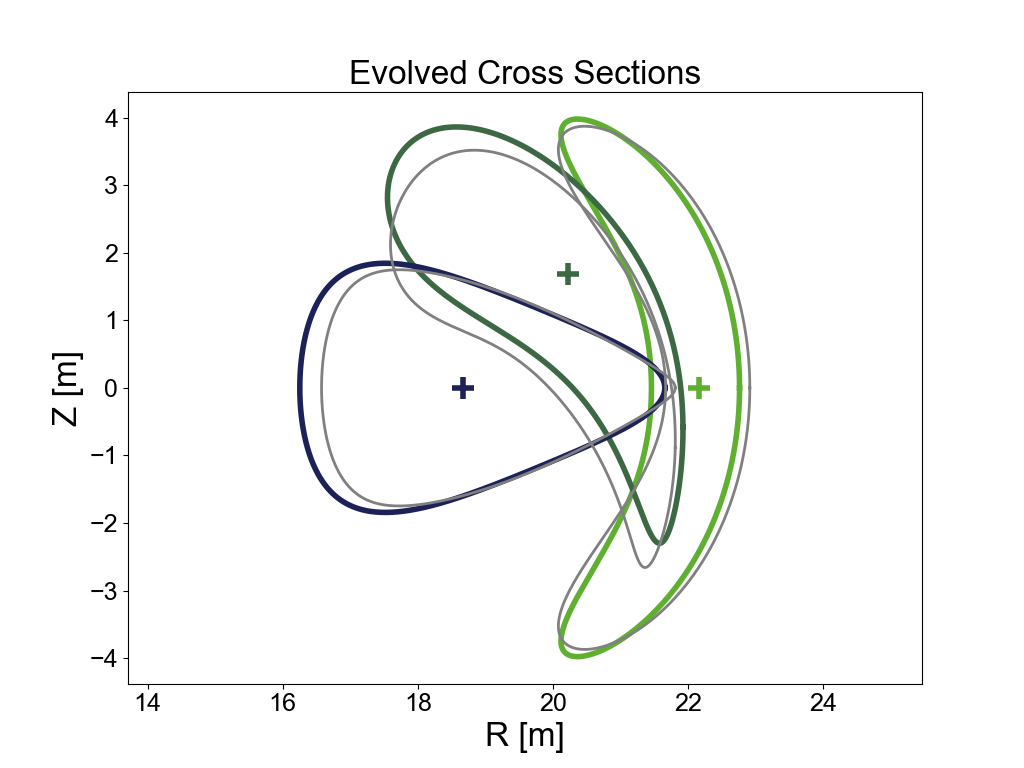}\includegraphics[width=0.49\textwidth]{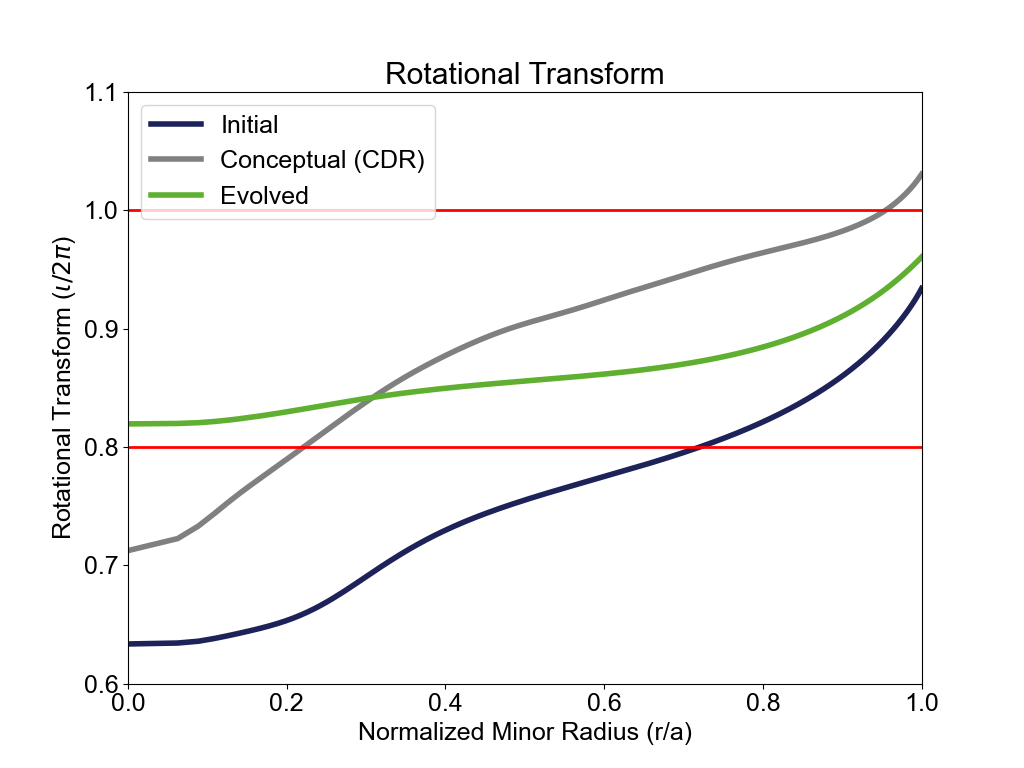}\\
        \includegraphics[width=0.49\textwidth]{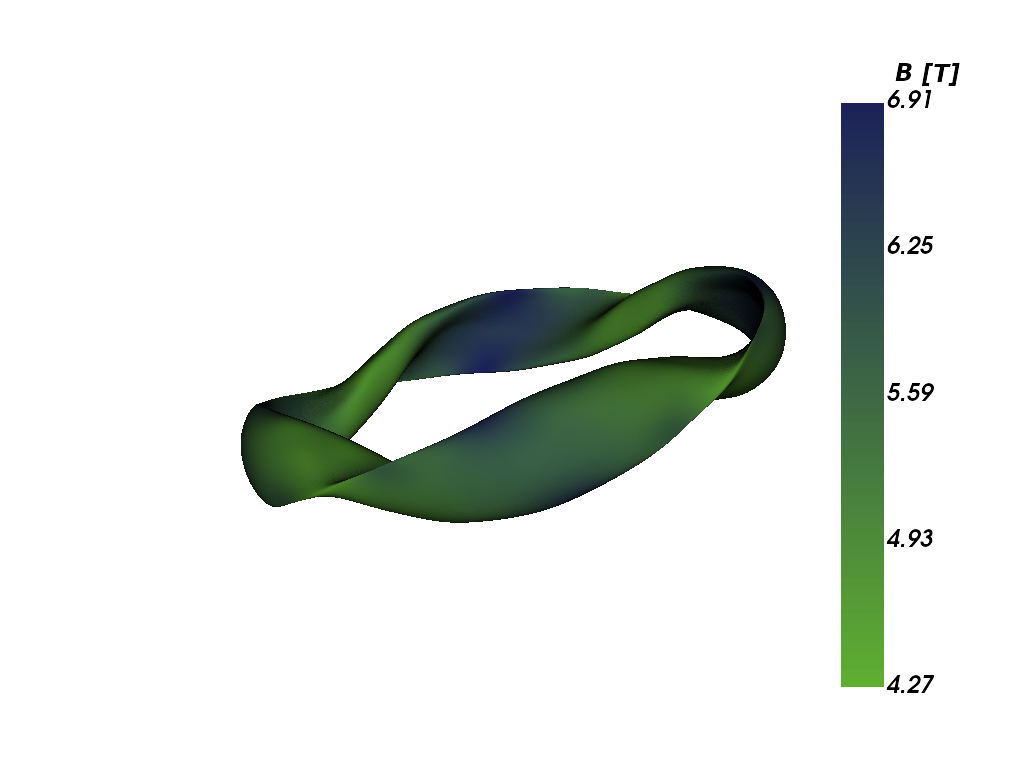}\includegraphics[width=0.49\textwidth]{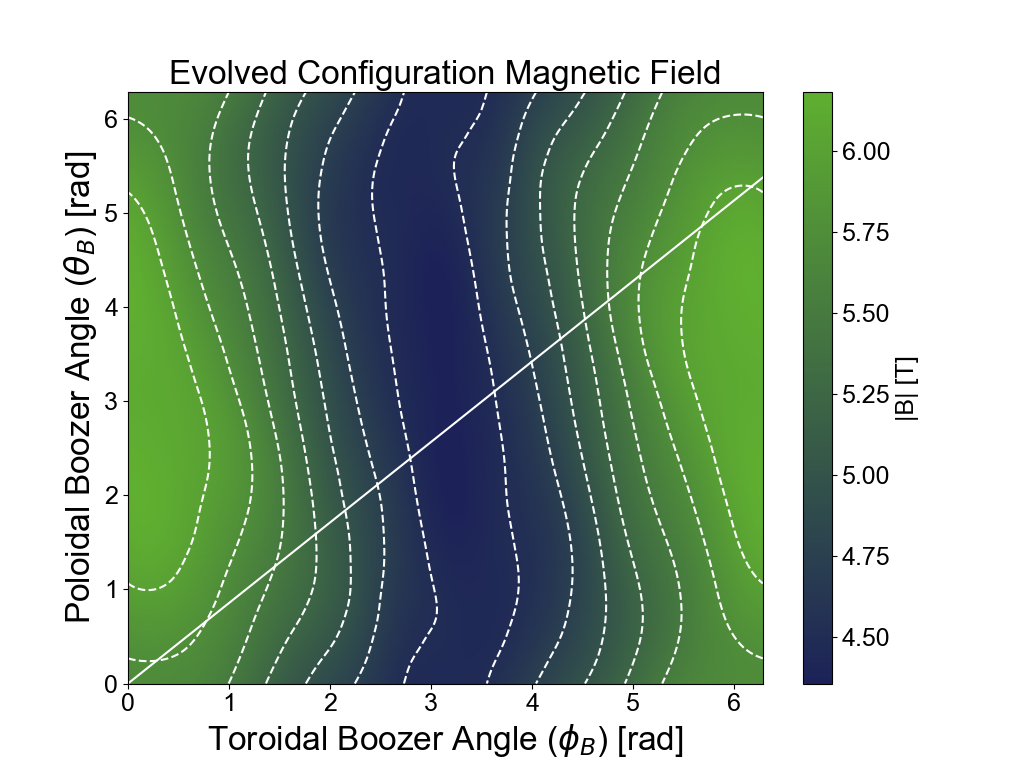}\\
 \caption{Boundary shape (upper left), rotational transform (upper right), 3D B-field pattern (lower left), and B-field pattern at mid radius (lower right) for the evolved equilibrium (GIGA\_v549). Cross sections shown at zero, quarter, and half field period. The conceptual design equilibrium is depicted in grey for reference. Red lines in rotational transform plot depict low order rationals. Solid white line in Boozer coordinate plot depicts the field line.}
\label{fig:GIGA_v549}
\end{figure}

The resulting evolved configuration is depicted in figure \ref{fig:GIGA_v549}.  
The depictions of the equilibrium cross sections show that this evolved configuration is only a subtle change from the conceptual design equilibrium from which it was developed. 
Comparing the contours of magnetic field strength at mid-radius, we find significant differences, despite the rather small boundary variation. 
This highlights how small geometric changes in the equilibrium can result in large changes to the magnetic field, and hence underlying physics. 
The on axis magnetic field decreased by approximately 25\% as we move toroidally.
The plot of the rotational transform shows the initial, conceptual, and evolved configurations including the bootstrap current as computed by the THRIFT code. 
The rotational transform of the evolved configuration avoids the $n/m=4/5$ resonance in the core while providing access to a $n/m=4/4$ edge resonance just outside the plasma.

Optimization of the evolved configuration from the conceptual design configuration involved a few changes to the optimization procedure. The largest of which was to switch from the GADE algorithm to the modified Levenberg-Marquardt. This relaxed the necessity to bound the input boundary harmonics. Additionally, after each optimization, the THRIFT code was used to compute the bootstrap current, which was then incorporated into the equilibrium computation of the next simulation. In order to better resolve the core region in PENTA calculations, the equilibrium radial resolution was increased from 128 to 256 radial grid points in the VMEC computations.

\begin{figure}
 \centering
        \includegraphics[width=0.49\textwidth]{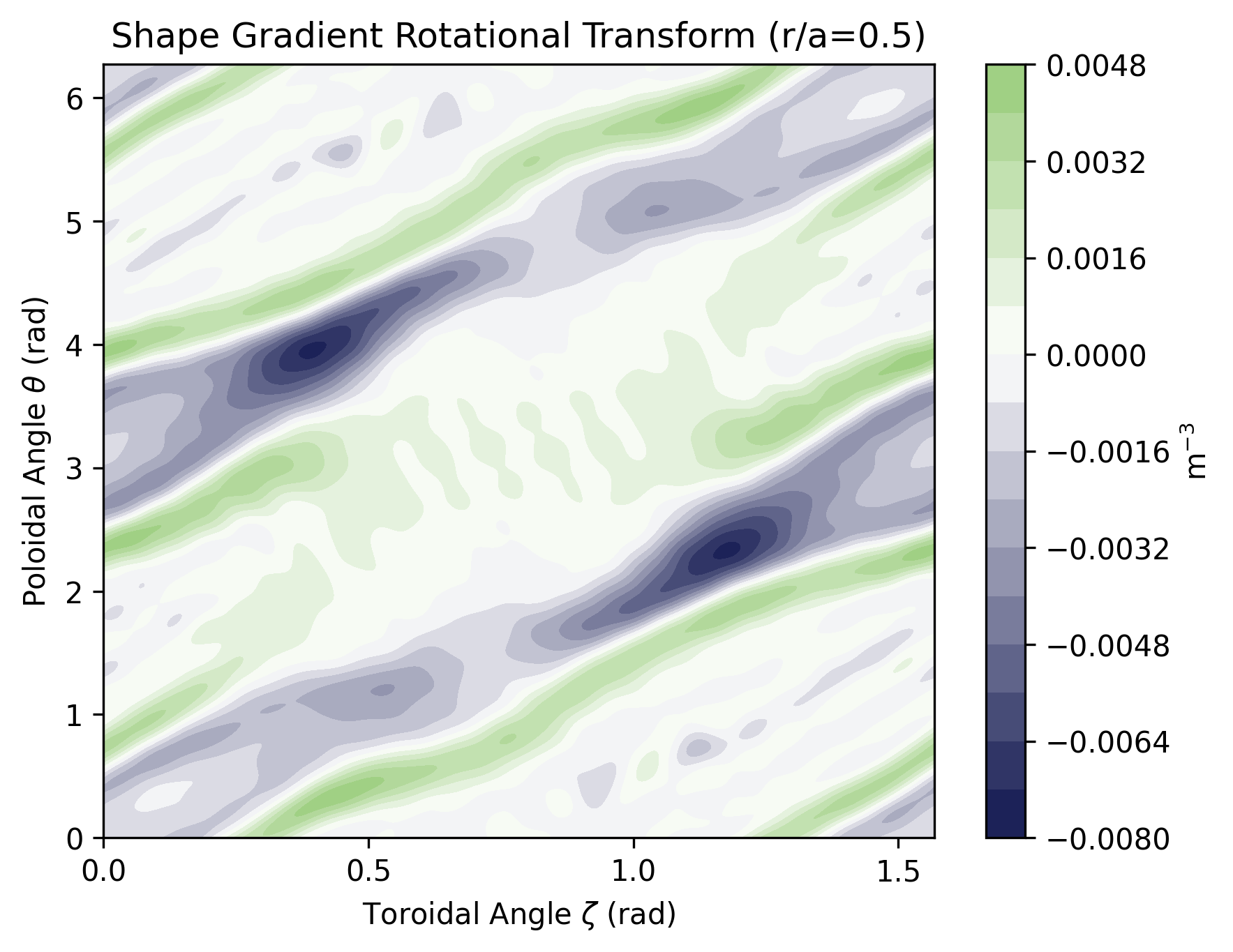}\includegraphics[width=0.49\textwidth]{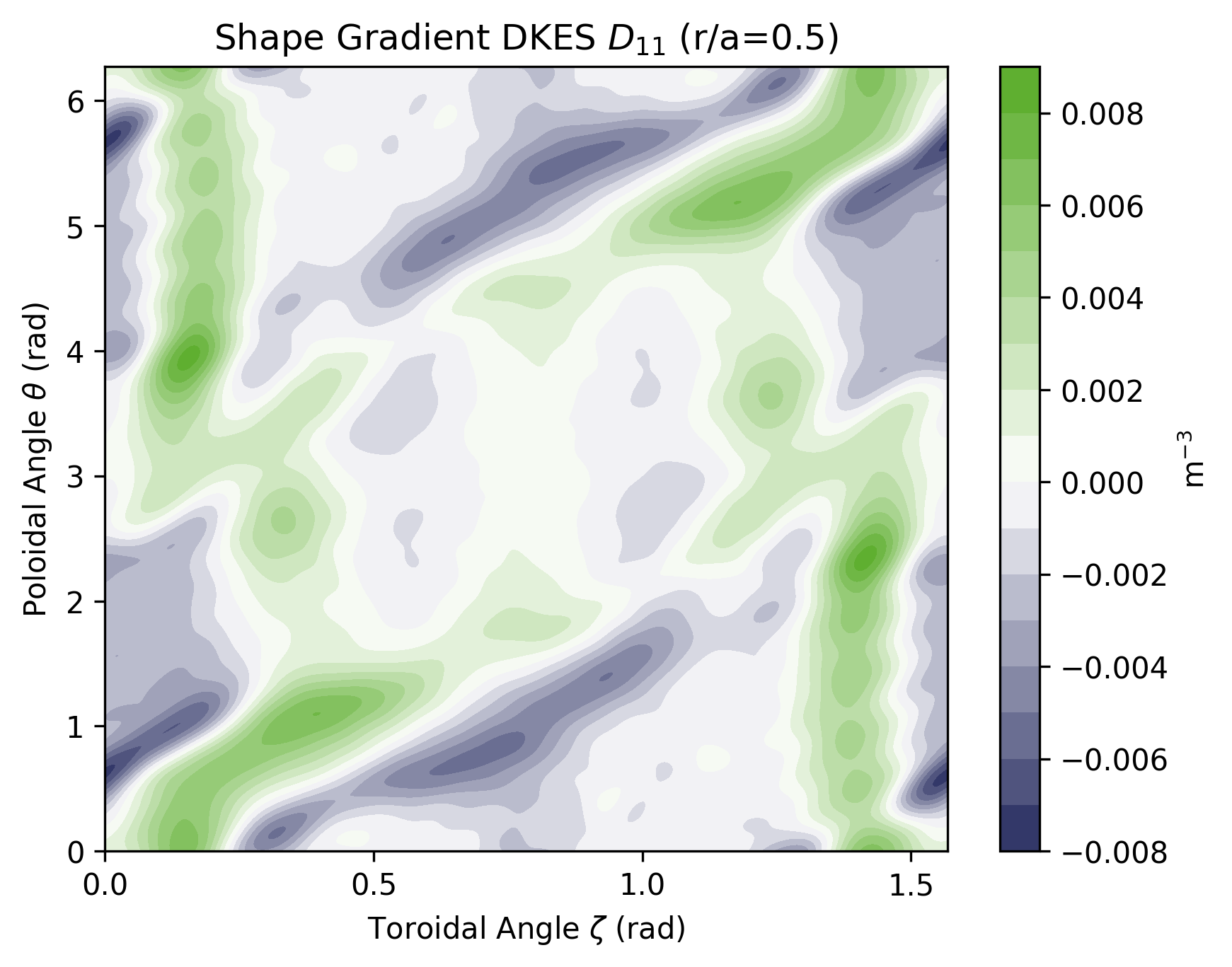}\\
        \includegraphics[width=0.49\textwidth]{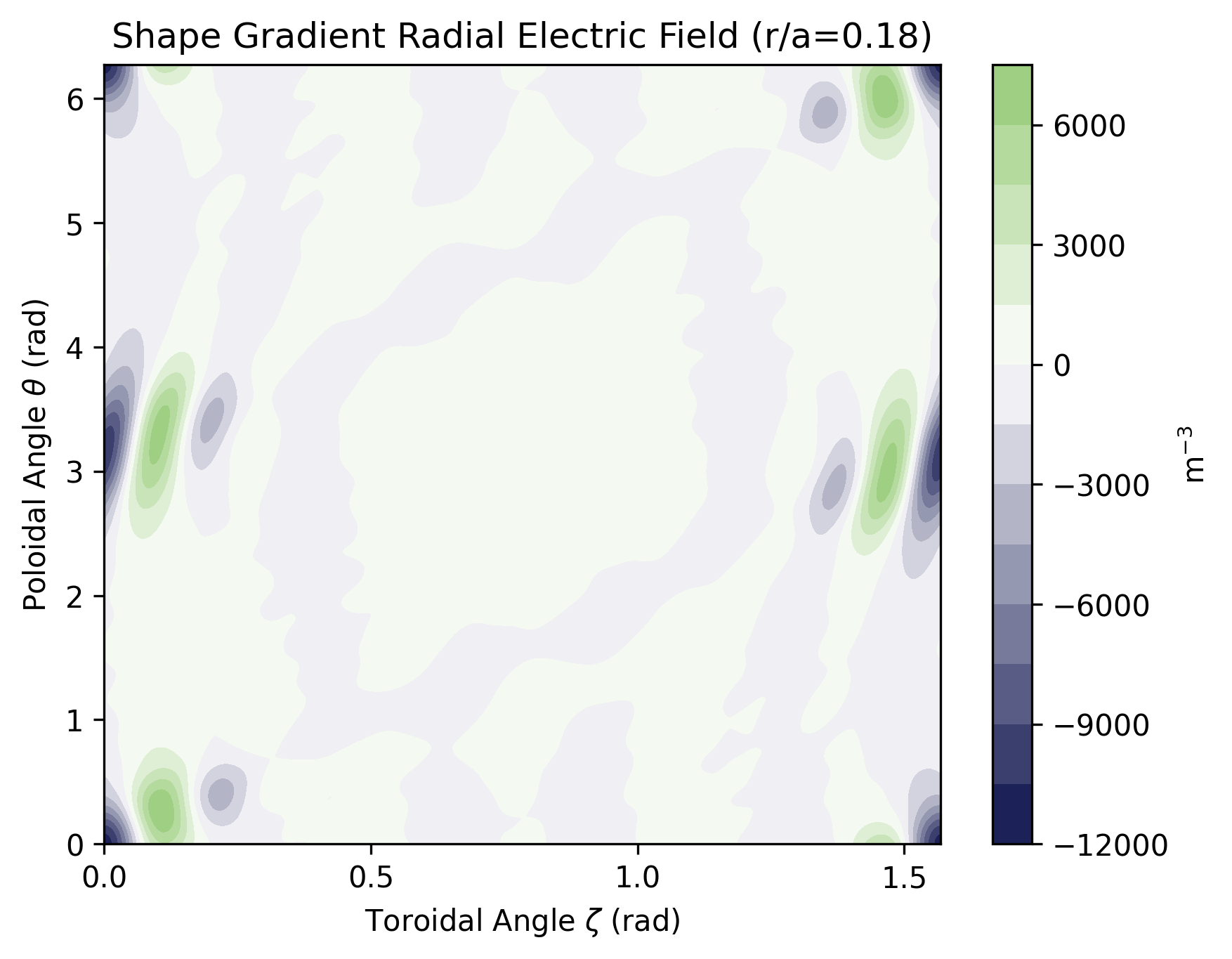}\includegraphics[width=0.49\textwidth]{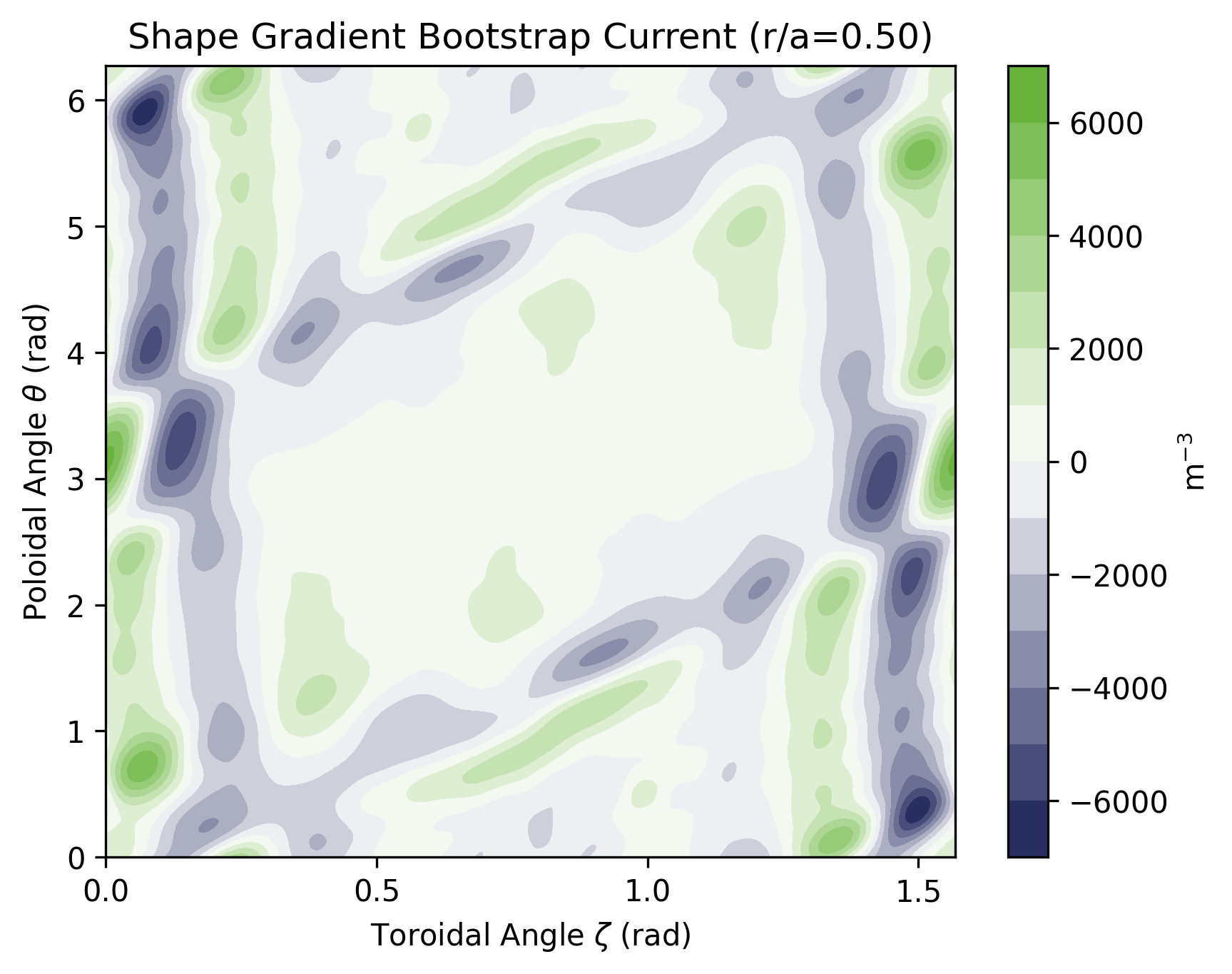}\\
 \caption{Shape gradients for the evolved equilibrium (GIGA\_v549). The toroidal angle ($\zeta=N_{fp}*\phi$) is defined over a field period.}
\label{fig:shapegrad}
\end{figure}

The shape gradients were computed for the evolved configuration to better understand how the shaping of the plasma impacts the plasma parameters \cite{landreman_computing_2018}. 
Shape gradients are a way of expressing how specific boundary variations relate to specific changes in a computed quantity, usually represented as radial displacements of the boundary.
The computation of the shape gradient made use of the STELLOPT Jacobian computation as generated by the Levenberg-Marquardt algorithm. 
First we note that not all quantities have a meaningful shape gradient, as some have a non-vanishing tangential displacement.
Such non-vanishing tangential displacements suggest that a given metric has sensitivity to the angle parameterization in VMEC.
The shape gradient appears to exist for the rotational transform at $r/a=0.5$, $0.70$, and $0.86$, while at the axis and edge there is a strong sensitivity to tangential variations.
The $\Gamma_C$ proxy, $g^{\rho\rho}$ proxy, ballooning stability, and neoclassical effective ripple all show a strong sensitivity to tangential variation.
The magnetic symmetry constraint shows some sensitivity to tangential variations but not nearly as strong when compared with the other metrics. 
Still a meaningful shape gradient could not be computed for the magnetic symmetry metric.
When a given shape gradient is insensitive to tangential variations, then we can understand the gradient in terms of radial pushes and pulls on the flux surface in a geometric fashion.
The presence of sensitivities to tangential variations is indicative that the angular parameterization of the boundary is in fact playing a role in the quantity computed.
Thus our simple push-pull picture of how the boundary affects the plasma physics breaks down.
The mono-energetic transport coefficient ($D_{11}$) shows no sensitivity to tangential variations allowing for a clear shape gradient to be computed (figure \ref{fig:shapegrad}).
Examining the radial electric field and bootstrap targets as computed by PENTA, we find that shape gradients exist for all but the inner most grid point. 

We can generate a few insights from the shape gradients in figure \ref{fig:shapegrad}. First, the rotational transform at mid-radius is sensitive to a helical pulling of the plasma. The helical stripe aligns with the points of maximum gradient, suggesting that increases in iota at mid-radius are most strongly driven by additional plasma elongation. The $D_{11}$ shape gradient suggests that a reduction in neoclassical transport can be achieved through a squeezing of the bean shaped cross section ($\zeta=0$). The core radial electric field appears sensitive to very high-order modes about the bean and triangular cross sections. The bootstrap current at mid-radius shows a similar sensitivity to the mono-energetic transport coefficient $D_{11}$, although slightly shifted in the toroidal angle. This is not surprising as both are neoclassical effects as measured at mid-radius.

\section{Qualification Record} \label{sec:qualirec}

\begin{table}
\caption{Design values of the GIGA equilibria showing their fulfillment of the requirements of table \ref{tab:targets}. The rotational transform is evaluated without bootstrap current.}
\centering
\begin{tabular}{ l | c | c c c}
\hline
Target & Unit & Initial Equilibrium & Conceptual Design & Evolved Equilibrium\\
\hline
Edge Rotational Transform & --- & 0.934 & 0.962 & 0.963 \\
Core Rotational Transform & --- & 0.634 & 0.777 & 0.818 \\
Alpha Power Confinement & MW & --- & 520 & $>520$ \\
Effective Helical Ripple ($\epsilon^{3/2}_{eff}$) & --- & 0.00163 & 0.00044 & 0.00111  \\
Maximum Heat Flux & $\textrm{kW/m}^2$ & --- & 370 & 430 \\
Net Toroidal Current & $kA$ & -1550 & -766 & 20  \\
Ballooning Stability & --- & Unstable & Marginal & Stable \\
Kink Stability ($n=0$) & --- & Unstable & Stable & Stable \\
Kink Stability ($n=1$) & --- & Unstable & Unstable & Stable \\
Kink Stability ($n=2$) & --- & Unstable & Unstable & Stable \\
Alfv\'enic Stability & ---  & No core-edge gaps & No core-edge gaps & No core-edge gaps  \\
Core Electric Field ($r/a=0.2$) & kV/m  & -5.3 & -4.4 & 7.4
\end{tabular}
\label{tab:achieved}
\end{table}

The qualification record of the GIGA equilibrium documents the evidence that the GIGA equilibrium has fulfilled the requirements set forth in its design (table \ref{tab:targets}).  Table \ref{tab:achieved} provides an overview of the parameters fulfilling these requirements, highlighting how the conceptual design equilibrium was an improvement over the initial equilibrium but did not fulfill all requirements.  It also highlights how the evolved equilibrium was able to fulfill all the design requirements. For the purposes of this section the same profiles used for equilibrium optimization are utilized in the analysis, unless otherwise noted. While the analyses which provide these values are the subject of future works, we provide an overview of each of these parameters in this section. The mirror term for the initial, conceptual and evolved equilibrium was 9\%, 6\% and 15\% respectively.  Here the mirror term is defined as $(B_0(\zeta=0)-B_0(\zeta=\pi))/(B_0(\zeta=0)-B_0(\zeta=\pi))$, where $B_0$ is the on axis magnetic field and $\zeta$ is a toroidal angle which runs from 0 to $2\pi$ over a field period. By comparison, the W7-X standard configuration has a mirror term around 4\% and the high-mirror configuration has 10\%.

\begin{figure}
 \centering
        \includegraphics[width=\textwidth]{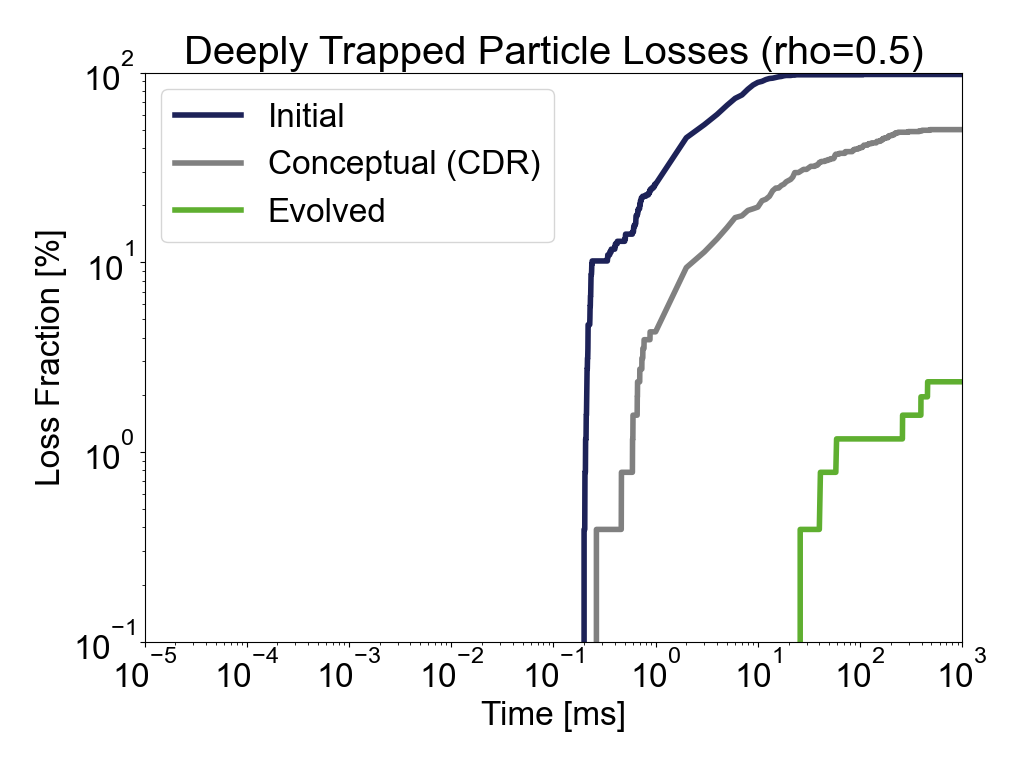}
 \caption{Time evolution of collisionless deeply trapped alpha particle losses from mid-radius. Losses for the deeply trapped particles in the evolved configuration are at $1\%$.}
\label{fig:alpha}
\end{figure}

That fusion alphas are well confined is critical to the success of a fusion power plant, both increasing the effective heating power and reducing fast ion wall loads. 
A detailed assessment of the conceptual design equilibrium was undertaken using the BEAMS3D and ASCOT5 codes \cite{vtt_2026}. 
Here BEAMS3D provided the birth profile for slowing down computations while ASCOT5 was used to simulate collisional gyro-center and gyro-orbit motions of the alpha particles. 
Such simulations showed that the conceptual design equilibria had only just achieved the requirement of $85\%$ alpha power confinement at full power.  
For comparison, a similarly scaled-up W7-X configuration had $25\%$ power loss and $33\%$ particle loss \cite{lazerson_simulating_2021}. 
It was noted that confinement was slightly worse (by a few percentage points) when assessing profiles based on the Cordey pass and burn datapoints from the 0.5D modeling.  
While the evolved configuration has yet to be modeled in such a way, collisionless simulations of deeply trapped particles show a significant improvement in confinement. 
Figure \ref{fig:alpha} depicts the time evolution of losses originating from the mid-radius trapped population. In this simulation a population of 256 alphas scanning $v_\parallel/v$ from -0.5 to 0.5 were followed. 
To sample the deeply trapped region the particles were launched from the point of minimum magnetic field on the mid-radius flux surface. 
Particles launched from $r/a=0.25$ and $r/a=0.75$ also showed similar improvements and the onset time for losses to occur was similarly shifted to longer timescales.
While not conclusive, such simulations imply that the alpha confinement should be much better for the evolved configuration as compared to the conceptual design equilibrium.
A discussion of wall loads is beyond the scope of this work as the edge magnetic structure, scrape-off-layer profiles, and wall shape play a strong role in localization of loads.

\begin{figure}
 \centering
        \includegraphics[width=0.32\textwidth]{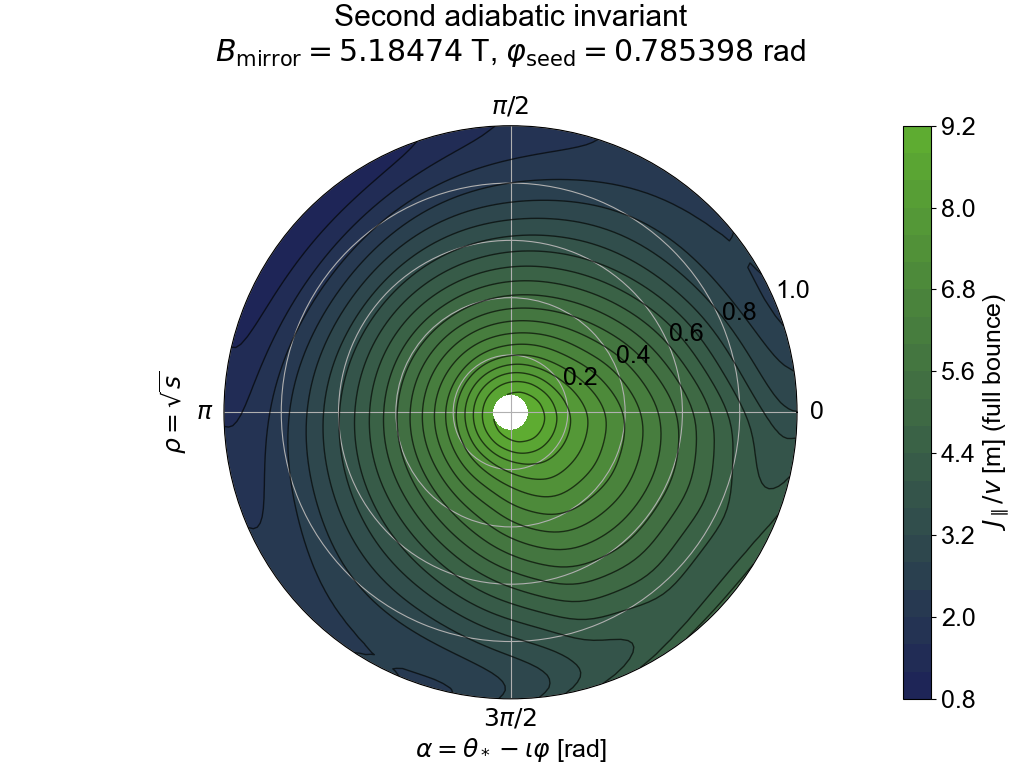}\includegraphics[width=0.32\textwidth]{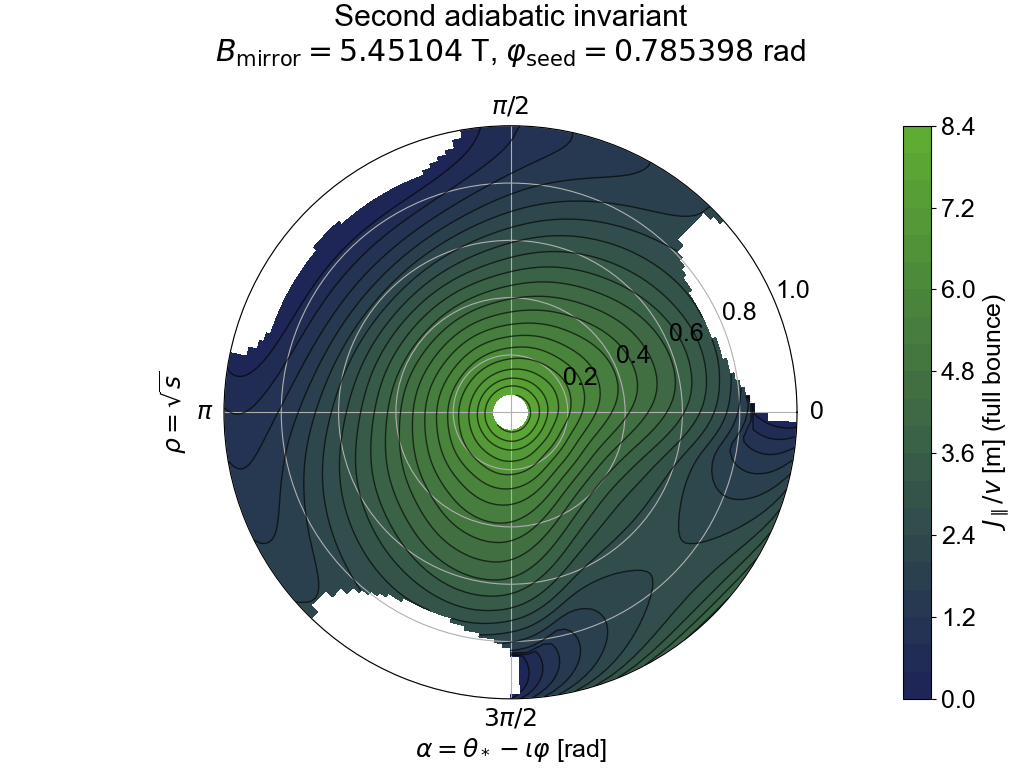}\includegraphics[width=0.32\textwidth]{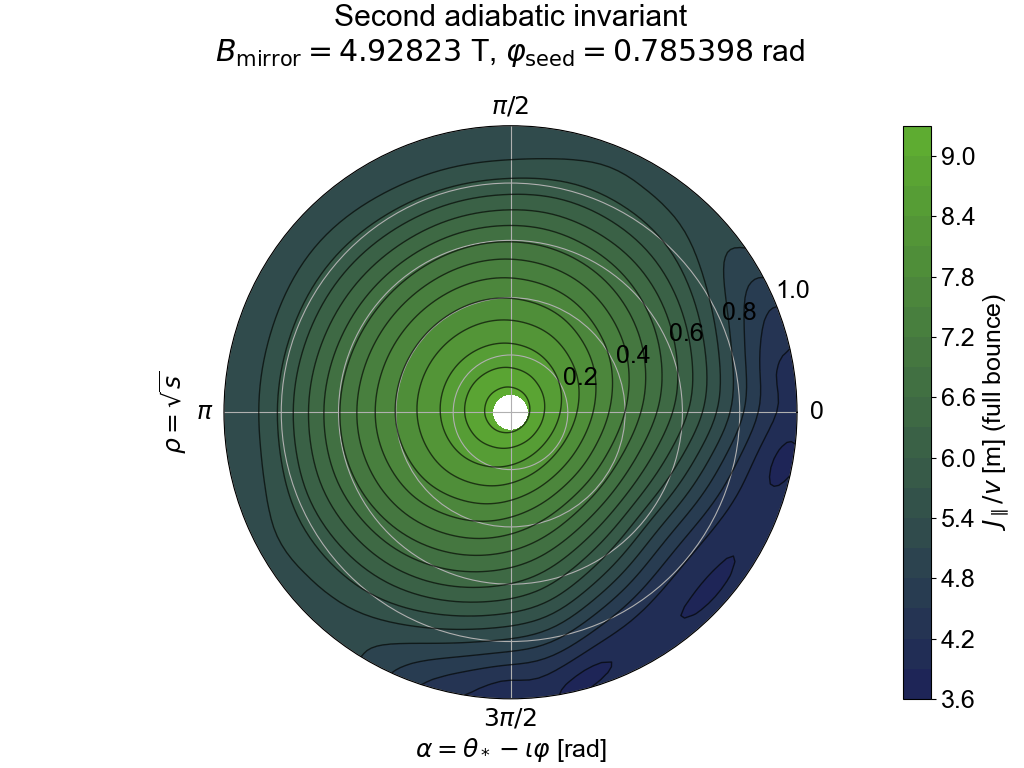}
 \caption{Contours of the 2nd adiabatic invariant for the initial (left), CDR (middle), and evolved (right) configurations. Mirroring magnetic field was chosen as 25\% the well depth. Regions where values are not plotted indicate an inability to compute the necessary value. This can be indicative of trapped-passing boundaries, but also numerical issues. The value on axis is no computed.}
\label{fig:2ndadia}
\end{figure}

In figure \ref{fig:2ndadia} we see a comparison of the 2nd adiabatic invariant for each configuration.
For reference, the 2nd adiabatic invariant is defined for a particle as
\begin{equation}
J_\parallel = \oint v_\parallel\mathrm{d}l = 2v \int_{l_1}^{l_2} \sqrt{1-\lambda B(s,\alpha,l)}\mathrm{d}l,
\qquad
\lambda = \frac{\mu}{E} = \frac{1}{B_{\mathrm{mirror}}}
\end{equation}  
where $v_\parallel$ is the parallel particle velocity, $l_{1,2}$ are the bounce points, $B$ the magnetic field strength, $\mu$ the particle magnetic moment, $E$ the particle energy, $B_{\mathrm{mirror}}$ the mirroring magnetic field strength, and the integral is over the particle bounce trajectory.
For the plots show the following equation was used
\begin{equation}
\frac{J_\parallel}{v}
= 2 \int_{l_1}^{l_2}
  \sqrt{1-\frac{B(s,\alpha,l)}{B_{\mathrm{mirror}}}}
  \,\mathrm{d}l.
\end{equation}
In making this comparison a choice must be made of magnetic mirror field to make comparable plots.
In this case the mirror magnetic field corresponding to 25\% the depth of the field variation was chosen.
Generally we see that each configuration has a higher value for $J/v_parallel$ in the core than at the plasma edge, implying a maximum-J like nature.
Although it is important to note that this feature is not well aligned to the flux surface everywhere and departures exist at the edge.
The initial configuration has open contours at $\rho=0.5$ which explains why deeply trapped particles are completely lost from this configuration.
The contours of the CDR configuration show some level of closure at mid radius, while some regions at mid-radius intersect possible trapped-passing boundaries.
The evolved configuration shows a large region of closed contours, with all contours inside of $\rho=0.6$ closing within the plasma boundary.
More importantly, there exist trapped orbits in the evolved configuration which close inside the equilibrium even at $\rho>0.8$ in nearly half the field line label domain ($\alpha$).
While only indicative of collisionless confinement, this comparison further supports the statement that fast ion confinement in the evolved configuration is better than that of the CDR configuration.

\begin{figure}
 \centering
        \includegraphics[width=\textwidth]{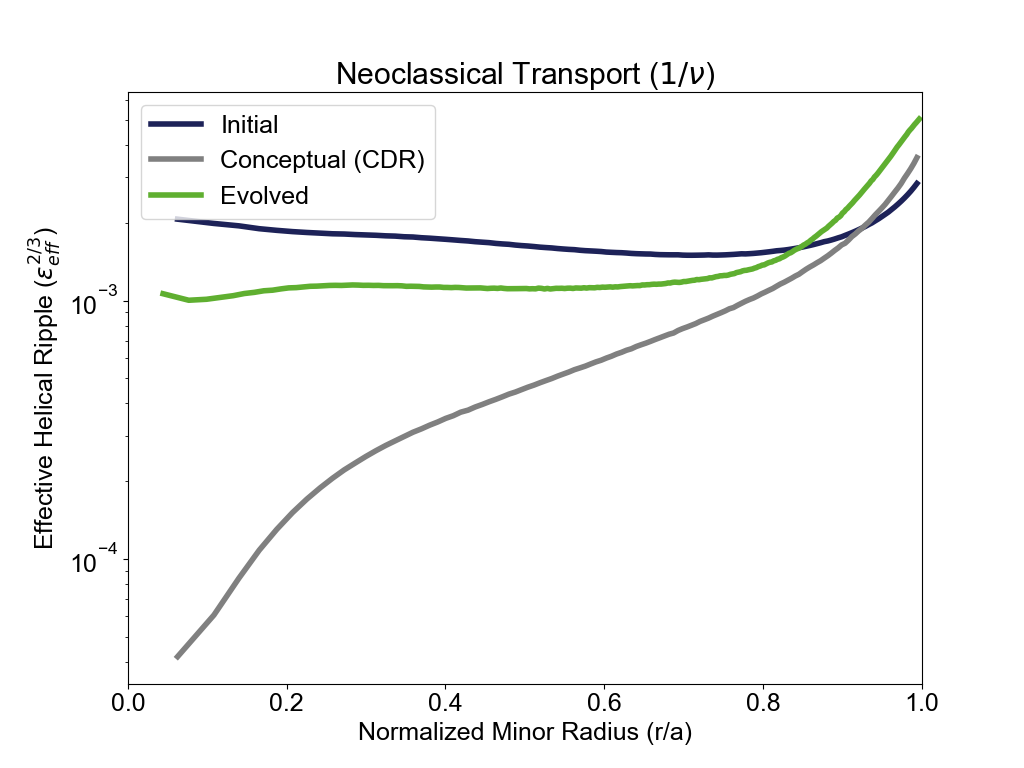}
 \caption{The profile of neoclassical effective helical ripple showing that all three configurations have suppressed neoclassical transport.}
\label{fig:neo}
\end{figure}

The neoclassical effective ripple provides a proxy for the $1/\nu$ collsionality regime and was generally adequate for all three configurations. Figure \ref{fig:neo} shows the neoclassical effective ripple for the three configurations as computed by the NEO code. The conceptual design equilibrium had the lowest neoclassical effective helical ripple, despite having rather large bootstrap current. The increase in the core neoclassical ripple when going from the conceptual to evolved configuration is attributed to attempts to achieve CERC conditions by holding $\epsilon_{eff}^{3/2}$ at a fixed value and decreasing the mono-energetic coefficients. 

\begin{figure}
 \centering
        \includegraphics[width=0.49\textwidth]{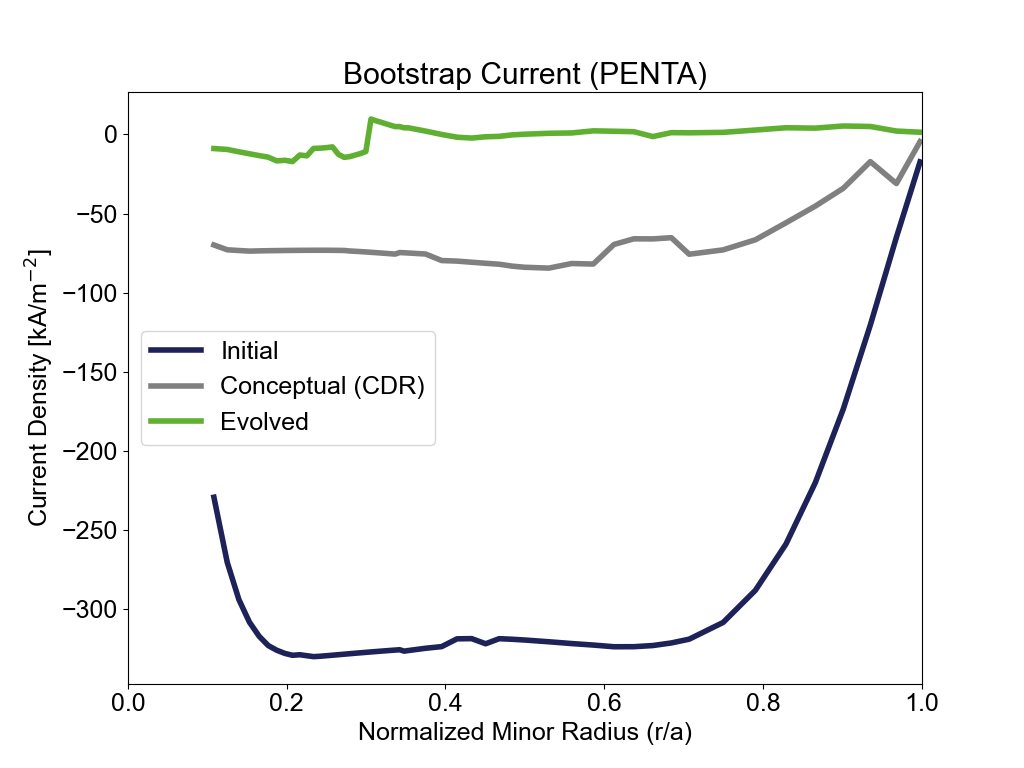}\includegraphics[width=0.49\textwidth]{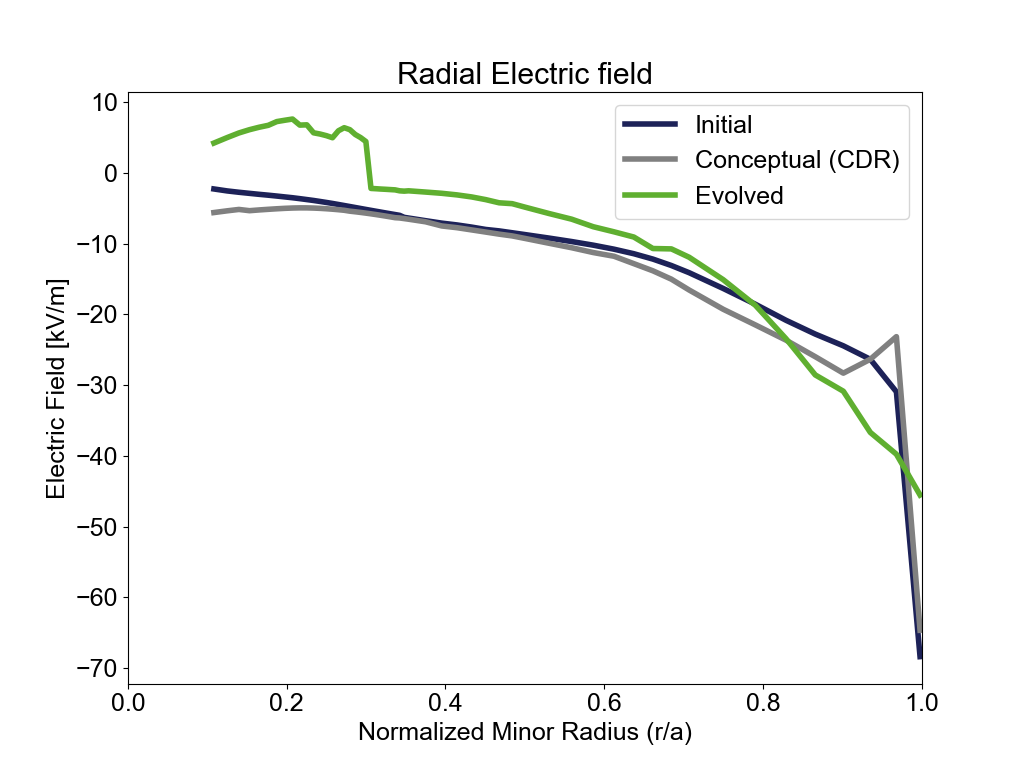}
 \caption{The boostrap current (right) and radial electric field (left) as computed by the coupled PENTA/DKES code. When possible the electron root is plotted in favor of the ion-root solution.}
\label{fig:PENTA}
\end{figure}

The reduction of bootstrap current was a major focus of the optimization of the GIGA evolved equilibrium.
Computations of the steady-state bootstrap current density at full power are depicted in figure \ref{fig:PENTA}. 
These computations were performed using the THRIFT code and its newly implemented DKES/PENTA neoclassical model. 
Time-dependent simulations imply that given the large physical size of GIGA, the inductive response of GIGA occurs on timescales of around 30 minutes. 
This is favorable as fast changes in heating power should not result in fast changes in the edge rotational transform, making divertor strike-line control much less demanding. 
Temperature and density profiles were held constant across configurations. 
It should be recognized that, in these simulations, species profiles play a strong role in determining the bootstrap current. 
This highlights the interdependency between the bootstrap current and the transport problem.

The core electric field was targeted during optimization to be positive through inclusion of the DKES and PENTA codes in STELLOPT. 
In order to provide access to electron root conditions, the temperature profiles were slightly modified away from $T_e=T_i$ to having the electron temperature 5\% higher than the ion temperature while maintaining the pressure profile.
Before additional optimization of the plasma boundary, this choice of kinetic profiles produced only ion root solutions.
As the boundary harmonics were modified through optimization with STELLOPT, electron root solutions became accessible.
During this optimization process the kinetic profiles were not modified.
The values shown here are indicative that CERC conditions are achievable in the evolved equilibrium. 
It should be noted that GIGA will only have electron heating (ECRH and alphas).
Thus, the assumption of $T_e>T_i$ in the core is a reasonable one to make.
We note that the temperature and density profiles play a strong role here, thus achievement of CERC in GIGA will be an integrated transport problem.
Additionally, since the geometry plays a role such analysis should be re-evaluated in free boundary.
Finally, we note that these results are indicative that CERC is possible in GIGA but not guaranteed to be present without additional electron heating (from ECRH for example).

\begin{figure}
 \centering
        \includegraphics[width=0.32\textwidth]{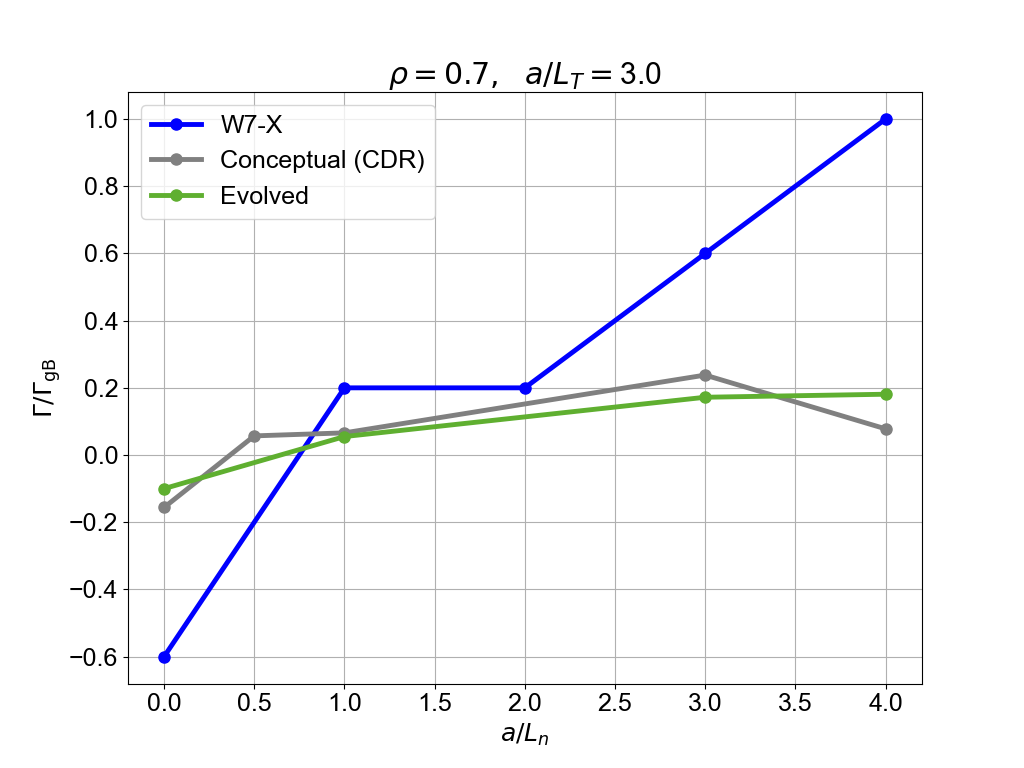}\includegraphics[width=0.32\textwidth]{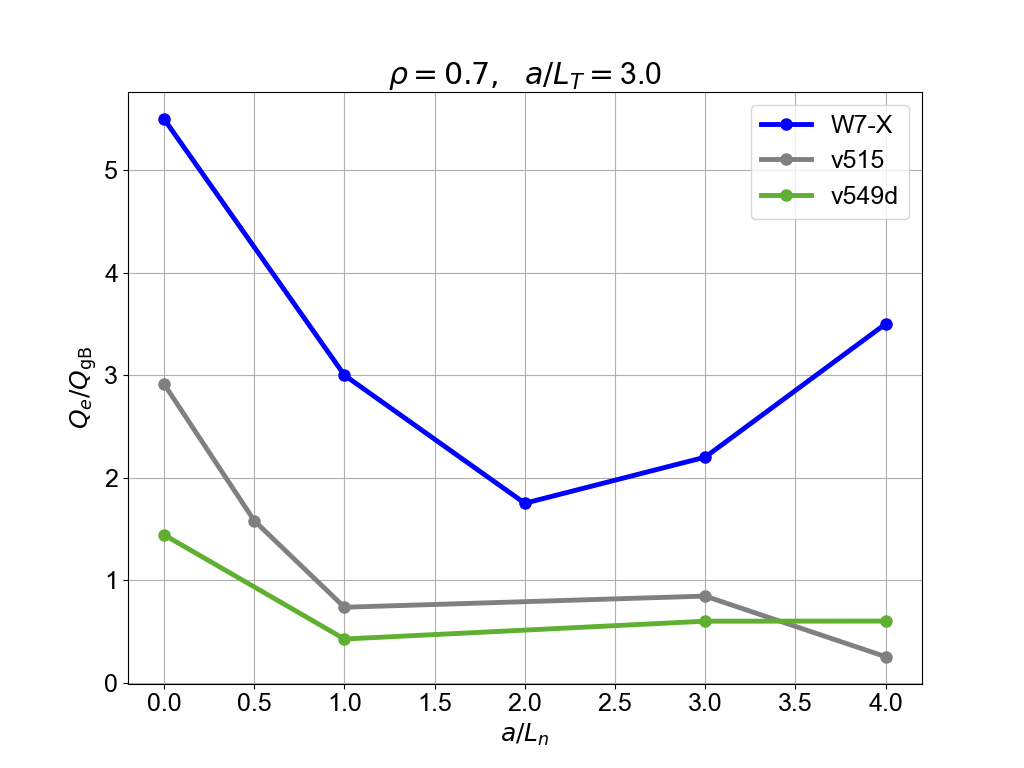}\includegraphics[width=0.32\textwidth]{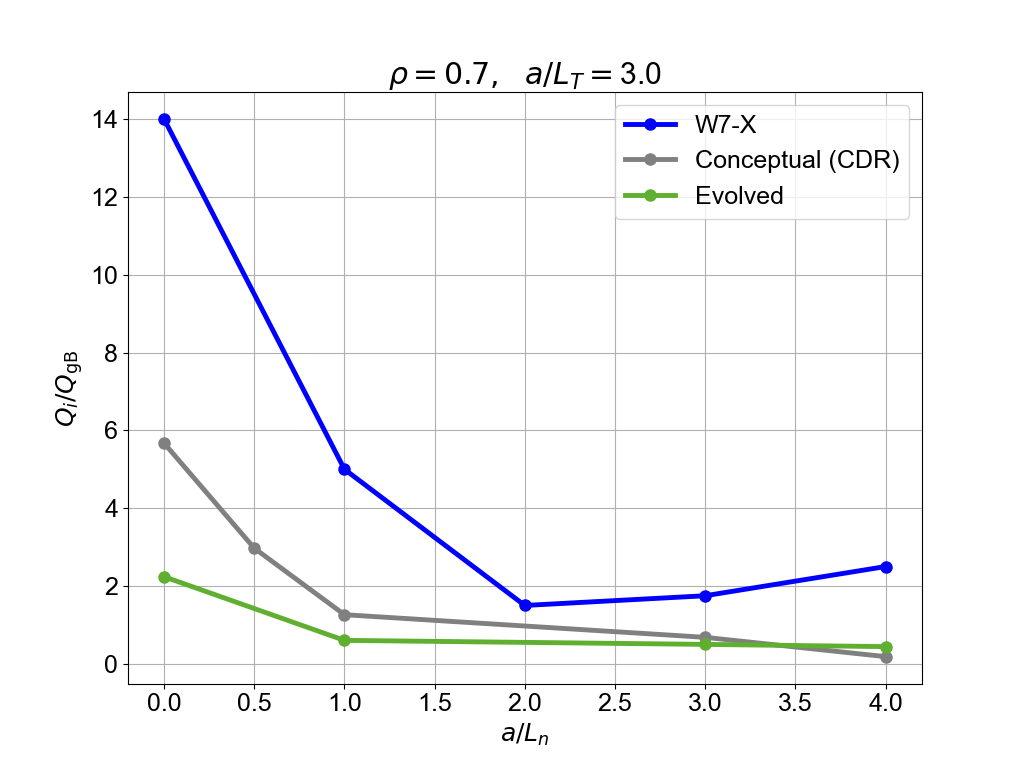}
 \caption{Turbulent ion particle (left), electron heat (center), and ion heat (right) flux as computed by STELLA for a fixed temperature gradient. The data for W7-X is taken from published results \cite{garcia-regana_reduced_2024}.}
\label{fig:txport}
\end{figure}

Figure \ref{fig:txport} depicts the electrostatic turbulent particle and heat fluxes for Wendelstein 7-X, the conceptual design equilibrium, and the evolved equilibrium. 
These electrostatic flux-tube simulations of GIGA were performed at fixed temperature gradient and varying density gradient considering both kinetic hydrogen ions and electrons. 
Overall the simulations show a reduced particle flux as compared to W7-X data in hydrogen \cite{garcia-regana_reduced_2024}. 
Additionally, we can see the effect of the maximum-J nature of all three configurations in the way density gradients suppress the electrostatic turbulence \cite{proll_turbulence_2022}.
We see the effect of the proxy function through a reduction in turbulent heat flux at fixed density gradient when comparing equilibria.
This is generally beneficial, although the turbulent pinch effect has also been reduced for low density gradients. 
The turbulent heat flux highlights the effect of the optimization with both the conceptual design and evolved design showing significantly lower heat fluxes at all density gradients. 
The electron heat flux shows a similar behavior.
Simulations holding the density gradient fixed and varying the temperature gradients have also been performed, confirming these results.
This generally suggests significantly improved turbulent transport characteristics as compared to W7-X.  
It should be noted that an electromagnetic treatment with collisions is necessary to confirm such claims.


\begin{figure}
 \centering
        \includegraphics[width=\textwidth]{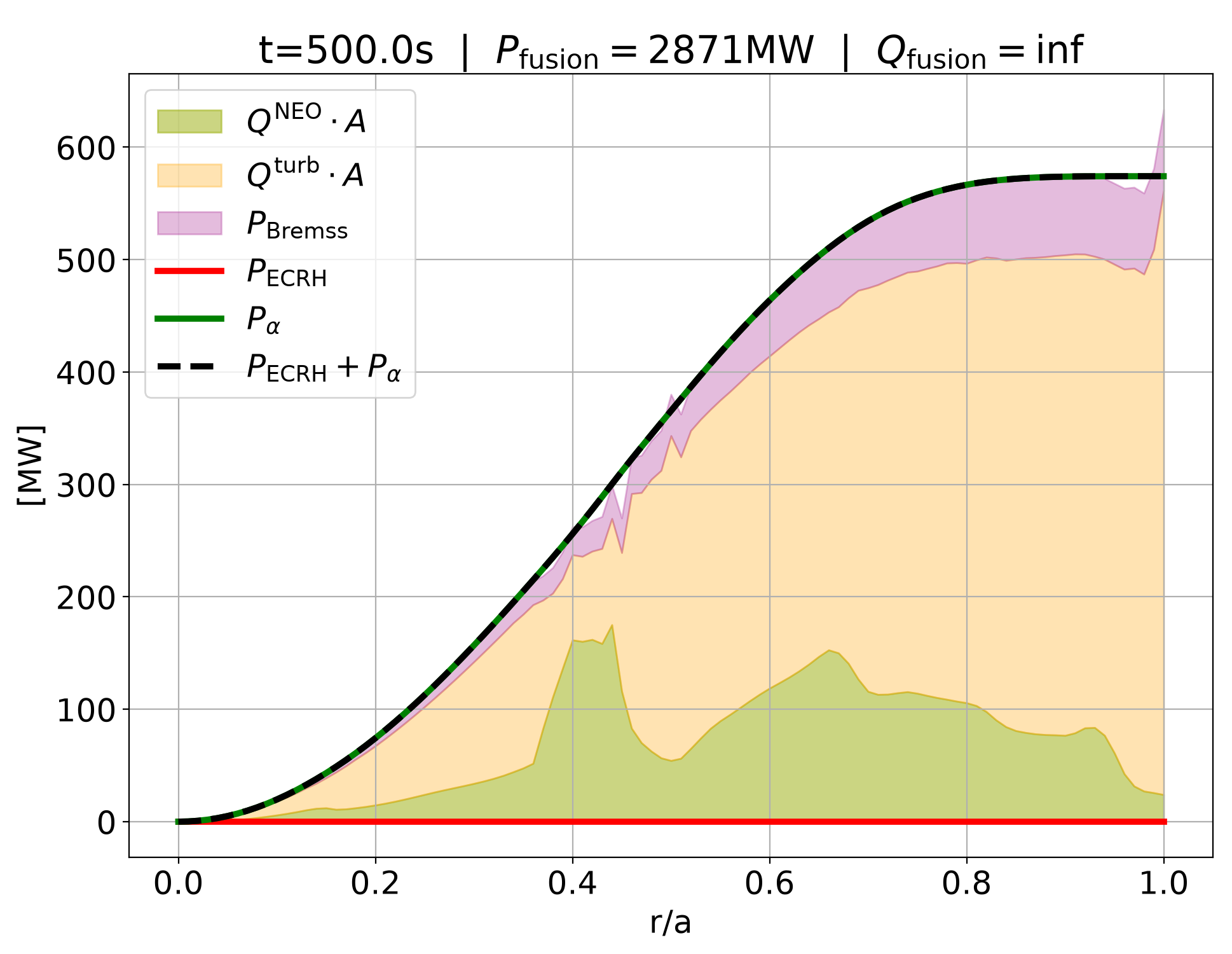}
 \caption{Steady-state integrated heat flux for the evolved configuration assuming diffusive particle transport and a critical gradient model for ion heat flux.  Electron heat diffusivities were assumed to be 0.7 $\textrm{m}^2/s$. No external heating is assumed in this simulation.}
\label{fig:txport_heatflux}
\end{figure}

Initial estimate of the steady-state transport based on the full power scenario predicted heat fluxes approaching $\sim 430~\textrm{kW/m}^2$.
This is below the upper limit of $600~\textrm{kW/m}^2$, suggesting that should temperature profiles be allowed to evolve, the temperatures would increase.
Assuming diffusive particle transport and a critical gradient based heat flux model \cite{beurskens_ion_2021}, the profiles were allowed to evolve.
Coefficients for the critical gradient model were generated from a database of turbulence runs scanning radial location, density gradient, and temperature gradients.
Meanwhile only heating from the alphas was considered, adjusting for temperatures and densities.
Figure \ref{fig:txport_heatflux} depicts the integrated heat flux for this scenario.
Such results show that at this level of transport optimization (turbulent and neoclassical) is sufficient to support a burning plasma state.

\begin{figure}
 \centering
        \includegraphics[width=0.32\textwidth]{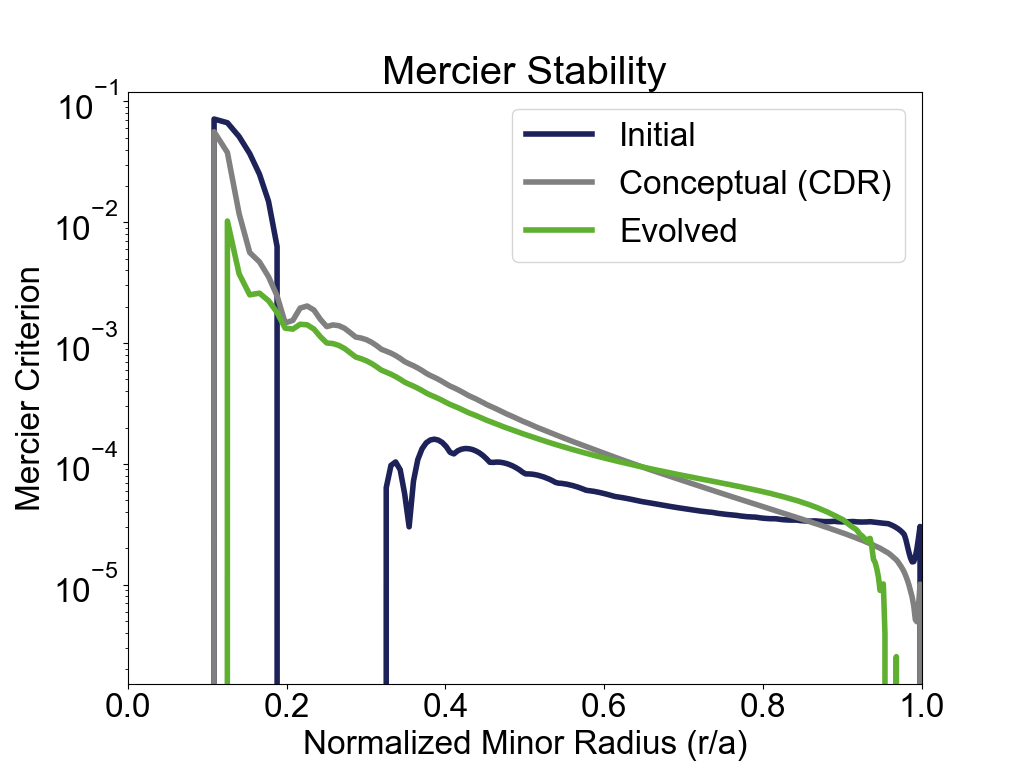}
        \includegraphics[width=0.32\textwidth]{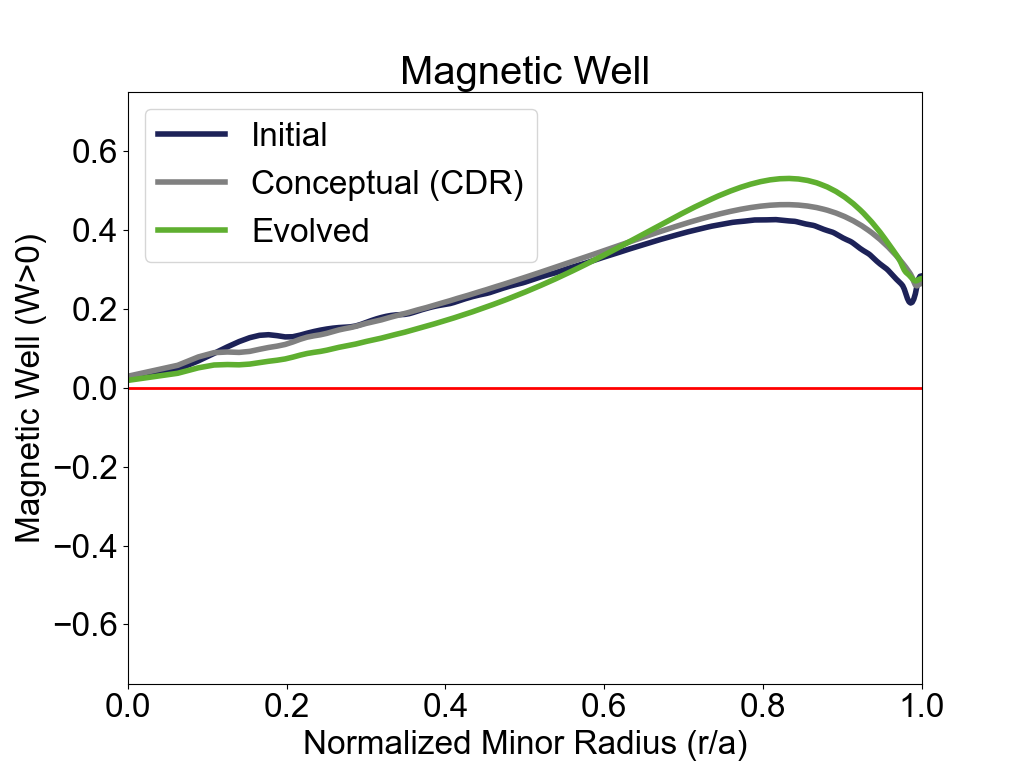}
        \includegraphics[width=0.32\textwidth]{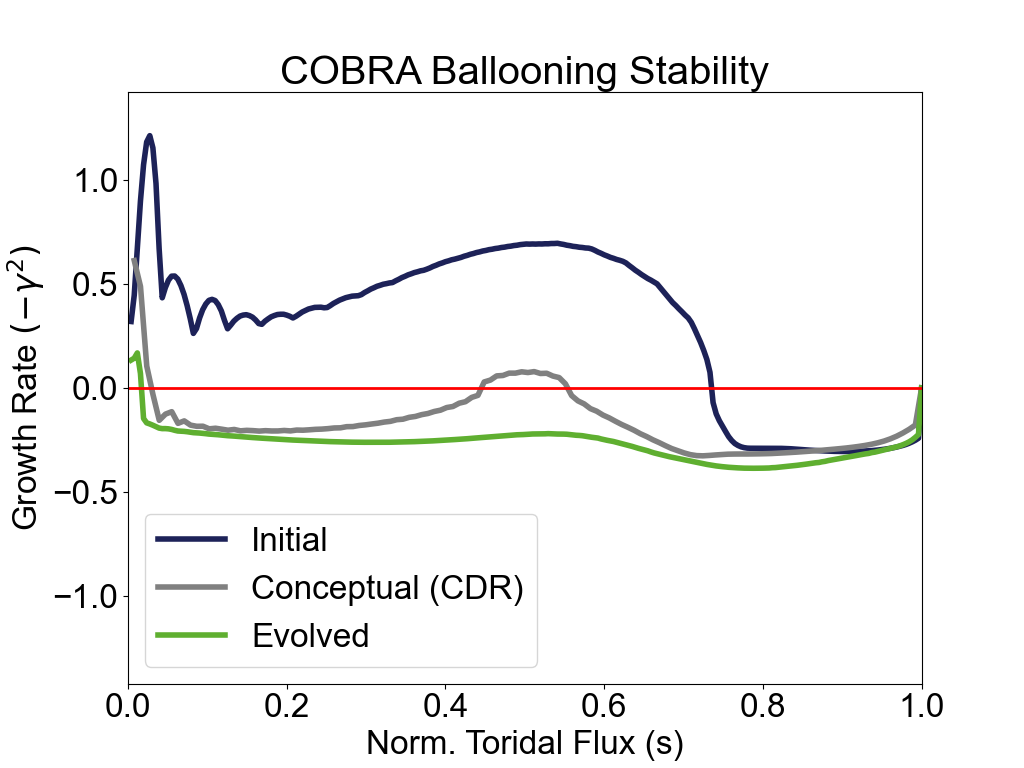}
 \caption{The Mercier criterion (left), Magnetic well/hill condition (center), and infinte-n ballooning stability (right) for the initial, conceptual, and evolved GIGA equilibria. Positive values indicate magnetic well and Mercier stability, while negative values imply ballooning stability.}
\label{fig:stability}
\end{figure}

The stability of the GIGA equilibria were evaluated through a combination of metrics (figure \ref{fig:stability}). 
The interchange stability is characterized by the Mercier criterion which is positive across most radii for all three equilibria.  
Here in all but the deep core, the evolved configuration is found to be stable. 
The full magnetic well condition was found for all radii across all configurations.  
While not explicitly targeted, a magnetic hill was never found at any point during the optimization.  
This suggests that such a property is fundamental to the general 3D shape of the plasma.  
The infinite-n stability (a directly optimized quantity) was evaluated using the COBRAVMEC code.  
The optimization of the conceptual design equilibrium targeted discrete radial points.  
This allowed a small region toward the plasma edge to become marginally unstable. 
A subsequent increase in the number of targeted surfaces improved this feature in the evolved configuration.
A scan of flux tubes at multiple radial locations indicated that the most unstable tube was always the one located at the outboard mid-plane of the bean shaped cross section.
This is consistent with arguments regarding bad curvature and ballooning stability.

Finite toroidal mode stability was evaluated with the TERPSICHORE code. 
The $n=0$, $n=1$, and $n=2$ mode families were investigated.
The Boozer mode spectrum considered extended from $m=[0,32]$ and $n=[-16,16]$.
The full poloidal spectrum was considered for all toroidal modes in a given family.
Generally it was found that inclusion of toroidal current in the equilibrium was destabilizing.
For the $n=0$ mode family such destabilization was core localized while the $n=1$ and $n=2$ mode families tended to have an edge character when destabilized.
Scans of the effect of toroidal current were performed for the conceptual design equilibrium suggesting that reduction in bootstrap could stabilize the plasma.
From this study it was suggested that lowering bootstrap current to 50 kA would be sufficient to stabilize the plasma.
Analysis of the evolved equilibria confirmed this, as it was found that the $n=0$, $n=1$, and $n=2$ mode families were stable.

In addition to ideal MHD stability, the Alfv\'enic stability of the plasma was considered through the application of the STELLGAP code. While STELLGAP does not directly compute the stability of Alfv\'en modes, it does compute the gap mode structure where such modes can exist and live. A lack of core-edge gaps implies that should modes be destabilized, they will be radially localized and not cause large core to edge transport. This was found to be the case for all three of the equilibria considered here. This is attributed to the larger toroidal modulation of magnetic field and the general behavior of stellarators to not have such strong core-edge gaps as compared to tokamaks \cite{kramer_mitigation_2016}. 

\section{Qualification Plan} \label{sec:qualiplan}
The results of the previous section provide a strong theoretical basis for the notion that the GIGA evolved equilibrium will enable the GIGA power plant to produce 1 GWe. However, one may ask how well any of the values predicted in this work can be trusted. In the context of systems engineering the concept of technical readiness levels (TRLs) is of great value. The TRL of a given system both helps us understand the maturity of the system and what steps are necessary to further raise the TRL of any given system. We will apply the concept of TRLs to each of the GIGA requirements to both understand what has been achieved by the scientific community at large and to define how a given plasma physics quantity can be further qualified.

\begin{table}
\caption{Technical readiness levels (TRLs) with proposed definitions as they apply to physical theories.}
\centering
\resizebox{\columnwidth}{!}{
\begin{tabular}{ r | c | l | l }
  & TRL & Engineering Definition & Physics Definition \\
\hline
\multirow{ 3}{*}{Power Plant} & 9 & System proven in operational environment & Proven in power plant \\
 & 8 & System complete and qualified & Qualified in alpha environment \\
 & 7 & Prototype demonstrated in operational environment & Demonstrated in alpha environment \\
 \hline
\multirow{ 3}{*}{Institutional} & 6 & Demonstrated in industrial environment & Demonstrated in integrated system (stellarator) \\
 & 5 & Validated in industrial environment & Controlled in integrated system (stellarator)  \\
 & 4 & Validated in lab & Controlled in experiment \\
 \hline
\multirow{ 3}{*}{Academic} & 3 & Experimental Proof of Concept & Measured in experiment \\
 & 2 & Technology Concept Formulated & Calculations performed \\
 & 1 & Basic Principles Observed  & Fundamental theory \\
\end{tabular}
}
\label{tab:trl}
\end{table}

Before beginning a discussion on this point it is worthwhile to have a short discussion on the definition of TRLs and how to relate the terminology of TRLs to plasma physics.  
Table \ref{tab:trl} provides an overview of the standard TRL definitions and our proposed definitions with respect to physical theories as they apply to fusion energy. 
This chart naturally divides into three regions. 
First from TRL 1 to 3 we see a pen and paper theory being developed, calculated and the physical quantity being measured in experiment. 
We should note that such a measurement does not require an integrated system like a stellarator or tokamak, but simply that the physically predicted phenomena have been measured. 
Of course, as is often the case, a phenomenon may be measured well before any theory has been formulated to predict the said phenomenon. 
Going from TRL 4 to TRL 6 we move from prediction to control in an integrated system, such as a stellarator or tokamak. 
From TRL 7 to TRL 9 we require the operational environment of the power plant, implying a predominantly alpha heated plasma.
As data from the TFTR and JET experiments is of limited value to stellarator theory, TRL 6 is the maximum achievable TRL for most plasma physics phenomena, in modern experiments. 
It should be noted that neither of the aforementioned tokamak experiments had high enough alpha particle populations to meet the `operational environment' condition.
 
Before moving on we note the importance of the operational environment by drawing the readers attention back to figure \ref{fig:POPCON} and the Cordey pass point. 
This is the point of maximum required ECRH on the optimal path to burn, requiring the least amount of installed auxiliary heating. 
It should be noted that this point does not occur at $Q=P_{fusion}/P_{aux}=1$ but rather between values of 15 and 20. 
Thus we argue that while press worthy, $Q=1$ is not enough to achieve TRL 7 when it comes to plasma physics.
One can argue that in a smaller machine, a $Q=5$ may achieve TRL 7 as it implies $P_{aux}=P_{alpha}$.
More importantly, it is possible in such a device for the fast ion beta to be equivalent to that of the larger power plant at full power.
We can now turn to a discussion about the TRLs of our physical requirements.

The fundamental theory of ideal magnetohydrodynamics is a well-tested and demonstrated theory. 
It is safe to assume that in the context of using a stellarator to generate fusion energy the ideal MHD equilibrium has reached TRL6. 
This is not to say that it is the applicable theory. 
In particular, it shields core islands explicitly through its formulation. 
This assumption may or may not be reasonable given the transport physics and the presence of local shielding currents and flows. 
The W7-X experiment provides a unique environment to explore this phenomenon in a controlled manner \cite{lazerson_fast_2024,kulla_effect_2026}.

The theory of energetic particle orbits in stellarators is quite advanced with collisionless, collisional, gyro center and gyro orbit calculations present. Experimentally, the lack of D-T stellarator experiments requires that proxy processes be used. These processes include neutral beam injection, ion-cyclotron resonance heating, and fast tritium burn-up. These mechanisms can provide us with a wealth of data for testing our models, but unfortunately never fill the same phase space as that of fusion alphas. Moreover, quantitative agreement between model predictions and experimental data is difficult to achieve as such measures require significant levels of integrated analysis on relatively small signals. Thus we argue that the TRL of alpha confinement is around TRL 5. We believe that should the W7-X experiment achieve its goal of demonstrating the improvement in fast ion confinement with increasing plasma beta, the TRL could advance to 6 \cite{lazerson_optemist_2024}.

The phenomenon of neoclassical transport and its effect on particle transport, heat transport, bootstrap current, and the radial electric field are sufficiently well advanced. Significant work on LHD has demonstrated the relationship between $\epsilon_{eff}^{3/2}$ and neoclassical transport. Furthermore, HSX and W7-X experiments have demonstrated that one can optimize configurations for low neoclassical transport. The achievement and prediction of core electron root conditions in stellarators has also been well studied.  The relationship between stellarator shaping and the minimization of bootstrap current has similarly been well established. For this reason neoclassical phenomena can be considered to have reached TRL 6.

Turbulence in stellarators has reached TRL 4 as numerical predictions of stellarator turbulence have only recently been possible.
The phenomenon of stellarator turbulence has been measured for a long time. 
The first predictions that stellarator shapes could be used to limit turbulence were made over 15 years ago.
To date, there has been no experiment to demonstrate turbulence optimization, as has been done for neoclassical optimization.
Recent work on W7-X shows some preliminary evidence of magnetic configuration affecting turbulent fluctuations levels \cite{bahner_magnetic_2025,banon_navarro_global_2026}.
Changes in rotational transform, associated with strong shaping changes in W7-X, do show correlated trends in turbulence both experimentally and theoretically.
However, in the same work changes in the magnetic field mirror term do not agree between experiment and theory.
Limits in the shape space of W7-X, imply that validation of turbulence optimization proxies may not be possible in that device.
Moreover, while fluctuations measurement can be used to measure turbulence, it is the change in the density and temperature profiles which provide a measurement of turbulent heat fluxes, a clearly integrated modeling problem.
The CFQS device may be able to provide data to advance the TRL of stellarator turbulence, however this device is early in its operational life, with such results most likely a few years away.  
Additionally, one could propose a new saddle coil set for W7-X to achieve shaping parameters predicted to reduce turbulence in the device. However, such ideas are costly and not straightforward to implement.

The stability of stellarator plasmas has received much attention in the form of both experimental and theoretical work, as plasma stability could ultimately limit the operational space. Experimental stability limits are generally considered `soft,' with mode activity being present at stability thresholds but ultimately benign \cite{zhou_benign_2024}. The W7-AS device has shown that in some cases beta limits are not limits at all but rather transient phenomena which can be exceeded, resulting in more quiescent plasma conditions \cite{weller_significance_2006}. The LHD device has demonstrated that beta limits which are not exceeded can result in losses of confinement but do not result in catastrophic phenomena (such as disruptions in tokamaks). The W7-X device has shown that radiative collapses brought on by density limits can even be recovered from through continual application of ECRH \cite{pandey_stable_2025}. From the theoretical side a multitude of predictive models for various instabilities have been developed for stellarators. However, validation has been difficult given the benign nature of the instabilities in stellarators. Most validation of such stability calculations has been achieved by applying the tools to tokamak plasmas. For these reasons, the TRL of stability is considered to be at a level of 5.

The existing stellarator experiments, worldwide, should be sufficient for achieving TRL 6 with regard to physical phenomena. 
It was argued during the conceptual design of GIGA that the power plant, through staged operation, would bring all technologies from TRL 6 through TRL 9. 
Such staged operation envisioned a non-nuclear phase, a breeding demonstration phase, and finally power operation. 
The breeding demonstration phase would not require tritium self-sufficiency while the power operation phase would.
While the most direct path to a power plant, such a development plan would preclude much scientific operation.
An alternative is now being considered, where a smaller industrial demonstrator with a $Q=5$ mission would seek to achieve TRL 7.
This could reduce the demonstrator burden on GIGA and has the potential to accelerate the development of a fusion power plant.

\section{Conclusion} \label{sec:conclusions}
The fixed boundary stellarator equilibrium basis for the GIGA power plant producing 1 GWe has been presented using a systems engineering framework.  
A plasma with $1500~\textrm{m}^3$ volume, major radius of $20$ m, minor radius of $\sim2$ m, peak on-axis magnetic field strength of 6 T, and edge rotational transform slightly less than unity has been developed.
The plasma optimization leveraged the STELLOPT code to provide multi-objective optimization of a W7-X high-iota and high mirror magnetic configuration.
The configuration was modified to meet the field periodicity, plasma volume, and field strength requirements of GIGA.
The requirements for the GIGA optimized equilibria were captured through a set of qualitative and quantitative measures which define the success of a given optimization.
This included rotational transform, total bootstrap current, neoclassical transport, fast ion confinement, radial electric field, MHD stability and turbulent transport.
Through successive optimizations a conceptual design plasma (GIGA\_v515) and evolved equilibrium (GIGA\_v549) were produced.
The conceptual design plasma met many of the requirements but had unacceptably high bootstrap current, resulting in mode destabilization, along with marginal alpha confinement.
This motivated the development of the evolved equilibrium which now serves as the equilibrium basis for GIGA.

The minimization of bootstrap current performed in this work is preliminary as free boundary effects may change the results.
It was found during optimization that inclusion of the predicted bootstrap in the equilibrium would change the equilibrium enough to modify the said prediction.
The minimization of bootstrap required some level of current-equilibrium self-consistency to achieve a minimized bootstrap current.
The fixed boundary nature of the equilibrium implies a vertical field is present to balance the hoop force generated by the net toroidal current.
While this force may be small, as the device is of large aspect ratio in nature, it could still result in additional changes to the geometry.
For this reason there may be a need to perform and integrated coil-plasma optimization once an initial coil-set is established.

Recent advances in superconducting technology are allowing for devices to be proposed with significantly higher magnetic fields than the 6 T GIGA device proposed here. 
This is advantageous as plasma confinement scales positively ($\propto B^{0.84}$) with magnetic field strength. 
Thus a smaller device could be envisioned without compromise in the energy confinement time.
For example, going from 6 T to 9 T would allow for a device with a 14\% smaller minor radius (or 40\% smaller major radius).
However, the volume (and therefore power) scales linearly with the major radius, and as the square of the minor radius.
The smaller machine would now produce less fusion power assuming the same profiles are achieved.
Assuming higher plasma performance could recover the power loss, but then the reduced surface area would imply higher wall loads.
Additionally, reductions in stellarator size usually come with added complexity due to a more compact build on the inboard side of the device.
Therefore, going to higher magnetic fields in the plasma does not necessarily solve any fundamental design issue of stellarators when considering the challenges of a first of a kind reactor design.
Nor does it reduce overall cost when assuming that power output is a fixed requirement.

With a fixed boundary equilibrium defined for GIGA many parallel avenues of development are underway. 
The preliminary radial build is allowing for nuclear modeling helping to gauge breeding and shielding needs.
Development of a coil which reproduces the fixed boundary equilibrium has also begun.
Integrated design of the divertor-coil-plasma system is undertaken in parallel.
In particular the question of edge island optimization while targeting a fixed boundary equilibrium is being explored.
Thus the fixed boundary equilibrium presented here lays the foundation for a significant amount of future works.

\ack{The authors would like to acknowledge the entire Gauss Fusion Team in their role supporting this work.
The authors would like to thank EPFL for access to the TERPSICHORE code and the entire fusion community for their efforts to open-source the tools used in this work and their continued development and maintenance.
The Qarnot HPC team \cite{QarnotCloudHPC} is thanked for providing all the computational architecture upon which this work was performed.}





\bibliographystyle{unsrtnat}
\bibliography{Lazerson_ISHW2026}

\end{document}